\documentstyle[emulateapj,psfig]{article}
\makeatletter

\newenvironment{inlinefigure}{%
\def\@captype{figure}%
\noindent\begin{minipage}{0.999\linewidth}\begin{center}}
{\end{center}\end{minipage}\smallskip}
\makeatother

\lefthead{Cowie, Barger, \& Hu}

\newcommand{\oiii}{[O\thinspace{III}]}

\begin{document}
\title{Ly$\alpha$ emitting galaxies as early stages in galaxy formation\altaffilmark{1,2,3}}
\author{
Lennox~L.~Cowie,$\!$\altaffilmark{4} 
Amy~J.~Barger,$\!$\altaffilmark{5,6,4}
Esther~M.~Hu$\!$\altaffilmark{4}
}

\altaffiltext{1}{Based in part on data obtained from the Multimission
Archive at the Space Telescope Science Institute (MAST).  STScI is
operated by the Association of Universities for Research in Astronomy,
Inc., under NASA contract NAS5-26555.  Support for MAST for non-HST
data is provided by the NASA Office of Space Science via grant
NAG5-7584 and by other grants and contracts.}
\altaffiltext{2}{Based in part on data obtained at the W. M. Keck
Observatory, which is operated as a scientific partnership among the
the California Institute of Technology, the University of
California, and NASA and was made possible by the generous financial
support of the W. M. Keck Foundation.}
\altaffiltext{3}{This research used the facilities of the Canadian Astronomy 
Data Centre operated by the National Research Council of 
Canada with the support of the Canadian Space Agency.}
\altaffiltext{4}{Institute for Astronomy, University of Hawaii,
2680 Woodlawn Drive, Honolulu, HI 96822.}
\altaffiltext{5}{Department of Astronomy, University of
Wisconsin-Madison, 475 North Charter Street, Madison, WI 53706.}
\altaffiltext{6}{Department of Physics and Astronomy,
University of Hawaii, 2505 Correa Road, Honolulu, HI 96822.}



\begin{abstract}
We present optical spectroscopy of two samples
of {\em GALEX\/} grism selected Ly$\alpha$ emitters (LAEs):  
one at $z=0.195-0.44$ and the other at $z=0.65-1.25$.
We have also observed a comparison sample of galaxies
in the same redshift intervals with the same UV magnitude distributions 
but with no detected Ly$\alpha$.  
We use the optical 
spectroscopy to eliminate active galactic nuclei (AGNs) and to 
obtain the optical emission-line properties of the 
samples. 
We compare the luminosities of the LAEs in the two 
redshift intervals and show that there is dramatic evolution 
in the maximum Ly$\alpha$ luminosity over $z=0-1$. 
Focusing on the $z=0.195-0.44$ samples alone, we show that there 
are tightly defined relations between all of the galaxy parameters 
and the rest-frame equivalent width (EW) of H$\alpha$. The 
higher EW(H$\alpha$) sources all have lower metallicities,
bluer colors, smaller sizes, and less extinction, consistent
with their being in the early stages of the galaxy formation
process. We find that $75\pm12\%$ of the LAEs
have EW(H$\alpha$)$>100$~\AA, and, conversely, that $31\pm13\%$ 
of galaxies with EW(H$\alpha$)$>100$~\AA\ are LAEs.
We correct the broadband magnitudes for
the emission-line contributions and use spectral
synthesis fits to estimate
the ages of the galaxies. We find a median age of $1.1\times10^{8}$~yr
for the LAE sample and $1.4\times10^{9}$~yr for the UV-continuum
sample without detected Ly$\alpha$. The median
metallicity of the LAE sample is $12+\log$(O/H)=8.24, or
about 0.4~dex lower than the UV-continuum sample. 
\end{abstract}


\section{Introduction}
\label{secintro}

Ly$\alpha$ emission-line searches have been widely used to find 
high-redshift galaxies and, for the highest redshift galaxies, this 
line is the only spectroscopic signature that can be used to confirm 
the redshift of a galaxy selected on the basis of its color properties. 
However, Ly$\alpha$ is a difficult line to interpret.
Because the line is resonantly scattered 
by neutral hydrogen, determining its escape path and 
hence its dust destruction is an extremely complex problem, 
both theoretically (e.g., Neufeld 1991; Finkelstein et al.\ 2007)
and observationally (e.g., Kunth et al.\ 2003; Schaerer \& Verhamme 2008;
\"{O}stlin et al.\ 2009).  
Thus, while we have empirical measurements that a significant fraction of
UV-continuum selected samples have Ly$\alpha$ lines
with rest-frame equivalent widths above 20~\AA\
over a wide range of redshifts from $z=0.3$ to $z=6.5$ 
(Shapley et al.\ 2003; Cowie et al.\ 2010, 2011; Stark et al.\ 2010),
our understanding of what determines this fraction is still weak. 
In particular, we would like to know whether the presence of Ly$\alpha$
emission is related to other properties of the galaxy, such as its
metallicity, extinction, morphology, or kinematics, 
and how the Ly$\alpha$ line escapes.

Due to the difficulty with observing in the UV, we currently have much more
information on the $z\sim2-3$ Ly$\alpha$ emitters (LAEs) and how their 
properties relate to those of other UV selected galaxies at these redshifts
(e.g., Shapley et al.\ 2003; Reddy et al.\ 2010; Kornei et al\ 2010) 
than we do on the local samples.  However, there has been considerable controversy 
in the interpretation of these high-redshift observations. The simplest 
interpretation is that the LAEs are younger, lower mass, and metal poor, 
representing early stages in galaxy evolution (e.g., Hu et al.\ 1998; 
Nilsson et al.\ 2007; Gawiser et al.\ 2007). 
However, other authors have argued that LAEs arise in 
relatively massive galaxies (e.g., Lai et al.\ 2008; Finkelstein et al.\ 2009c) 
with ages of around a Gyr, that young galaxies have weaker Ly$\alpha$ than old 
galaxies (Shapley et al.\ 2001), and, more recently, that LAEs are older, less 
dusty, and in a later stage of galaxy evolution than sources with weaker 
Ly$\alpha$ emission (Kornei et al.\ 2010).

The Ly$\alpha$ signature can be produced by a range of galaxy types and 
masses, including even the most ultraluminous infrared galaxies (ULIRGs) 
(e.g., Chapman et al.\ 2005; Nilsson \& M{\o}ller 2009), so some level 
of heterogeneity must be expected.  However, all of the results that 
argue for the LAEs being predominantly old are based on spectral synthesis 
fitting, and most of the old ages inferred are almost certainly mis-estimates
arising from the presence of very strong optical emission lines in the 
LAEs (e.g., Schaerer \& deBarros 2009).

Until recently the only low-redshift Ly$\alpha$ emitting sources that
could be studied in detail were the local blue compact galaxies.
However, these generally have much lower Ly$\alpha$ luminosities than
the high-redshift LAEs, and, while some of the blue compact galaxies have
been studied in exquisite detail (e.g., \"{O}stlin et al.\ 2009) on an 
individual basis, it has not been easy to form large, uniformly selected
samples that can be statistically analyzed. Thus, the recent determination
that substantial $z\sim0.2-0.4$ samples of LAEs can be found 
(Deharveng et al.\ 2008) with the {\em Galaxy Evolution Explorer (GALEX)\/} 
(Martin et al.\ 2005) grism spectrographs has
enabled a new approach to the subject (Atek et al.\ 2009a;
Finkelstein et al.\ 2009a, 2009b; Scarlata et al.\ 2009; Cowie et al.\ 2010).

The low-redshift LAE samples have many
advantages. The galaxies are bright and can be easily
studied at other wavelengths, but perhaps even more
importantly, they can be integrated into comprehensive
studies of galaxies at the same redshifts to understand the
selection biases intrinsic to the samples.
Early papers on {\em GALEX\/} LAEs worked with relatively small samples, 
but the general conclusions are that low-redshift LAEs are somewhat 
heterogeneous yet more weighted to low metallicities and extinctions
and more likely to be small, compact galaxies when compared to
UV-continuum selected galaxies without detected Ly$\alpha$ with the 
same luminosities in the same redshift interval. 

Here we study larger and more optically spectroscopically complete 
samples of LAEs at $z=0.195-0.44$ and $z=0.65-1.25$, together with 
comparison samples of UV-continuum selected galaxies without detected 
Ly$\alpha$ in the same redshift intervals.
In Section~\ref{secopt} we present our optical spectroscopy of
all the samples obtained with the DEep Imaging Multi-Object Spectrograph 
(DEIMOS; Faber et al.\ 2003) on the Keck~II 10~m telescope. 
In Section~\ref{secagngal} we use the optical data to remove the small 
number of active galactic nuclei (AGNs) that were not previously 
identified from the UV spectra and then provide our final sample of
candidate LAE galaxies in the two redshift intervals. We also use 
the spectra to measure the optical line fluxes and equivalent
widths and to determine the metallicities of the galaxies.
In Section~\ref{lae_lum_evol} we determine how the Ly$\alpha$
luminosity evolves with redshift. 
In Section~\ref{la_metal} we consider the overall properties 
of the LAEs and the UV-continuum selected galaxies without detected
Ly$\alpha$. 
Since the sample of $z\sim1$ LAEs is small (only eight objects) 
we only consider the $z=0.195-0.44$ sample in this section.
We use the spectra to determine
the emission-line contributions to the broadband
fluxes and to show that these corrections must be included
if spectral synthesis fitting is to give accurate
ages for the youngest ($<10^{9}$ yr) galaxies. 
In Section~\ref{secint} we interpret the results in terms of a constant 
star formation rate (SFR) model, comparing SFRs derived from the H$\alpha$, 
UV-continuum, and 20~cm observations and discussing the limits of 
validity for the UV-continuum determined SFRs. We also interpret the
metallicity evolution.
We summarize our results in Section~\ref{secdisc}. 
We use a standard $H_{0}$ = 70~km~s$^{-1}$~Mpc$^{-1}$, 
$\Omega_{\rm M}$ = 0.3, $\Omega_{\Lambda}$ = 0.7 
cosmology.


\section{The {\em GALEX\/} Sample}
\label{secopt}

For the present study we use the LAE samples with rest-frame
EW(Ly$\alpha$)$>15$~\AA\ from 
Cowie et al.\ (2010; their Tables~15 and 16), chosen from the nine 
blank high galactic latitude fields with the deepest {\em GALEX\/} 
grism spectroscopic observations (Morrisey et al.\ 2007).
The area covered is just over 8~deg$^2$. The samples consist of
sources whose grism UV spectra have a detectable single emission 
line, which we assume to be Ly$\alpha$. In order to eliminate
sources that are clearly AGNs based on their UV spectra, we do
not include any sources that have high-ionization lines, and we
require the FWHM of the Ly$\alpha$ lines to be less than 15~\AA\ 
in the {\em GALEX\/} FUV spectra ($z=0.195-0.44$) or less than 
30~\AA\ in the {\em GALEX\/} NUV spectra ($z=0.65-1.25$). 
In the redshift interval $z=0.65-1.25$ we also eliminate sources
that do not show the break between the FUV and NUV bands. Such a
break would be expected to be produced by the Lyman continuum
edge. (See Section~4 of Cowie et al.\ 2010 for details on 
constructing the samples.)
For the CDFS~00 field we have added additional sources
from a deeper grism spectroscopic exposure than was used in
Cowie et al.\ (2010). The full Ly$\alpha$
selected sample in the CDFS~00 is given in the Appendix 
(i.e., this is an update of Table~4 in Cowie et al.\ 2010, but
it also includes the optical redshifts, where available; see
Section~\ref{secspec}).

We summarize the final candidate LAE galaxy sample in the redshift 
interval $z=0.195-0.44$ in Table~\ref{tab1}, sorted 
by the rest-frame EW(Ly$\alpha$). 
Through most of the paper we restrict 
our analysis to sources with rest-frame EW(Ly$\alpha)\ge20$~\AA, so
we list these sources in the main body of the table.
However, for completeness, we also give the results for 
the lower EW(Ly$\alpha$) sources in a supplement to the table.
For each source in Table~\ref{tab1} we give the {\em GALEX\/} 
name, the J2000 right ascension and declination, the 
redshift from the {\em GALEX\/} UV spectrum, 
the logarithm of the bolometric continuum luminosity, $L_{\nu}\nu$, above 
1216~\AA, the logarithm of the Ly$\alpha$ luminosity, the 
rest-frame EW(Ly$\alpha$), the optical ground-based redshift, 
the rest-frame EW(H$\alpha$), the [NII]$\lambda6584$/H$\alpha$, 
[OIII]$\lambda5007$/H$\beta$, and [OIII]$\lambda4363$/H$\gamma$ ratios, 
and the logarithm of the H$\alpha$ and H$\beta$ fluxes for the sources
with optical continuum magnitudes in the Sloan Digital Sky Survey
(SDSS) (see Section~\ref{secother}).

We summarize the final candidate LAE galaxy sample in the redshift
interval $z=0.65-1.25$ in Table~\ref{tab2}, also sorted by the rest-frame
EW(Ly$\alpha$).  There are 8 sources in this sample. We list
the properties of these sources using the same format as in Table~\ref{tab1}, 
except here we give the rest-frame EW(H$\beta$) in place of the rest-frame 
EW(H$\alpha$), and we do not include the column for the H$\alpha$ and
H$\beta$ fluxes. We also do not split the sample into a main body of 
the table and a supplement to the table, since there are so few 
sources. Six of the eight sources satisfy the EW(Ly$\alpha)>20$~\AA\ 
criterion.

Although the $z\sim1$ sample is small, it is important in that it
demonstrates the presence of these LAEs, and it allows us to 
make a first estimate of the 
luminosity function at these redshifts.  However, most of our
analysis in this paper will focus on the $z\sim0.3$ sample.
For each source with $z<0.3$ we computed the UV spectral index
from the {\em GALEX\/} spectra using the Calzetti et al.\ (1994) 
prescription, which fits a power law to the spectrum
in a set of wavelength windows chosen to
minimize the effects of absorption and emission lines.
Since we have only short wavelength data in the 
rest frame, we corrected our measured index with the offset of $-0.16$ 
given by Meurer et al.\ (1999) to provide our final index. We did 
not compute the UV spectral index for the sources with $z>0.3$, since 
at those redshifts many of the shortest wavelength windows in 
Calzetti et al.\ (1994) lie in the noisy regions between the 
{\em GALEX\/} FUV and NUV spectra.

\subsection{Imaging Data}
\label{secother}

We compiled ancillary data from various archival sources. Just over 
$82\%$ of the area is covered by deep $U$-band images obtained with
the MegaPrime camera on the 3.6~m Canada France Hawaii Telescope (CFHT).
We took the reduced images from the CADC 
pipeline reduction, which gives the 5$\sigma$ AB magnitude limits.
These range from 26.5 to 27.5. We used these images to measure the 
$U$-band magnitudes and the FWHM sizes of the galaxies at this 
wavelength. For the fields covered by the SDSS we 
also compiled the $u',g',r',i',$ and $z'$ 
SDSS model C magnitudes ($2\sigma$ AB magnitude limits of 22.0, 22.2,
22.2, 21.3, and 20.5, respectively) 
from the DR6 release
(Adelman-McCarthy et al.\ 2008). These cover
roughly half of the observed targets. For the fields
with publicly available 20~cm images 
(COSMOS~00: Schinnerer et al.\ 2007, $1\sigma$ of $10.5~\mu$Jy~beam$^{-1}$; 
SIRTFFL~01: Condon et al.\ 2003, $1\sigma$ of $23~\mu$Jy~beam$^{-1}$; 
HDFN~00: Morrison et al.\ 2010, $1\sigma$ ranges from $3.9~\mu$Jy~beam$^{-1}$
near the center to $8~\mu$Jy~beam$^{-1}$ at $15'$), 
we measured the 20~cm fluxes from the images following the usual
methods (e.g., Morrison et al.\ 2010).
Finally, for the CDFS~00 targets covered by the 
Galaxy Evolution from Morphologies and SEDs (GEMS)
survey (Rix et al.\ 2004), we compiled thumbnails of the
galaxies in the F606W ($5\sigma$ AB magnitude limit of 28.25) and 
F850LP ($5\sigma$ AB magnitude limit of 27.10) filters from
the {\em Hubble Space Telescope (HST)\/} observations with the 
Advanced Camera for Surveys (ACS).

\subsection{Optical Spectroscopy}
\label{secspec}

We obtained optical spectroscopy for a large fraction of the two 
{\em GALEX\/} LAE samples using the DEIMOS spectrograph on Keck~II
with the ZD600 $\ell$/mm grating
in observing runs throughout 2009 and 2010. 
The $\sim 0.5-1.0$~hr spectra have a resolution of 4.5~\AA\ 
and a wavelength range of approximately 5000~\AA\ centered at 6800~\AA. 
More details of the observing and reduction procedures can be found in 
Cowie et al.\ (1996, 2010).

We observed 77 of the 91 $z=0.195-0.44$ LAEs in Table~\ref{tab1}
(this includes the supplement) and all 8 $z=0.65-1.25$ LAEs in 
Table~\ref{tab2}. 
We give the optical redshifts in Tables~\ref{tab1} 
and \ref{tab2}, where they can be compared with the UV redshifts from 
the {\em GALEX\/} data. For the 77 $z=0.195-0.44$ LAEs with optical
redshifts, we confirmed the UV redshifts for 70.
In Table~\ref{tab1} we show the optical redshifts in parentheses
for the sources where we did not confirm the UV redshifts.
We eliminated these sources from further consideration.
We confirmed the UV redshifts for all 8 $z=0.65-1.25$ LAEs 
with the optical redshifts.  
We show the UV and optical spectra for these sources
in Figures~\ref{uv_spectra} and \ref{optical_spectra}. 

In order to understand how the {\em GALEX\/} LAEs 
relate to other UV-continuum selected sources with the same 
UV-continuum magnitude distributions, we also observed at optical 
wavelengths sources randomly chosen from the {\em GALEX\/} UV 
spectroscopic sample that did not have UV spectral line identifications. 
The latter is not a complete sample
of all such sources in the field, but rather a randomly chosen
subsample with an NUV magnitude selection function that is similar 
to that for the sources with strong Ly$\alpha$ emission. Thus, 
it may be directly compared to the LAE sample. 

We have obtained optical spectra for 450 of
the 5642 objects in the total UV-continuum selected
sample without detected Ly$\alpha$. 114 of these
sources lie in the redshift interval $z=0.195-0.44$, 
and 5 lie in the redshift interval $z=0.65-1.25$.
We will hereafter refer to these sources as our {\em UV-continuum sample}.
These sources may be viewed as analogs of high-redshift Lyman 
break galaxies (LBGs) without strong Ly$\alpha$ emission and 
can be combined with the LAE sample to understand how
Ly$\alpha$ galaxies are related to the more general
UV-continuum selected population.
In order to compare the numbers of UV-continuum sources with a given
property with the LAE sample we need to correct for the fraction
of UV-continuum sources that were observed.
There is a weak dependence of the observed fraction on the 
NUV magnitude, ranging from $6\%$ observed at NUV=20 to $11\%$ 
observed at NUV=22. We use the
inverse of this fraction as a function of the NUV magnitude
to provide the weighting to the UV-continuum sample.

The new optical spectroscopic data presented in this paper
more than doubles the optical spectroscopic data used in 
Cowie et al.\ (2010). 
There we observed with DEIMOS (including four literature redshifts) 
34 of the 62 $z=0.195-0.44$ LAEs given in their Table 15 (this also 
includes EW(Ly$\alpha)< 20$~\AA\ sources) and 1 of the 4
$z=0.65-1.25$ LAEs given in their Table 16.  We
also observed with DEIMOS 124 UV-continuum selected sources
without detected Ly$\alpha$, of which 46 were in the redshift
interval $z=0.195-0.44$ and none were in the redshift interval
$z=0.65-1.25$.

\begin{inlinefigure}
\includegraphics[width=3.0in,angle=0,scale=1]{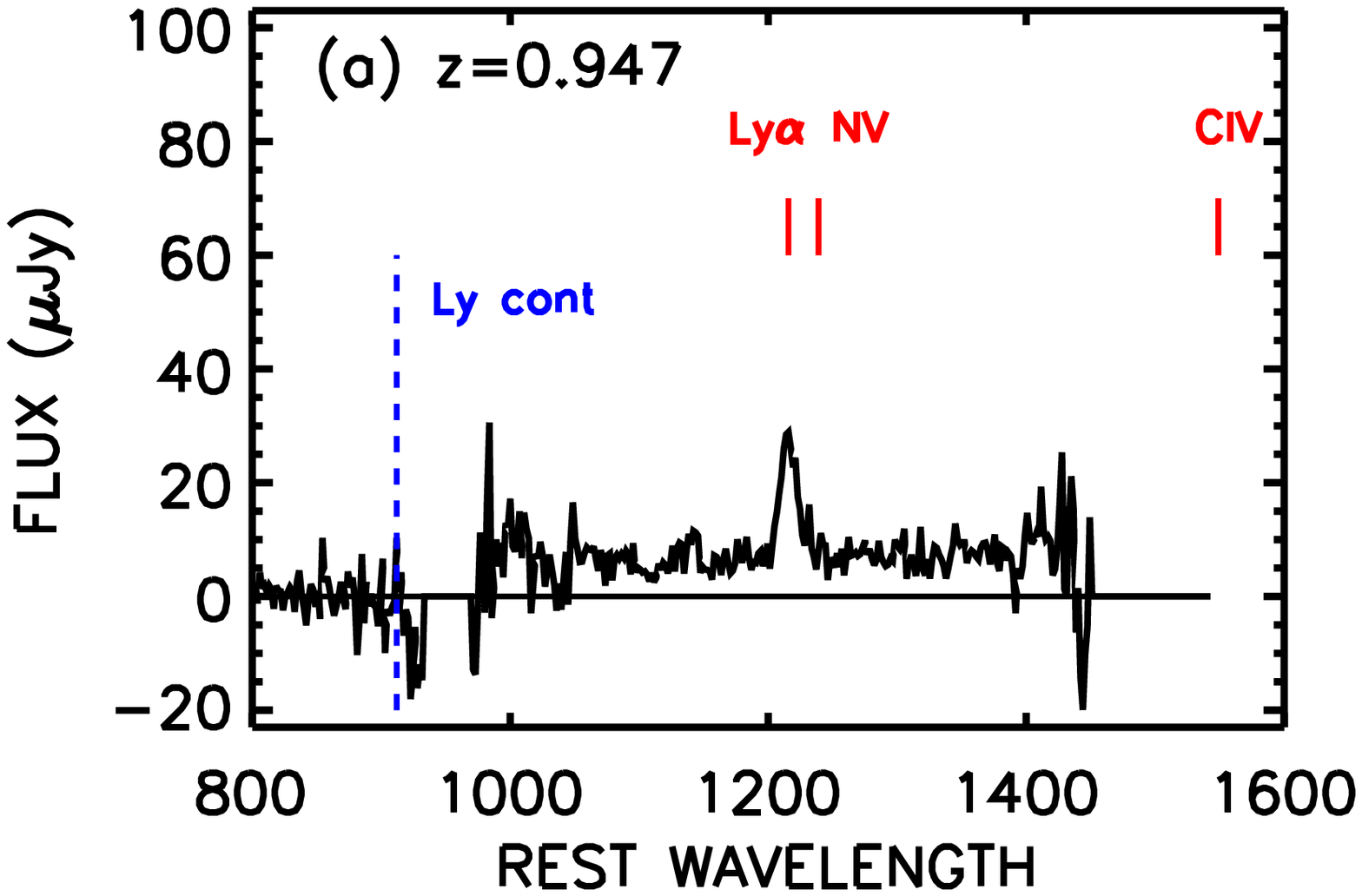}
\includegraphics[width=3.0in,angle=0,scale=1]{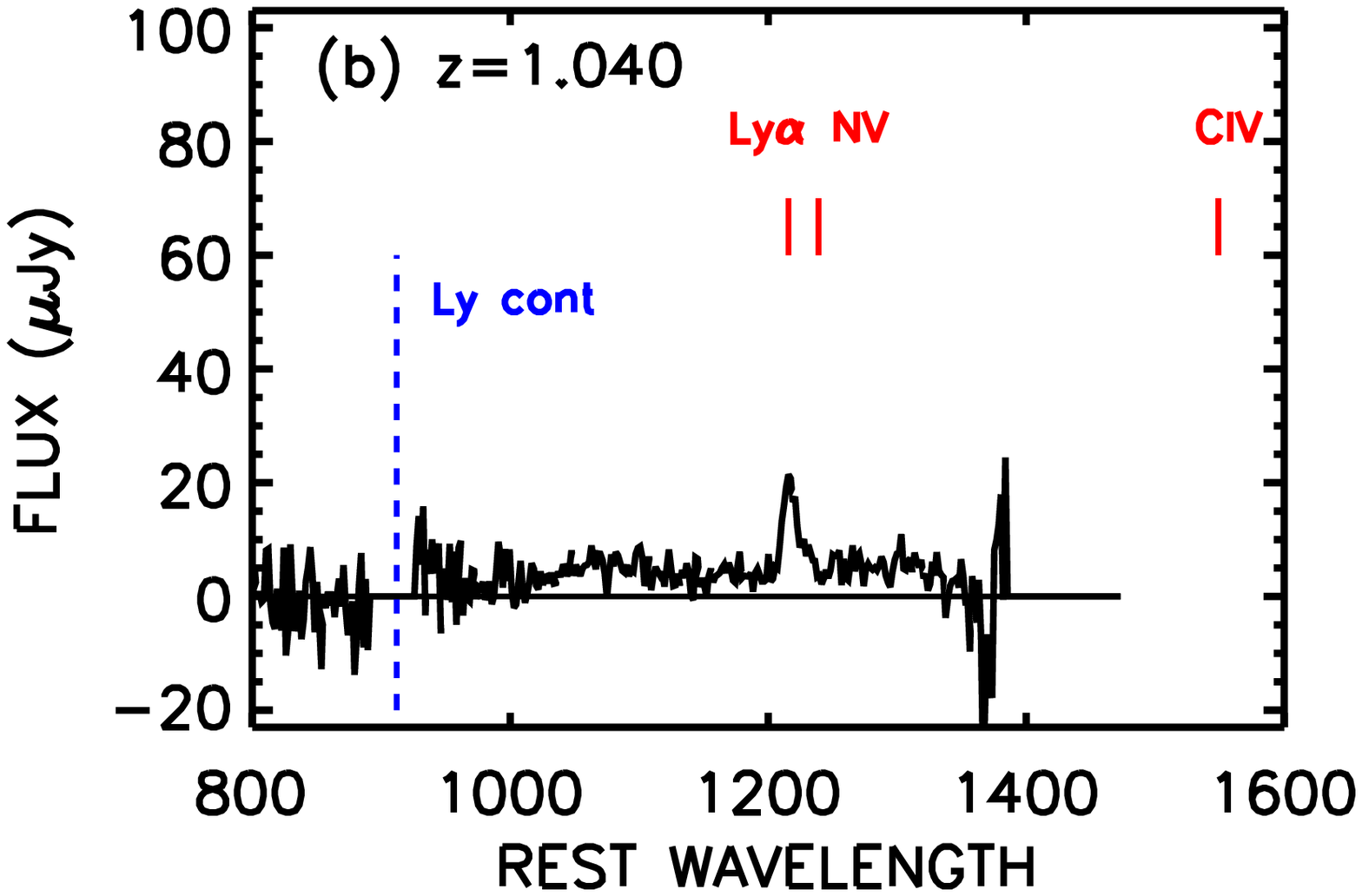}
\includegraphics[width=3.0in,angle=0,scale=1]{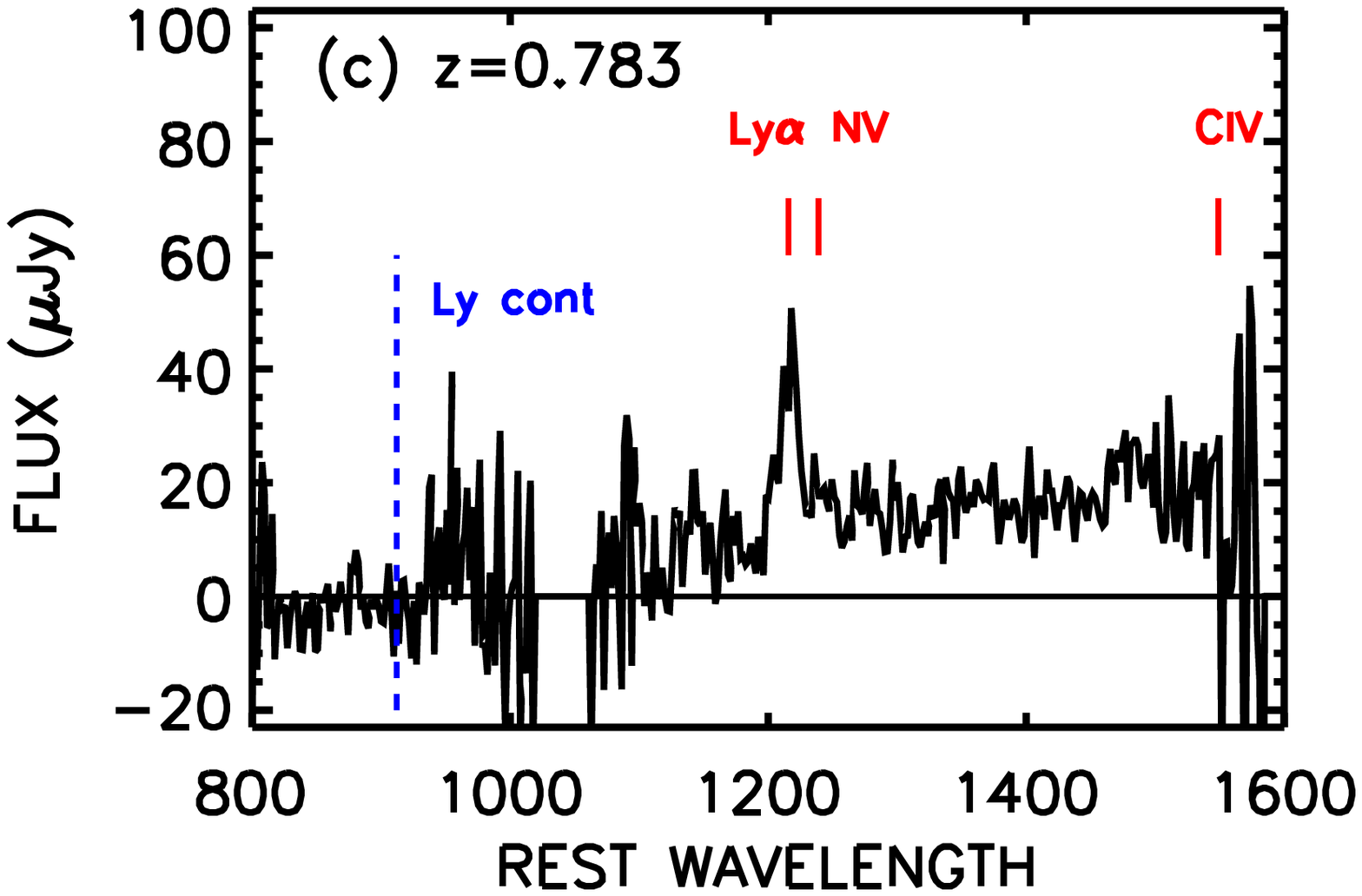}
\includegraphics[width=3.0in,angle=0,scale=1]{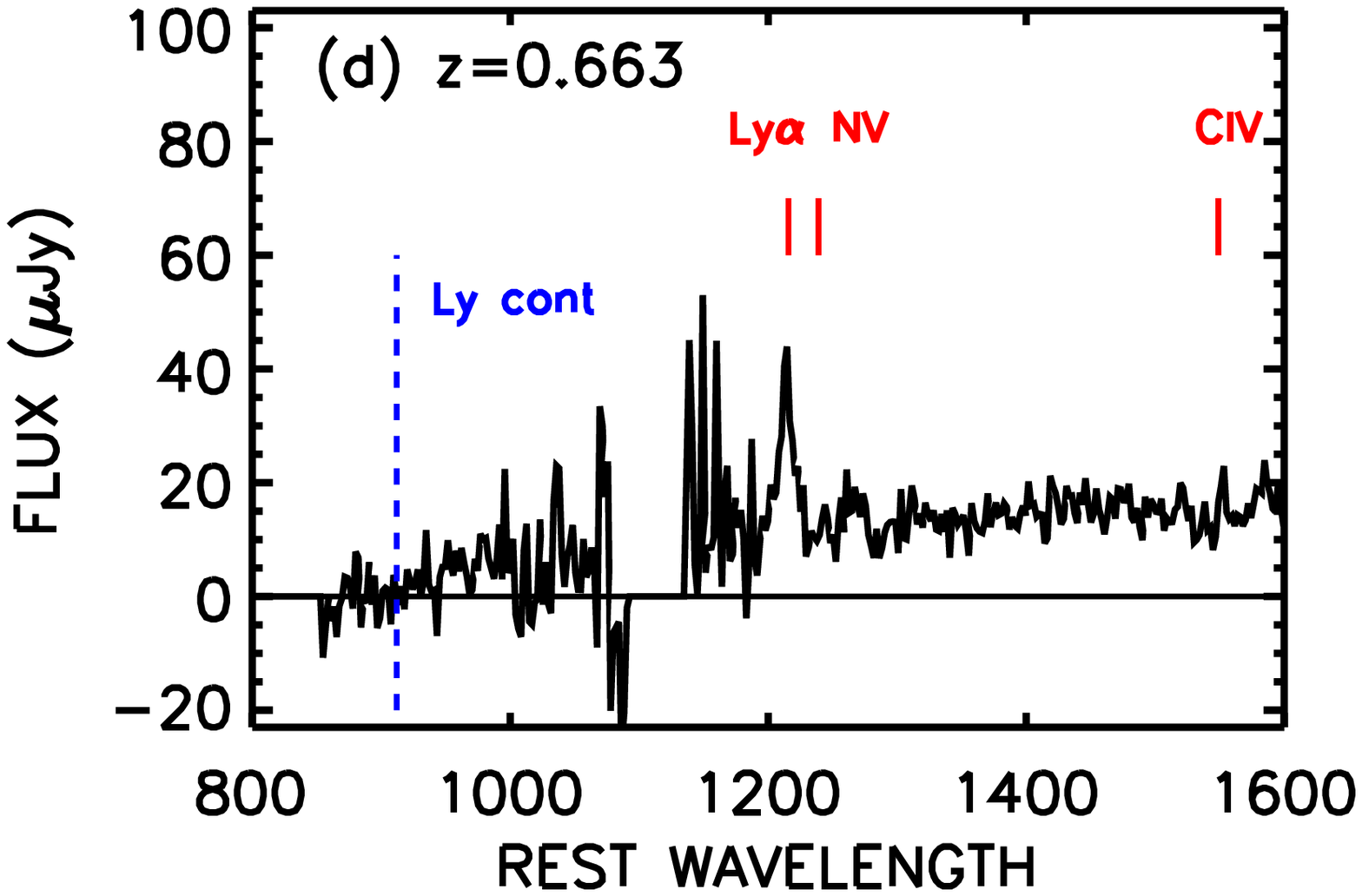}
\caption{{\em GALEX\/} spectra of the first four $z=0.65-1.25$ 
LAEs in Table~\ref{tab2}. 
For each source we give the redshift measured from the UV spectrum.
The Ly$\alpha$ line is marked, as are the positions where CIV and NV 
would fall. The blue vertical dashed line shows the position of 
the Lyman continuum edge at the galaxy redshift. 
This is generally slightly redward of the more accurate optical 
redshift (see Cowie et al.\ 2010).
\label{uv_spectra}
}
\end{inlinefigure}

\begin{inlinefigure}
\includegraphics[width=3.0in,angle=0,scale=1]{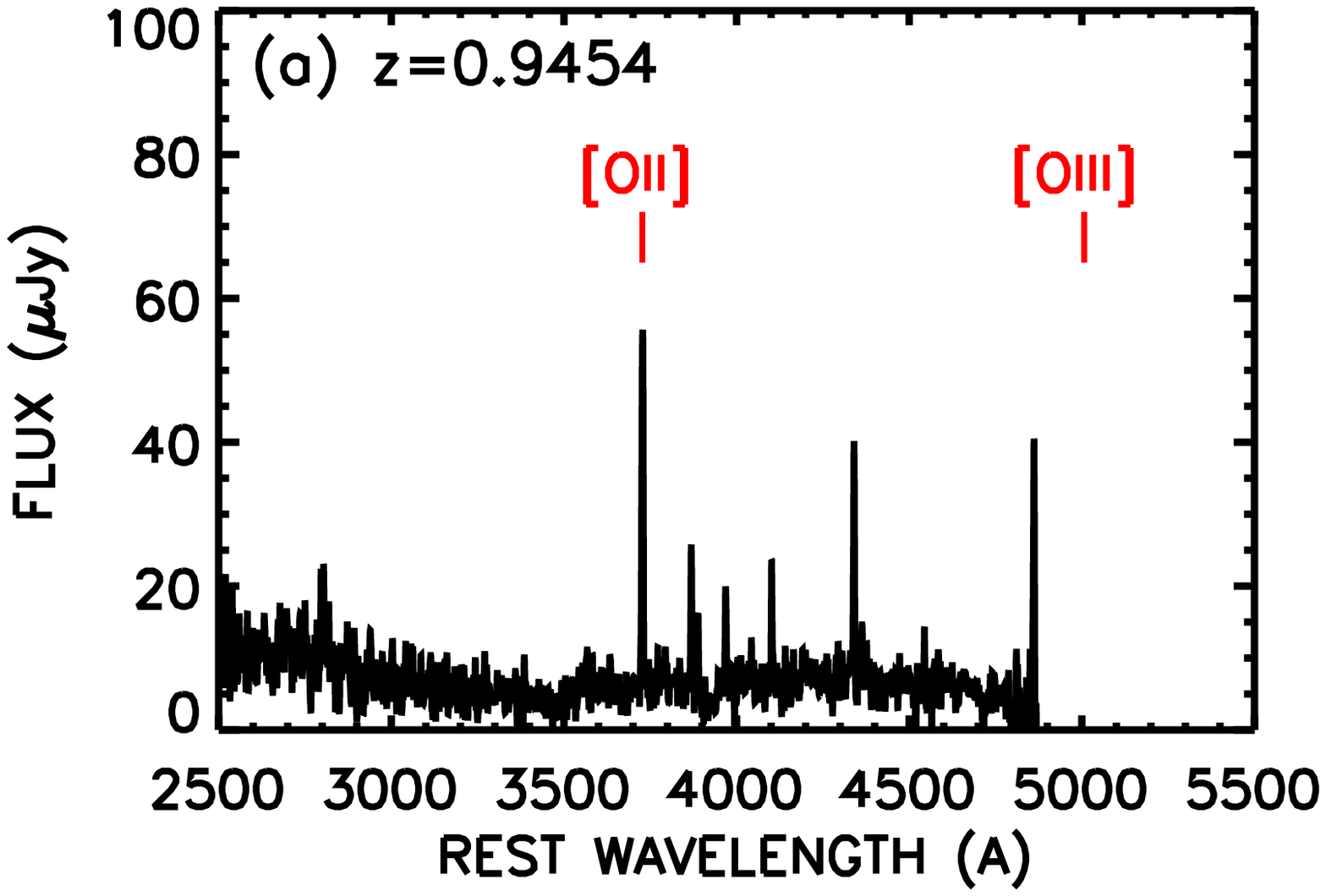}
\includegraphics[width=3.0in,angle=0,scale=1]{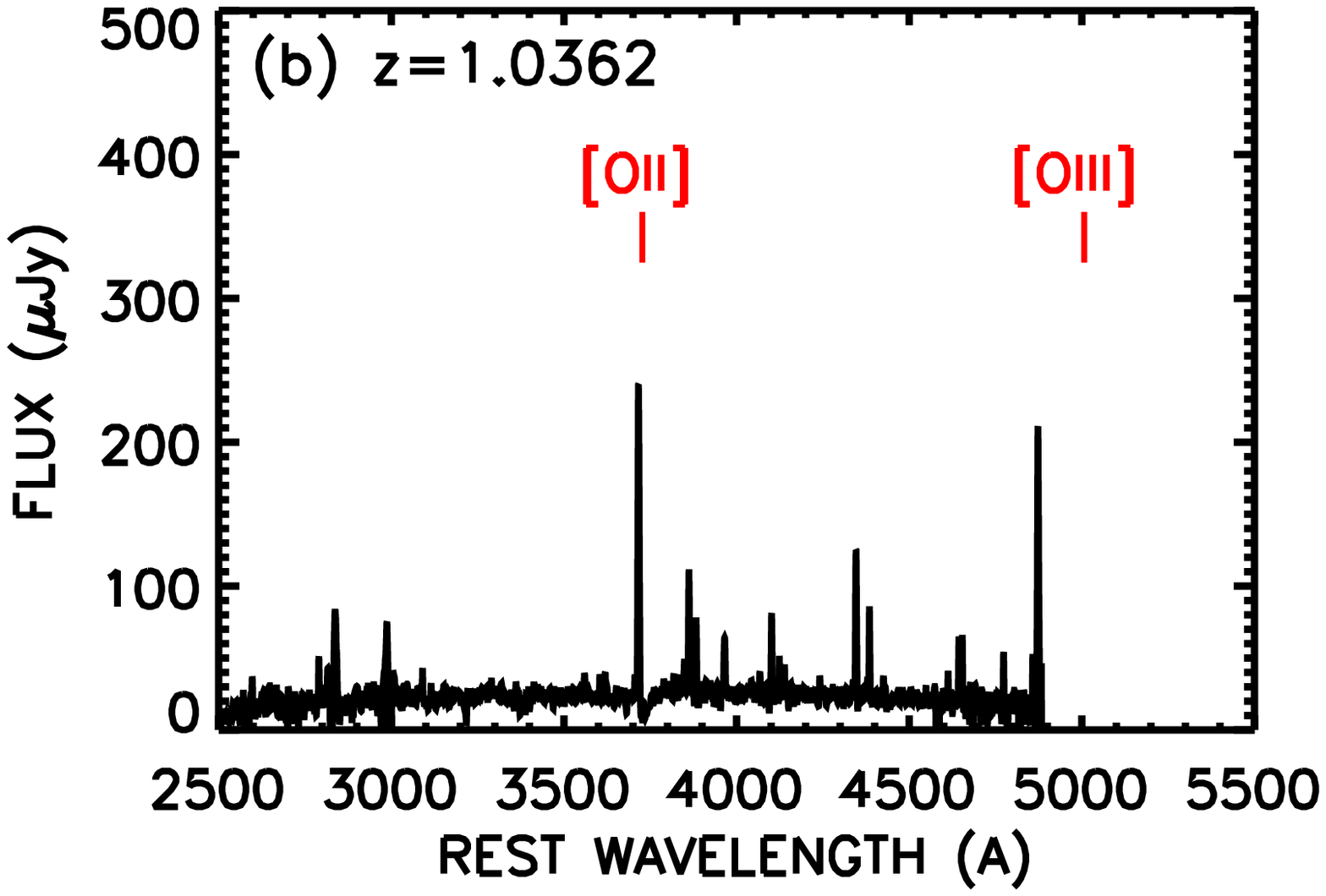}
\includegraphics[width=3.0in,angle=0,scale=1]{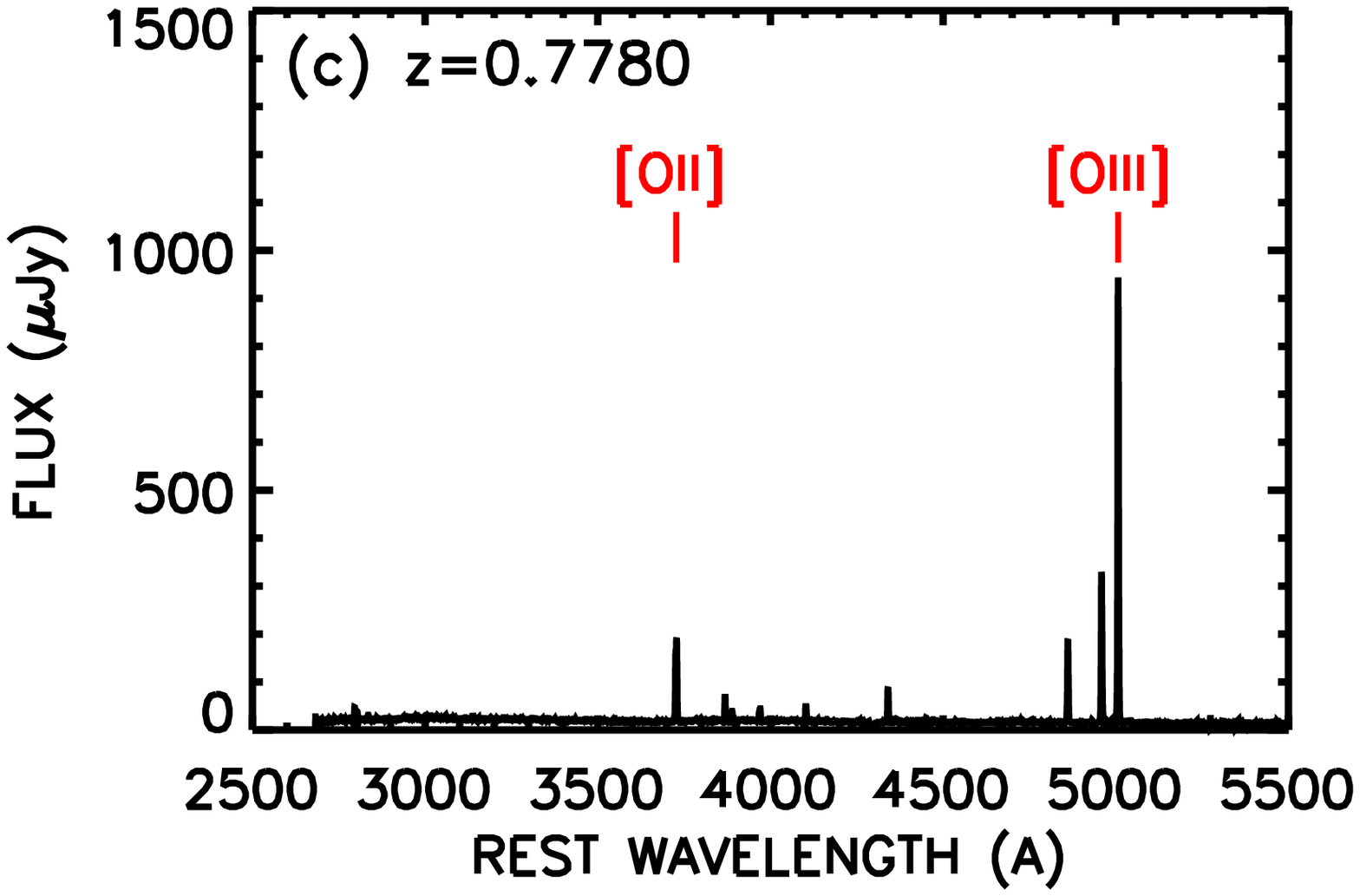}
\includegraphics[width=3.0in,angle=0,scale=1]{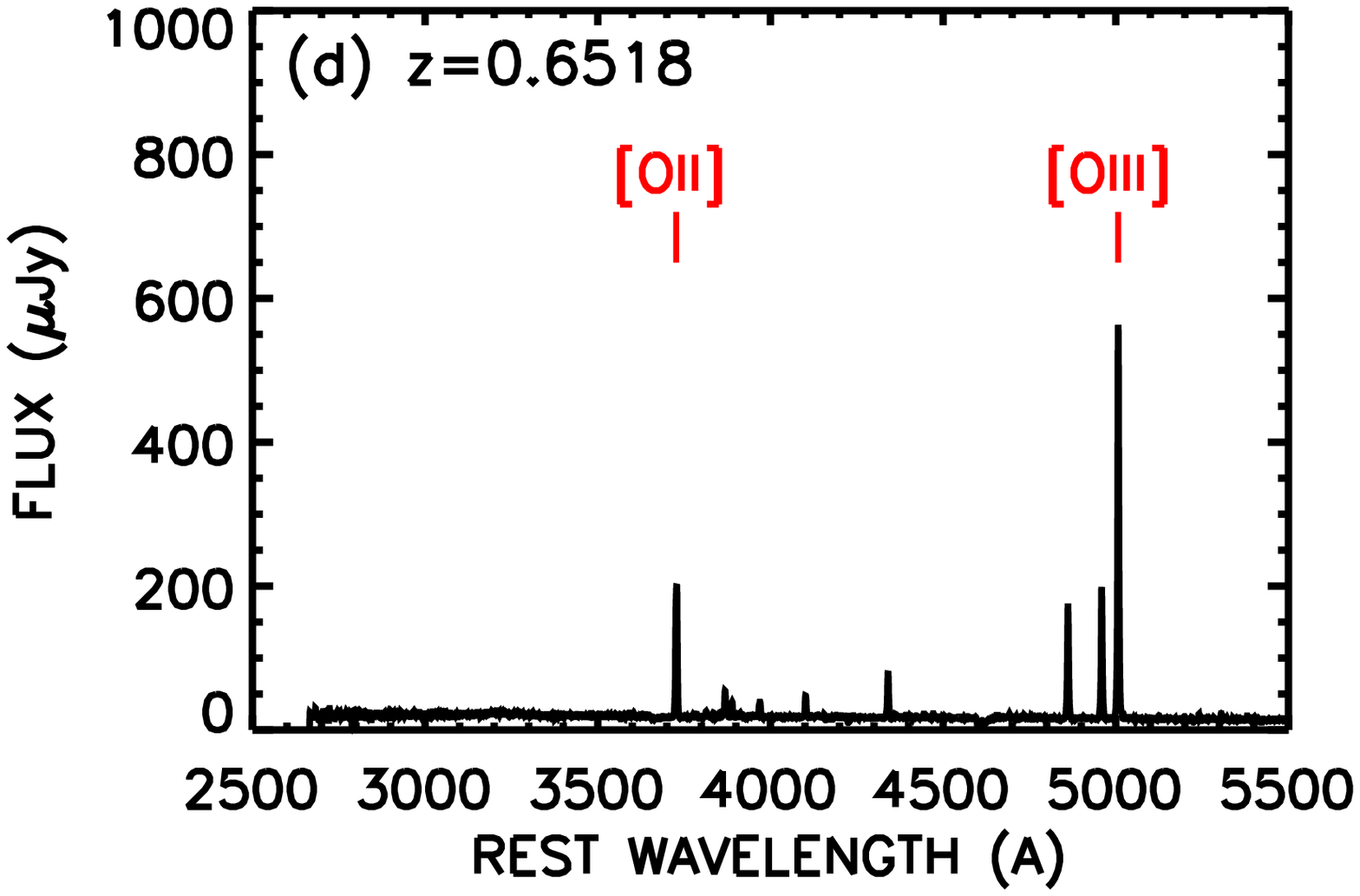}
\caption{DEIMOS optical spectra of the first four $z=0.65-1.25$ LAEs
in Table~\ref{tab2}. For each source we give the redshift measured from the 
optical spectrum. The positions of the [OII]$\lambda3727$ and [OIII]$\lambda5007$ 
lines are marked. Where continuum magnitudes are available 
we have normalized the spectrum to match these. None was
available for the source in (b), so its normalization should be
considered to be much more uncertain.
\label{optical_spectra}
}
\end{inlinefigure}

\begin{inlinefigure}
\figurenum{1 (cont)}
\includegraphics[width=3.0in,angle=0,scale=1]{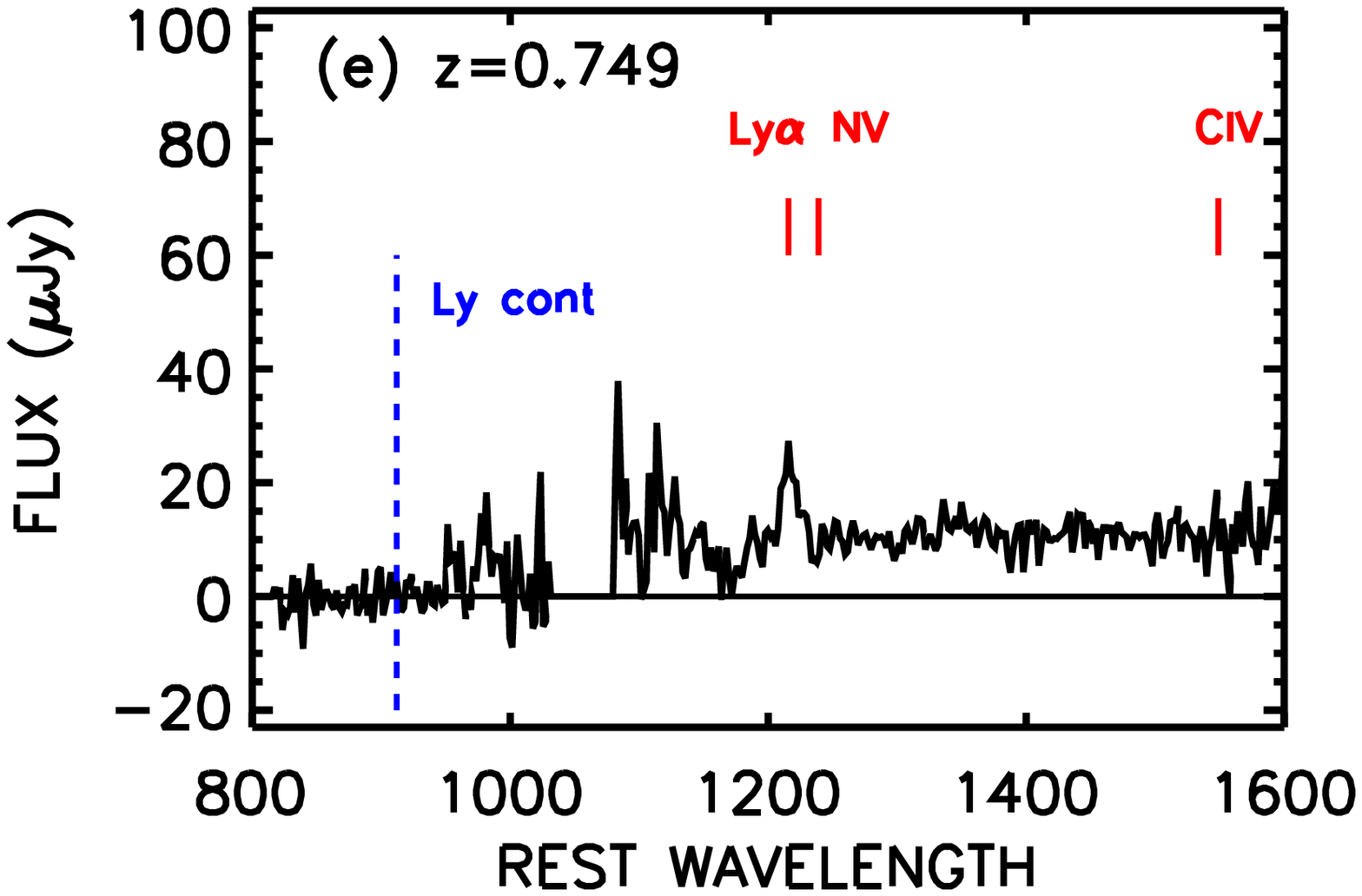}
\includegraphics[width=3.0in,angle=0,scale=1]{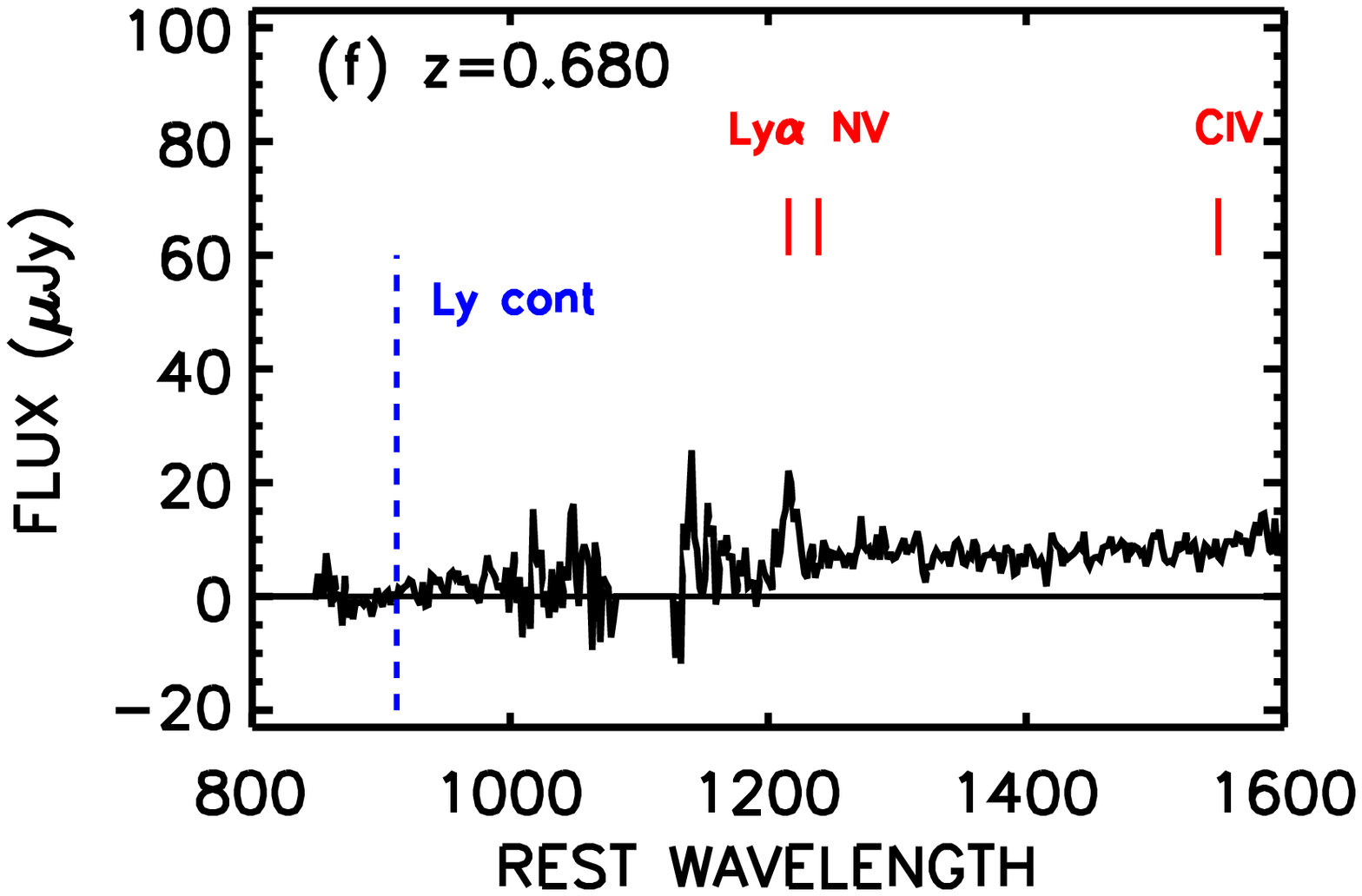}
\includegraphics[width=3.0in,angle=0,scale=1]{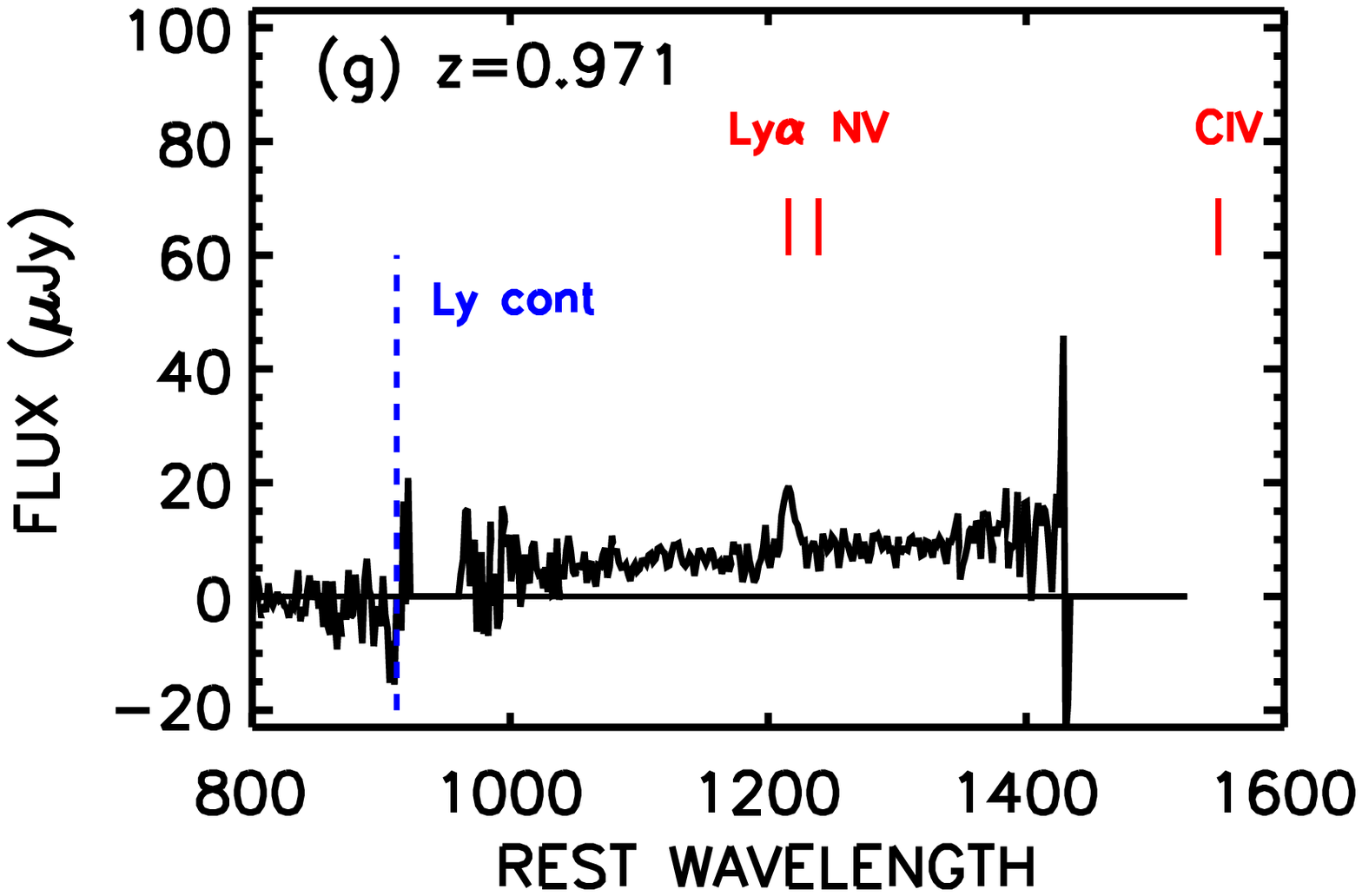}
\includegraphics[width=3.0in,angle=0,scale=1]{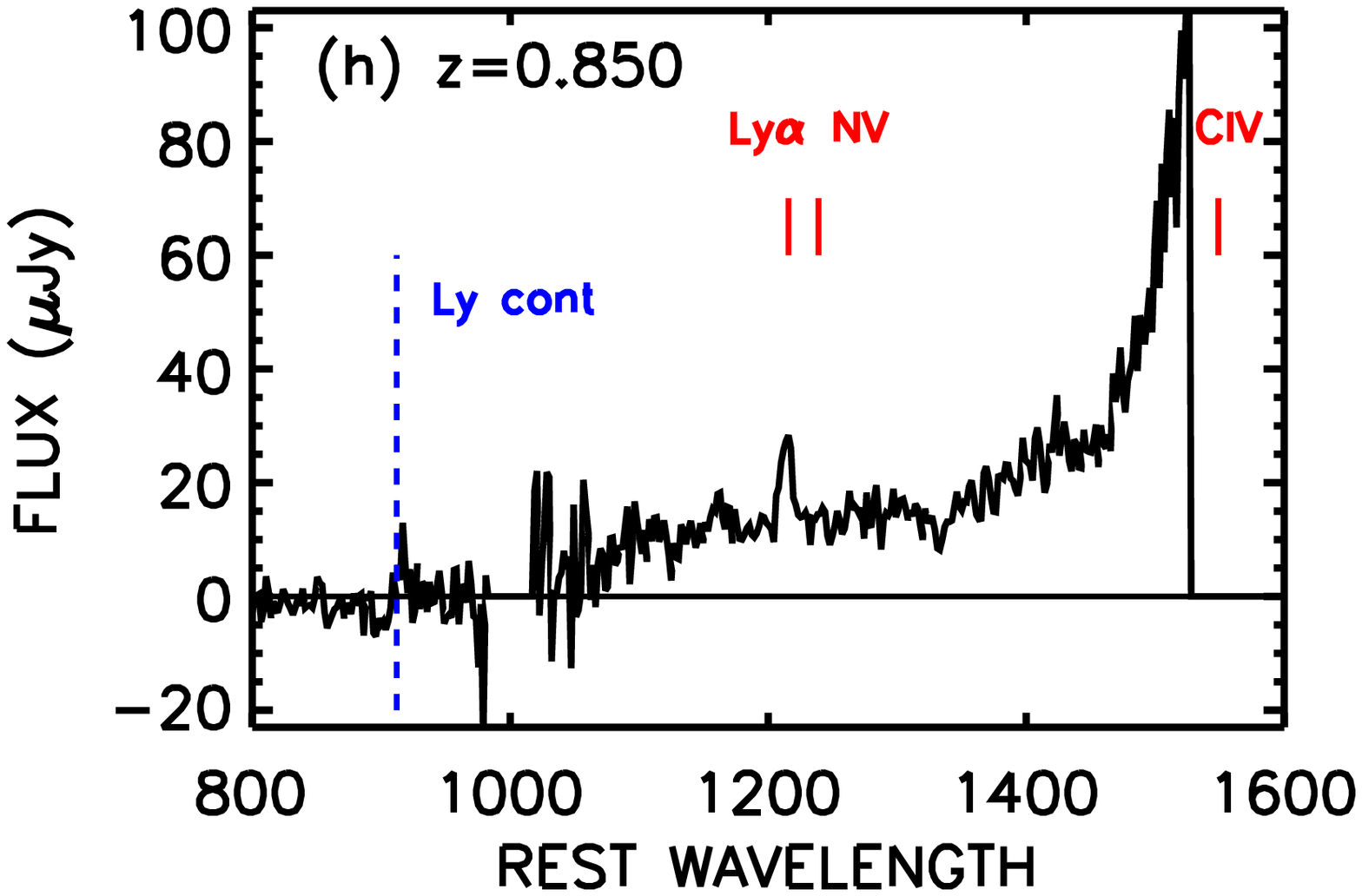}
\caption{{\em GALEX\/} spectra of the last four $z=0.65-1.25$ LAEs
in Table~\ref{tab2}. 
For each source we give the redshift measured from the UV spectrum.
The Ly$\alpha$ line is marked, as are the positions where CIV and NV 
would fall. The blue vertical dashed line shows the position of the 
Lyman continuum edge at the galaxy redshift. 
This is generally slightly redward of the more accurate optical 
redshift (see Cowie et al.\ 2010).
\label{uv_spectra2}
}
\end{inlinefigure}

\begin{inlinefigure}
\figurenum{2 (cont)}
\includegraphics[width=3.0in,angle=0,scale=1]{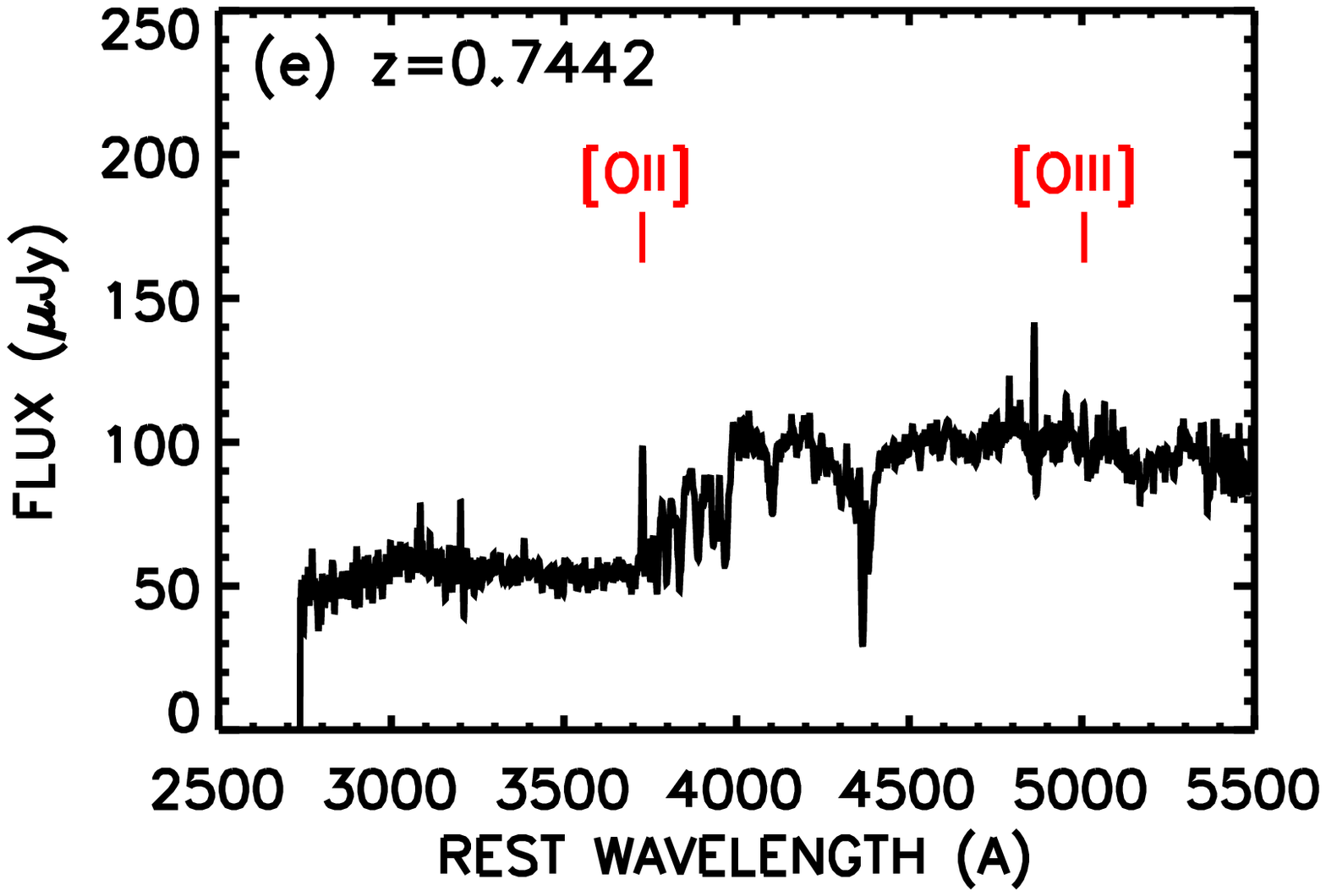}
\includegraphics[width=3.0in,angle=0,scale=1]{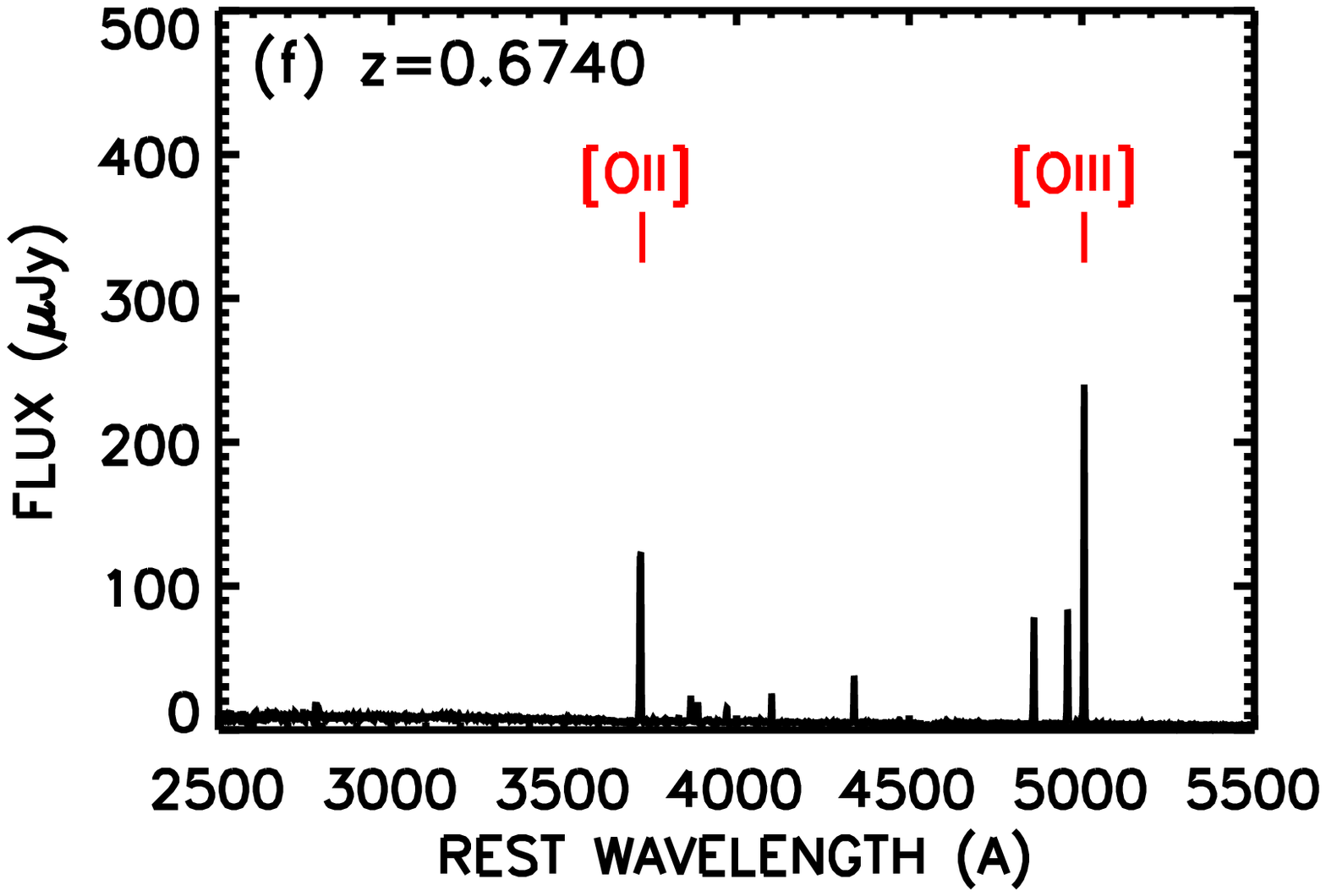}
\includegraphics[width=3.0in,angle=0,scale=1]{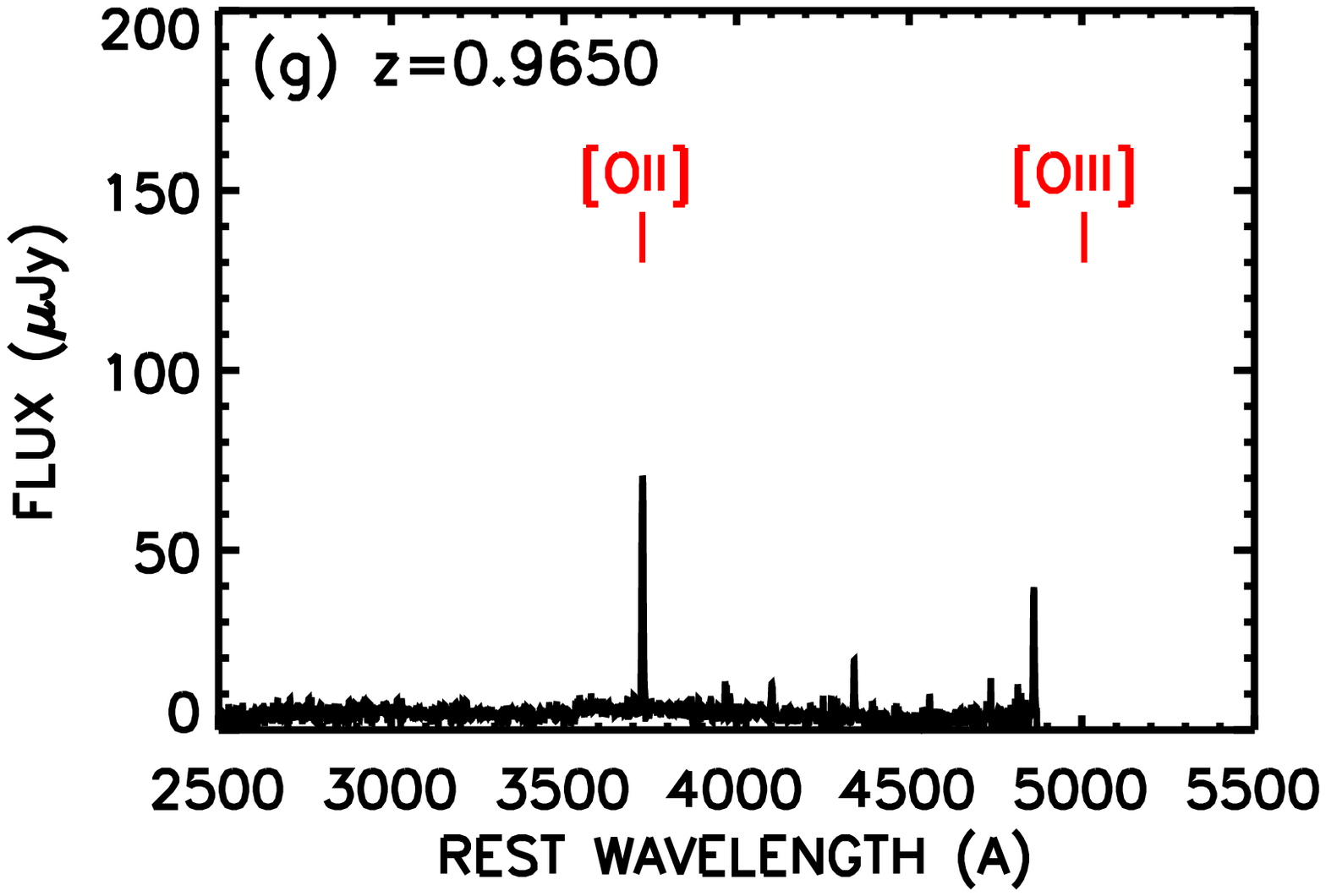}
\includegraphics[width=3.0in,angle=0,scale=1]{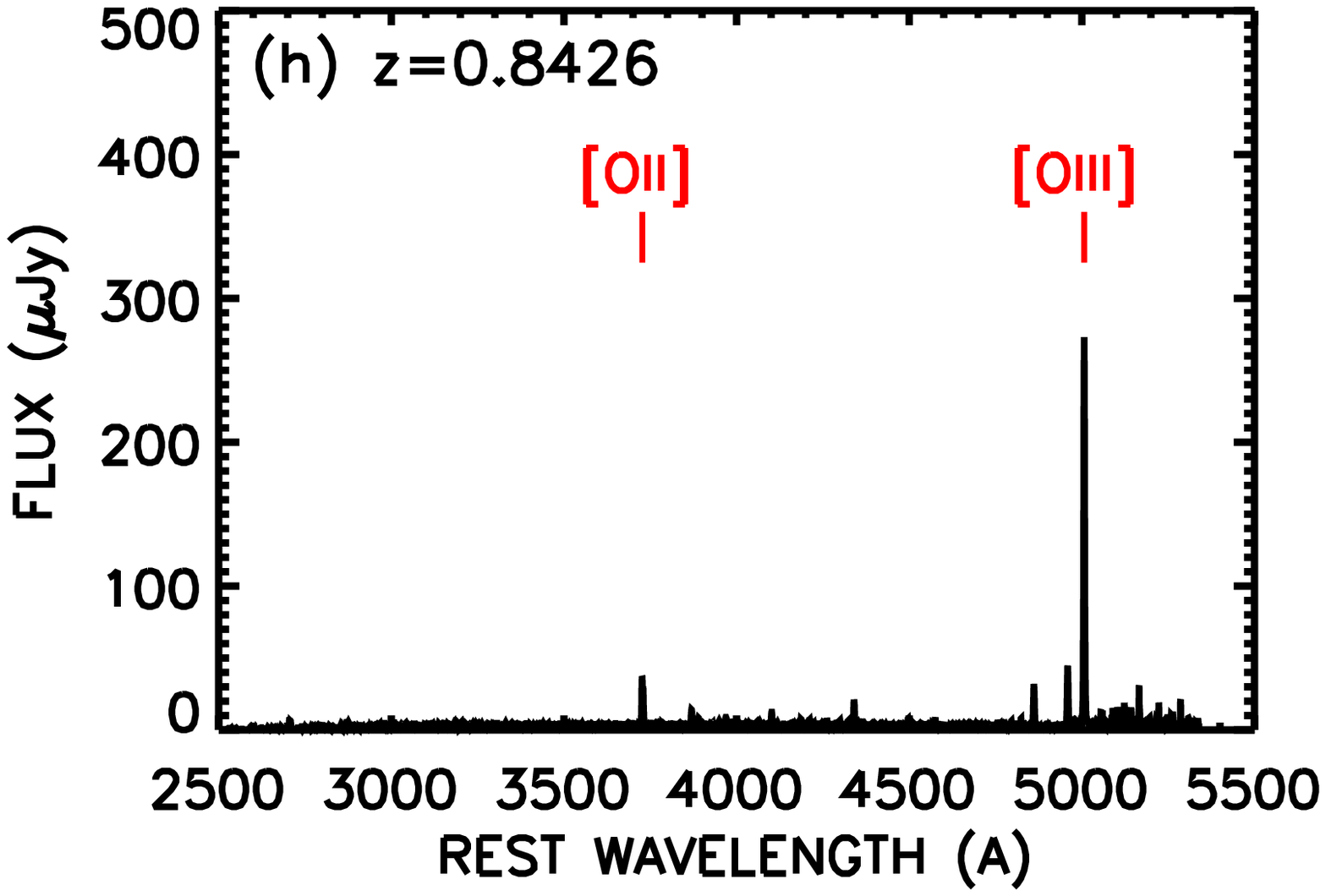}
\caption{DEIMOS optical spectra of the last four $z=0.65-1.25$ LAEs
in Table~\ref{tab2}. 
For each source we give the redshift measured from the optical spectrum. 
The positions of the [OII]$\lambda3727$ and [OIII]$\lambda5007$ 
lines are marked. Where continuum magnitudes are available 
we have normalized the spectrum to match these. None
was available for the source in (f), so its normalization should be
considered to be much more uncertain.
\label{optical_spectra2}
}
\end{inlinefigure}

\subsection{Line Fluxes}
\label{subsecfluxes}

Most of the spectra were obtained at a near-parallactic angle
to minimize atmospheric refraction effects. However,
in some cases the mask configuration did not allow for this.
In addition, a small number of the spectra were obtained during periods of
varying extinction.
All of the spectra were relatively calibrated using 
the measured response from  observed
calibration stars. However, relative slit losses always pose a 
problem, and special care must be taken for the flux calibration. 

For each spectrum we fitted all of the emission lines using the
IDL MPFIT program of Markwardt (2009). We used simultaneous Gaussian 
fits to neighboring lines together with a linear fit to the continuum 
baseline. For weaker lines we held the full width constant, using 
the value measured in the stronger lines, and set the central wavelength 
to the measured redshift. We also measured the noise as a function of 
wavelength by fitting to random positions in neighboring portions of the 
spectrum and computing the dispersion in the results.

For the AGN classifications and the metallicity measurements we used 
only the ratios of the emission lines that are close in wavelength.
These line flux ratios can be robustly measured from the spectra without 
even the relative flux calibration, since for these neighboring lines the 
DEIMOS response is essentially constant. 
For the primary lines used in our analysis we give the
line ratios and their $1\sigma$ errors in Tables~\ref{tab1} and \ref{tab2}.

\begin{inlinefigure}
\includegraphics[width=3.6in,angle=0,scale=1]{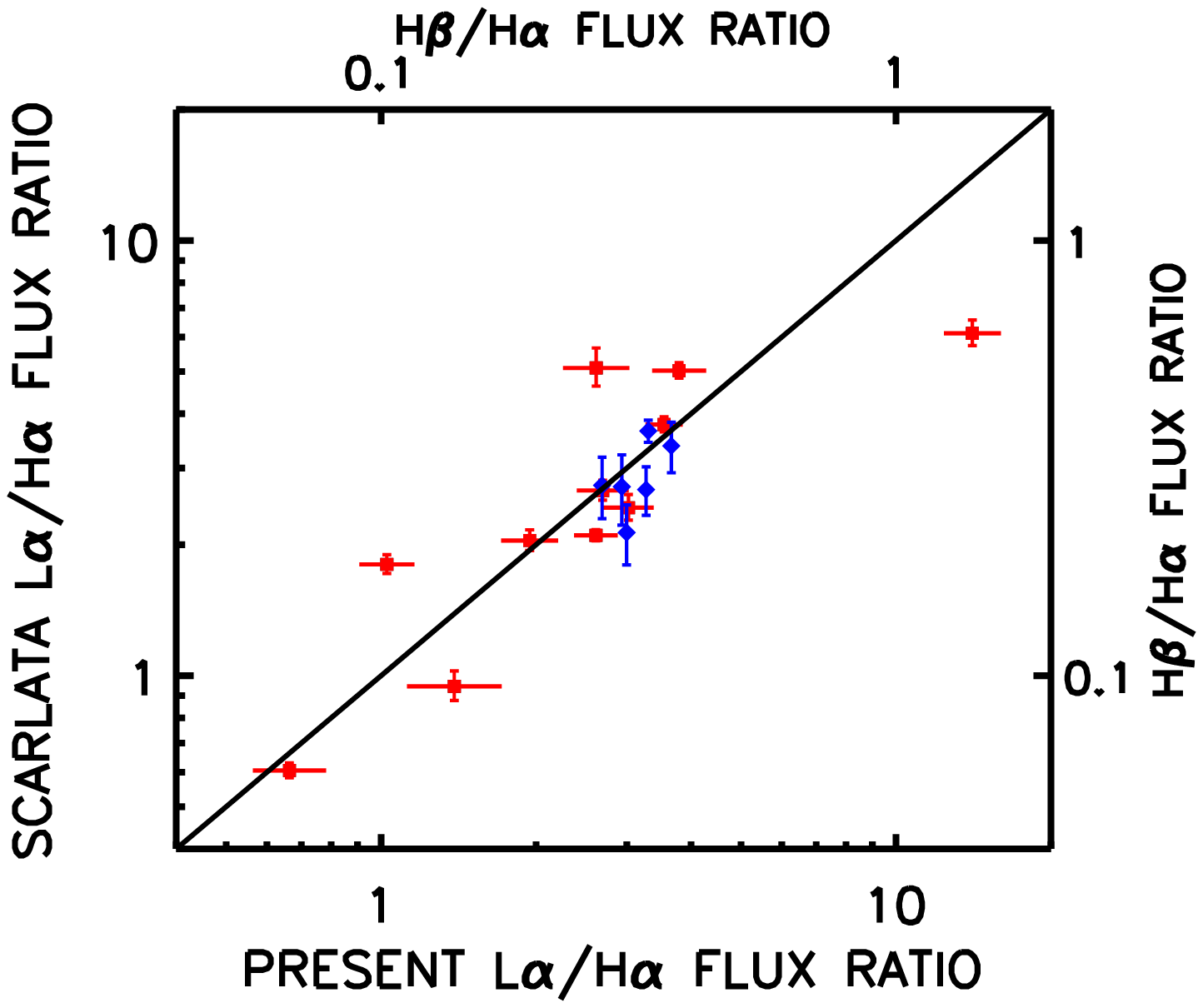}
\caption{Comparison of the $f$(Ly$\alpha$)/$f$(H$\alpha$) ratios
(red squares) and the $f$(H$\beta$)/$f$(H$\alpha$) ratios (blue diamonds)
measured from the present data with the ratios measured by
Scarlata et al.\ (2009). The scale for the $f$(Ly$\alpha$)/$f$(H$\alpha$) 
ratios is shown on the bottom and left axes and that for the
$f$(H$\beta$)/$f$(H$\alpha$) ratios on the right and top
axes. For the $f$(Ly$\alpha$)/$f$(H$\alpha$) ratios
the vertical error bars only include the uncertainties in
$f$(H$\alpha$), since the uncertainties in $f$(Ly$\alpha$) are
not given in Scarlata et al.'s table.
\label{error_claudia}
}
\end{inlinefigure}

In order to measure the absolute H$\alpha$ and H$\beta$ line fluxes, we 
used the measured EW(H$\alpha$) and EW(H$\beta$) in combination with 
the measured continuum fluxes at the appropriate wavelength.  
We first determined which SDSS filter the emission line lay in and
then integrated through the transmission of this filter
to obtain the averaged spectral flux. We then renormalized
the spectrum to match the corresponding SDSS model C
magnitude from the DR6 release
(Adelman-McCarthy et al.\ 2008). We then determined
the continuum flux and multiplied this by the
observed-frame EW to obtain the corresponding line flux. This
is an approximation, since it assumes that the measured
EW is representative of the value averaged over the total
light of the galaxy, including regions outside the slit.
However, for the photometric cases we derive very similar
values directly from the calibrated spectra, including
calibrated spectra that we obtained using larger 
($2''$) slit widths for a small subset of the galaxies.
This suggests the procedure is relatively robust. 

However, the relative apertures used in measuring the {\em GALEX\/} 
Ly$\alpha$ fluxes and the optical emission line fluxes may introduce 
systematic variations reflecting the relative geometries
of the emission. Thus, the $f$(Ly$\alpha$)/$f$(H$\alpha$) ratios
are more reliably obtained in large ($5''$)
aperture measurements, such as those used in Atek et al.\ (2009a).
In the present paper we are primarily concerned with determining
the optical continuum and emission-line properties that result
in a galaxy showing a strong Ly$\alpha$ emission line and not 
so much with the relative fluxes of the optical and Ly$\alpha$ 
emission lines, making this less of a concern. In the discussion 
we will note where aperture effects might play a role.

In order to check the calibrations and aperture effects,
we tested our measured fluxes against the independent measurements 
by Scarlata et al.\ (2009) for the subset of galaxies that overlap
with our sample. The Scarlata et al.\ data were obtained with a 
larger $1\farcs5$ slit. For eight cases
they compared their results with those made in a $5''$ slit, finding
only small differences between the two measurements. 
Scarlata et al.'s procedure for measuring the fluxes relied on the 
relative accuracy of the spectrophotometry, using the $i'$ magnitude 
from the SDSS data to set the absolute flux normalization.
This differs from our procedure, where we normalized
each region of the spectrum separately to the
SDSS magnitudes. We compare the two results in
Figure~\ref{error_claudia}, where we show
both the $f$(Ly$\alpha$)/$f$(H$\alpha$) (red squares)
and the $f$(H$\beta$)/$f$(H$\alpha$) ratios (blue diamonds). 
The overall agreement is very good, but in a small number of 
cases the disagreement is much larger than the statistical
error in the $f$(Ly$\alpha$)/$f$(H$\alpha$) ratio. In
the most extreme case the ratios differ by a multiplicative
factor of just over two. We take this to be an estimate of the
potential systematic errors.

%
%
\begin{deluxetable}{lccccccccccccc}
\renewcommand\baselinestretch{1.0}
\small\addtolength{\tabcolsep}{-3pt}
\tablenum{1}
\tablewidth{0pt}
\tablecaption{z$=0.195-0.44$ LAE Sample}
\scriptsize
\tablehead{NAME & R.A. & Decl. & $z_{\rm galex}$ & $\log L$ & $\log L$ & EW &  $z_{\rm opt}$\tablenotemark{1} & EW & [NII]/H$\alpha$ & [OIII]/H$\beta$ & [OIII]/H$\gamma$ & $\log f$ \\ & (J2000.0) & (J2000.0) & & $L_{\nu}\nu$ & $L_\alpha$ & Ly$\alpha$ & &H$\alpha$ & 6584/6563 & 5007/4861 & 4363/4341 & $f$(H$\alpha$), $f$(H$\beta$) \\ 
&  & & & (erg/s) & (erg/s) & (\AA) & & (\AA) & & & & (erg/cm$^2$/s)  \\ 
(1) & (2) & (3) & (4)  & (5) & (6) & (7) & (8) & (9) & (10) & (11) & (12) & (13) }
\startdata
GALEX0332-2748    &  53.221958   & -27.809139   &   0.2277 &    43.14 &     42.17 &    148$\pm$12 &          0.2265 &    317$\pm$2. &  0.047$\pm$0.004 &   5.64$\pm$0.007  &  0.101$\pm$0.016  &  \nodata,\nodata \cr
GALEX1718+5949    & 259.748260   &  59.825752   &   0.3717 &    43.70 &     42.24 &    114$\pm$49 &        (0.5416) &       \nodata &          \nodata &          \nodata  &          \nodata  &  \nodata,\nodata \cr
GALEX1417+5228    & 214.430923   &  52.468307   &   0.2104 &    43.01 &     42.01 &    100$\pm$9. &          0.2083 &    110$\pm$2. &          .001$\pm$0.001 &   7.17$\pm$0.004  &  0.468$\pm$0.017  &    -14.6,  -15.1 \cr
GALEX0333-2733    &  53.340416   & -27.560862   &   0.2767 &    43.22 &     42.08 &      99$\pm$9 &          0.2758 &    315$\pm$2. &  0.049$\pm$0.006 &   5.81$\pm$0.049  &  0.105$\pm$0.044  &  \nodata,\nodata \cr
GALEX0332-2747    &  53.195126   & -27.787361   &   0.2277 &    42.90 &     41.86 &     90$\pm$11 &          0.2266 &    300$\pm$2. &  0.023$\pm$0.005 &   5.18$\pm$0.057  &  0.275$\pm$0.073  &  \nodata,\nodata \cr
GALEX1436+3525    & 219.193008   &  35.417641   &   0.3697 &    43.71 &     42.32 &     88$\pm$20 &     0.3681(Sy1.8) &       \nodata &          \nodata &          \nodata  &          \nodata  &  \nodata,\nodata \cr
GALEX0331-2814    &  52.976501   & -28.238640   &   0.2824 &    43.44 &     42.18 &      72$\pm$5 &          0.2803 &     25$\pm$.5 &  0.561$\pm$0.012 &  0.794$\pm$0.041  &          \nodata  &  \nodata,\nodata \cr
GALEX0333-2757    &  53.496414   & -27.966471   &   0.3630 &    43.66 &     42.21 &     71$\pm$12 &     0.3578(B) &     26$\pm$1. &  0.956$\pm$0.015 &   5.87$\pm$ 0.11  &  0.417$\pm$ 0.37  &  \nodata,\nodata \cr
GALEX1422+5252    & 215.533752   &  52.873722   &   0.3055 &    43.34 &     42.01 &      71$\pm$8 &     0.3016(Sy1.5) &       \nodata &          \nodata &          \nodata  &          \nodata  &  \nodata,\nodata \cr
GALEX0332-2801    &  53.049751   & -28.025057   &   0.2191 &    43.13 &     41.87 &      69$\pm$6 &          0.2155 &    387$\pm$1. &  0.059$\pm$0.002 &   4.57$\pm$0.005  &  0.018$\pm$0.016  &  \nodata,\nodata \cr
GALEX1240+6233    & 190.041428   &  62.561668   &   0.2104 &    43.34 &     42.01 &     60$\pm$10 &     0.2068(B) &    133$\pm$1. &  0.255$\pm$0.008 &   7.52$\pm$0.012  &  0.186$\pm$0.058  &    -13.9,  -14.5 \cr
GALEX0332-2810    &  53.001919   & -28.182611   &   0.2795 &    43.12 &     41.76 &      55$\pm$9 &          0.2775 &    110$\pm$3. &  0.071$\pm$0.028 &   3.04$\pm$ 0.14  &  0.480$\pm$ 0.41  &  \nodata,\nodata \cr
GALEX1418+5307    & 214.728928   &  53.130028   &   0.2047 &    43.02 &     41.73 &      53$\pm$5 &          0.2034 &    301$\pm$2. &  0.012$\pm$0.007 &   5.59$\pm$0.020  &  0.143$\pm$0.092  &    -14.9,  -15.5 \cr
GALEX1438+3457    & 219.746078   &  34.960636   &   0.3740 &    43.79 &     42.27 &     50$\pm$12 &     0.3690(B) &     6.$\pm$1. &  0.419$\pm$0.068 &          \nodata  &          \nodata  &    -15.3,\nodata \cr
GALEX1421+5249    & 215.463165   &  52.830891   &   0.2018 &    42.95 &     41.50 &      49$\pm$9 &     0.2021(Sy1.5) &       \nodata &          \nodata &          \nodata  &          \nodata  &  \nodata,\nodata \cr
GALEX0332-2811    &  53.174255   & -28.190306   &   0.2076 &    43.74 &     42.36 &      48$\pm$1 &          0.2043 &    334$\pm$1. &  0.017$\pm$0.001 &   6.47$\pm$0.003  &  0.146$\pm$0.011  &  \nodata,\nodata \cr
GALEX1419+5315    & 214.794739   &  53.265835   &   0.2651 &    43.37 &     41.85 &      48$\pm$6 &          0.2637 &     11$\pm$1. &  0.403$\pm$0.035 &   1.84$\pm$ 0.31  &          \nodata  &    -14.4,  -15.3 \cr
GALEX1435+3449    & 218.764587   &  34.821220   &   0.3782 &    43.72 &     42.03 &     46$\pm$17 &        (0.4984) &       \nodata &          \nodata &          \nodata  &          \nodata  &  \nodata,\nodata \cr
GALEX0332-2722    &  53.045292   & -27.378277   &   0.3083 &    43.49 &     41.94 &      45$\pm$5 &         \nodata &       \nodata &          \nodata &          \nodata  &          \nodata  &  \nodata,\nodata \cr
GALEX1001+0220    & 150.294785   &   2.346944   &   0.2508 &    43.50 &     42.05 &      45$\pm$3 &          0.2481 &    462$\pm$2. &  0.025$\pm$0.004 &   6.44$\pm$0.040  &  0.190$\pm$0.027  &    -14.5,  -15.0 \cr
GALEX1000+0157    & 150.116043   &   1.951000   &   0.2680 &    43.80 &     42.42 &      43$\pm$2 &          0.2647 &    423$\pm$1. &  0.024$\pm$0.002 &   5.61$\pm$0.042  &  0.171$\pm$0.022  &    -14.1,  -14.7 \cr
GALEX0334-2815    &  53.541294   & -28.255499   &   0.3430 &    43.41 &     41.83 &     41$\pm$11 &          0.3370(A) &     .7$\pm$.5 &          \nodata &          \nodata  &          \nodata  &  \nodata,\nodata \cr
GALEX1436+3459    & 219.100540   &  34.993530   &   0.2148 &    43.27 &     41.73 &      40$\pm$5 &          0.2131 &     60$\pm$1. &  0.121$\pm$0.018 &   3.76$\pm$ 0.13  &  0.687$\pm$ 0.34  &    -15.5,\nodata \cr
GALEX0334-2752    &  53.711498   & -27.876585   &   0.3371 &    43.54 &     42.11 &      39$\pm$5 &     0.3336(B) &     6.$\pm$1. &  0.655$\pm$ 0.14 &   1.72$\pm$ 0.31  &          \nodata  &  \nodata,\nodata \cr
GALEX1714+5949    & 258.591766   &  59.833332   &   0.2306 &    43.34 &     41.94 &      38$\pm$5 &          0.2298 &     85$\pm$1. &  0.071$\pm$0.010 &   3.91$\pm$0.053  &          \nodata  &    -14.8,  -15.4 \cr
GALEX0330-2744    &  52.574123   & -27.741528   &   0.2623 &    43.02 &     41.52 &      36$\pm$7 &         \nodata &       \nodata &          \nodata &          \nodata  &          \nodata  &  \nodata,\nodata \cr
GALEX0332-2809    &  53.025002   & -28.161083   &   0.2219 &    42.98 &     41.30 &      35$\pm$9 &          0.2148(A) &     0.$\pm$.5 &          \nodata &          \nodata  &  0.527$\pm$ 0.25  &  \nodata,\nodata \cr
GALEX0333-2756    &  53.389545   & -27.946222   &   0.4293 &    43.91 &     42.31 &      34$\pm$7 &         \nodata &       \nodata &          \nodata &          \nodata  &          \nodata  &  \nodata,\nodata \cr
GALEX1418+5259    & 214.733200   &  52.992195   &   0.2882 &    43.38 &     41.80 &      34$\pm$4 &          0.2871 &    150$\pm$2. &  0.092$\pm$0.009 &   3.79$\pm$0.033  &  0.082$\pm$0.039  &    -14.5,  -15.2 \cr
GALEX0332-2823    &  53.240459   & -28.388306   &   0.2191 &    43.45 &     41.87 &      34$\pm$3 &          0.2136 &     64$\pm$.5 &  0.396$\pm$0.005 &   1.10$\pm$0.027  &          \nodata  &  \nodata,\nodata \cr
GALEX0331-2817    &  52.869667   & -28.283443   &   0.2162 &    42.96 &     41.26 &      33$\pm$9 &          0.2149(A) &     2.$\pm$.5 &  0.431$\pm$ 0.13 &          \nodata  &          \nodata  &  \nodata,\nodata \cr
GALEX0038-4352    &   9.520541   & -43.874584   &   0.2220 &    43.46 &     41.89 &      32$\pm$3 &          0.2190 &     12$\pm$1. &  0.499$\pm$0.037 &   1.77$\pm$ 0.29  &          \nodata  &  \nodata,\nodata \cr
GALEX1417+5246    & 214.438126   &  52.771694   &   0.2479 &    43.15 &     41.57 &      32$\pm$5 &          0.2441 &    146$\pm$2. &  0.068$\pm$0.009 &   3.74$\pm$0.042  &  0.125$\pm$0.069  &    -15.0,  -15.6 \cr
GALEX1001+0233    & 150.413208   &   2.563583   &   0.3889 &    43.87 &     42.42 &      32$\pm$5 &          0.3824 &    313$\pm$1. &  0.026$\pm$0.004 &   5.41$\pm$0.019  &  0.155$\pm$0.011  &    -14.4,  -14.9 \cr
GALEX1437+3445    & 219.334000   &  34.757637   &   0.3285 &    43.74 &     42.23 &      31$\pm$3 &          0.3237 &    137$\pm$1. &  0.095$\pm$0.006 &   3.50$\pm$0.026  &  0.118$\pm$0.025  &    -14.8,  -15.2 \cr
GALEX1436+3440    & 219.086914   &  34.672001   &   0.3740 &    43.93 &     42.08 &      29$\pm$9 &        (0.3311) &       \nodata &          \nodata &          \nodata  &          \nodata  &  \nodata,\nodata \cr
GALEX1437+3441    & 219.356308   &  34.685585   &   0.2944 &    43.74 &     41.96 &      28$\pm$5 &         0.2900(A) &     .7$\pm$.5 &  0.752$\pm$ 0.25 &          \nodata  &          \nodata  &  \nodata,\nodata \cr
GALEX1436+3456    & 219.092041   &  34.942139   &   0.2716 &    43.62 &     41.91 &      27$\pm$3 &          0.2684 &     67$\pm$2. &  0.156$\pm$0.038 &   2.52$\pm$ 0.11  &          \nodata  &    -14.8,  -15.3 \cr
GALEX0331-2814    &  52.978458   & -28.235861   &   0.3198 &    43.72 &     41.89 &      27$\pm$3 &          0.3163 &     66$\pm$1. &  0.411$\pm$0.008 &  0.386$\pm$0.020  &  0.063$\pm$0.053  &  \nodata,\nodata \cr
GALEX0332-2758    &  53.102917   & -27.977055   &   0.3803 &    43.70 &     41.93 &      26$\pm$7 &        (0.1231) &       \nodata &          \nodata &          \nodata  &          \nodata  &  \nodata,\nodata \cr
GALEX1418+5218    & 214.593506   &  52.306747   &   0.2392 &    43.19 &     41.44 &      26$\pm$4 &          0.2388 &    170$\pm$2. &  0.038$\pm$0.012 &   4.41$\pm$0.064  &  0.213$\pm$ 0.14  &    -14.8,  -15.3 \cr
GALEX0332-2811    &  53.061584   & -28.186527   &   0.2651 &    43.42 &     41.72 &      25$\pm$3 &     0.2611(B) &     .7$\pm$.5 &   1.87$\pm$ 0.20 &          \nodata  &          \nodata  &  \nodata,\nodata \cr
GALEX1417+5305    & 214.291290   &  53.086723   &   0.2680 &    43.45 &     41.65 &      25$\pm$3 &          0.2671 &     42$\pm$1. &  0.261$\pm$0.019 &   1.16$\pm$0.039  &          \nodata  &    -15.0,  -15.5 \cr
GALEX0331-2811    &  52.962204   & -28.188999   &   0.2162 &    43.52 &     41.82 &      25$\pm$2 &     0.2130(B) &     2.$\pm$.5 &  0.858$\pm$ 0.14 &          \nodata  &          \nodata  &  \nodata,\nodata \cr
GALEX0334-2743    &  53.711334   & -27.729334   &   0.3227 &    43.90 &     42.09 &      24$\pm$3 &         \nodata &       \nodata &          \nodata &          \nodata  &          \nodata  &  \nodata,\nodata \cr
GALEX0959+0149    & 149.873795   &   1.832889   &   0.2076 &    43.05 &     41.27 &      24$\pm$6 &          0.2051 &     25$\pm$.5 &  0.358$\pm$0.010 &  0.636$\pm$0.055  &          \nodata  &    -14.9,  -15.5 \cr
GALEX1434+3532    & 218.718338   &  35.547527   &   0.1977 &    43.13 &     41.45 &      24$\pm$5 &          0.1946 &    238$\pm$2. &  0.051$\pm$0.005 &   4.25$\pm$0.015  &          \nodata  &    -14.6,  -15.2 \cr
GALEX0334-2803    &  53.586754   & -28.065695   &   0.3602 &    43.73 &     41.95 &      24$\pm$5 &         \nodata &       \nodata &          \nodata &          \nodata  &          \nodata  &  \nodata,\nodata \cr
GALEX0332-2810    &  53.155334   & -28.177500   &   0.2076 &    43.30 &     41.58 &      24$\pm$3 &          0.2035 &     17$\pm$.5 &  0.355$\pm$0.015 &  0.972$\pm$0.095  &          \nodata  &  \nodata,\nodata \cr
GALEX0330-2816    &  52.737499   & -28.279444   &   0.2853 &    43.53 &     41.89 &      24$\pm$2 &          0.2813 &    233$\pm$1. &  0.151$\pm$0.004 &   1.81$\pm$0.018  &          \nodata  &  \nodata,\nodata \cr
GALEX0332-2750    &  53.102707   & -27.847721   &   0.3803 &    43.67 &     41.92 &      23$\pm$6 &          (STAR) &       \nodata &          \nodata &          \nodata  &          \nodata  &  \nodata,\nodata \cr
GALEX1423+5246    & 215.775833   &  52.779697   &   0.3458 &    43.56 &     41.86 &      22$\pm$4 &          0.3431 &    595$\pm$1. &  0.040$\pm$0.003 &   5.94$\pm$0.020  &  0.135$\pm$0.018  &    -14.3,  -14.9 \cr
GALEX1421+5239    & 215.352737   &  52.655472   &   0.2565 &    43.16 &     41.48 &      22$\pm$3 &          0.2594 &     33$\pm$1. &  0.109$\pm$0.029 &   2.21$\pm$0.076  &          \nodata  &    -13.9,  -15.1 \cr
GALEX0959+0151    & 149.918091   &   1.855917   &   0.2536 &    43.46 &     41.77 &      21$\pm$3 &          0.2506 &    811$\pm$2. &  0.051$\pm$0.001 &   6.04$\pm$0.034  &  0.096$\pm$0.008  &    -13.9,  -14.6 \cr
GALEX1420+5306    & 215.170456   &  53.114002   &   0.1989 &    43.12 &     41.32 &      21$\pm$4 &          0.1990(A) &     2.$\pm$78 &          \nodata &          \nodata  &          \nodata  &    -15.0,  -14.7 \cr
GALEX1714+5956    & 258.683746   &  59.947247   &   0.2220 &    43.28 &     41.53 &      21$\pm$5 &          0.2157 &     27$\pm$1. &  0.100$\pm$0.032 &   5.08$\pm$0.038  &          \nodata  &    -15.2,  -15.2 \cr
GALEX1420+5247    & 215.132080   &  52.799389   &   0.2565 &    43.60 &     41.83 &      20$\pm$2 &          0.2525 &     72$\pm$13 &          \nodata &   9.41$\pm$ 0.78  &          \nodata  &  \nodata,\nodata \cr
GALEX1725+5920    & 261.344940   &  59.345528   &   0.3861 &    44.28 &     42.43 &      20$\pm$6 &        (.03086) &       \nodata &          \nodata &          \nodata  &          \nodata  &  \nodata,\nodata \cr
GALEX0331-2820    &  52.897709   & -28.345751   &   0.2651 &    43.29 &     41.48 &      20$\pm$4 &         \nodata &       \nodata &          \nodata &          \nodata  &          \nodata  &  \nodata,\nodata \cr
GALEX0334-2752    &  53.554749   & -27.880083   &   0.2364 &    43.25 &     41.52 &      20$\pm$3 &          0.2333 &    300$\pm$2. &  0.030$\pm$0.005 &   5.45$\pm$0.052  &          \nodata  &  \nodata,\nodata \cr
GALEX1717+5944    & 259.465027   &  59.748554   &   0.1960 &    43.24 &     41.40 &      20$\pm$6 &          0.1979 &    112$\pm$1. &  0.172$\pm$0.010 &   1.68$\pm$0.037  &  0.094$\pm$0.058  &    -14.4,  -15.1 \cr
\enddata
\tablenotetext{1}{The optical redshift is shown in parentheses when it does not agree with the {\em GALEX\/} 
redshift. Intermediate-type Seyferts are labeled as Sy1.5 or Sy1.8, \\while AGNs identified in the BPT diagram 
are labeled as B. Galaxies with weak optical emission lines are labeled as A.\\}
\label{tab1}
\end{deluxetable}

%
%
\begin{deluxetable}{lcccccccccccc}
\renewcommand\baselinestretch{1.0}
\small\addtolength{\tabcolsep}{-3pt}
\tablewidth{0pt}
\tablecaption{z$=0.195-0.44$ LAE Sample (Supplement)}
\scriptsize
\tablehead{NAME & R.A. & Decl. & $z_{\rm galex}$ & $\log L$ & $\log L$ & EW &  $z_{\rm opt}$\tablenotemark{1} & EW & [NII]/H$\alpha$ & [OIII]/H$\beta$ & [OIII]/H$\gamma$ &  $\log f$ \\ 
& (J2000.0) & (J2000.0) & & $L_{\nu}\nu$ & $L_\alpha$ & Ly$\alpha$ & & H$\alpha$ & 6584/6563 & 5007/4861 & 4363/4341 & $f$(H$\alpha$), $f$(H$\beta$)\\ 
&  & & & (erg/s) & (erg/s) & (\AA) & & (\AA) & & & & (erg/cm$^2$/s)  \\ 
(1) & (2) & (3) & (4)  & (5) & (6) & (7) & (8) & (9) & (10) & (11) & (12) & (13) }
\startdata
GALEX1001+0145    & 150.396500   &   1.756222   &   0.2191 &    43.14 &     41.40 &      19$\pm$4 &          0.2192 &     68$\pm$1. &  0.061$\pm$0.014 &   2.54$\pm$0.083  &          \nodata  &    -15.2,  -15.8 \cr
GALEX1423+5244    & 215.842850   &  52.742474   &   0.2795 &    43.29 &     41.46 &      19$\pm$5 &          0.2771 &     39$\pm$2. &  0.297$\pm$0.037 &  0.853$\pm$ 0.11  &          \nodata  &    -15.1,  -15.6 \cr
GALEX0331-2724    &  52.949291   & -27.402834   &   0.3688 &    43.82 &     41.80 &      19$\pm$5 &         \nodata &       \nodata &          \nodata &          \nodata  &          \nodata  &  \nodata,\nodata \cr
GALEX0334-2748    &  53.729584   & -27.800833   &   0.3141 &    44.20 &     42.34 &      19$\pm$1 &         \nodata &       \nodata &          \nodata &          \nodata  &          \nodata  &  \nodata,\nodata \cr
GALEX1421+5224    & 215.477295   &  52.406723   &   0.3515 &    44.05 &     42.12 &      19$\pm$2 &          0.3454 &     26$\pm$.5 &  0.558$\pm$0.011 &  0.326$\pm$0.068  &  0.421$\pm$ 0.17  &    -14.2,  -14.9 \cr
GALEX1716+5932    & 259.178070   &  59.548889   &   0.3600 &    43.76 &     41.99 &      19$\pm$6 &         \nodata &       \nodata &          \nodata &          \nodata  &          \nodata  &  \nodata,\nodata \cr
GALEX1000+0201    & 150.149017   &   2.020417   &   0.2709 &    43.84 &     41.99 &      19$\pm$2 &          0.2653 &    195$\pm$.5 &  0.060$\pm$0.001 &   4.71$\pm$0.031  &  0.119$\pm$0.026  &    -14.3,  -14.9 \cr
GALEX1419+5221    & 214.970749   &  52.350250   &   0.2680 &    43.46 &     41.57 &      19$\pm$3 &     0.2647(B) &     3.$\pm$.5 &  0.571$\pm$0.088 &          \nodata  &          \nodata  &    -15.4,\nodata \cr
GALEX1418+5245    & 214.521545   &  52.752110   &   0.2450 &    43.51 &     41.60 &      19$\pm$2 &          0.2445 &    285$\pm$2. &  0.031$\pm$0.005 &   5.82$\pm$0.039  &  0.113$\pm$0.046  &    -14.4,  -15.0 \cr
GALEX1419+5223    & 214.811249   &  52.390808   &   0.2565 &    43.14 &     41.35 &      19$\pm$4 &          0.2541(A) &     76$\pm$5. &  0.159$\pm$0.067 &   3.37$\pm$ 0.69  &          \nodata  &    -14.1,  -14.7 \cr
GALEX0333-2820    &  53.286755   & -28.333389   &   0.3697 &    43.65 &     41.80 &      19$\pm$6 &         \nodata &       \nodata &          \nodata &          \nodata  &          \nodata  &  \nodata,\nodata \cr
GALEX1418+5217    & 214.701202   &  52.299168   &   0.2421 &    43.31 &     41.42 &      18$\pm$4 &          0.2398 &     46$\pm$2. &  0.062$\pm$0.033 &   3.80$\pm$ 0.22  &          \nodata  &    -15.1,  -15.9 \cr
GALEX0330-2801    &  52.551250   & -28.025417   &   0.2479 &    43.21 &     41.47 &      18$\pm$3 &         \nodata &       \nodata &          \nodata &          \nodata  &          \nodata  &  \nodata,\nodata \cr
GALEX0330-2801    &  52.550297   & -28.028778   &   0.2479 &    43.21 &     41.38 &      18$\pm$3 &         \nodata &       \nodata &          \nodata &          \nodata  &          \nodata  &  \nodata,\nodata \cr
GALEX1418+5249    & 214.732056   &  52.824558   &   0.2680 &    43.43 &     41.48 &      18$\pm$3 &     0.2634(B) &     1.$\pm$.5 &   1.44$\pm$ 0.25 &          \nodata  &          \nodata  &    -15.6,\nodata \cr
GALEX0333-2801    &  53.301208   & -28.022917   &   0.2911 &    43.31 &     41.52 &      17$\pm$4 &         \nodata &       \nodata &          \nodata &          \nodata  &          \nodata  &  \nodata,\nodata \cr
GALEX1717+5929    & 259.379578   &  59.487473   &   0.2997 &    43.91 &     41.99 &      17$\pm$3 &          0.2988 &     34$\pm$1. &  0.423$\pm$0.024 &   1.09$\pm$ 0.10  &          \nodata  &    -14.3,  -15.0 \cr
GALEX1436+3527    & 219.026245   &  35.458553   &   0.2517 &    43.65 &     41.79 &      17$\pm$2 &          0.2495 &     74$\pm$2. &  0.103$\pm$0.016 &   2.53$\pm$0.066  &          \nodata  &    -14.9,  -15.4 \cr
GALEX1420+5243    & 215.180359   &  52.718750   &   0.2508 &    43.45 &     41.63 &      17$\pm$2 &          0.2470 &     70$\pm$2. &          \nodata &   4.83$\pm$ 0.10  &          \nodata  &    -15.1,  -15.6 \cr
GALEX0332-2744    &  53.080002   & -27.745890   &   0.2191 &    43.26 &     41.38 &      17$\pm$3 &          0.2174 &    101$\pm$1. &  0.086$\pm$0.006 &   2.95$\pm$0.053  &  0.115$\pm$0.060  &  \nodata,\nodata \cr
GALEX0331-2809    &  52.999332   & -28.164444   &   0.2392 &    43.46 &     41.62 &      17$\pm$2 &          0.2363 &    188$\pm$1. &  0.146$\pm$0.004 &   2.57$\pm$0.025  &  0.048$\pm$0.036  &  \nodata,\nodata \cr
GALEX1420+5250    & 215.186218   &  52.835056   &   0.2536 &    43.39 &     41.58 &      17$\pm$2 &          0.2519 &     66$\pm$2. &  0.168$\pm$0.038 &   2.38$\pm$ 0.11  &          \nodata  &    -14.8,  -15.1 \cr
GALEX0331-2809    &  52.999332   & -28.164444   &   0.2392 &    43.46 &     41.62 &      17$\pm$2 &          0.2363 &    188$\pm$1. &  0.146$\pm$0.004 &   2.57$\pm$0.025  &  0.048$\pm$0.036  &  \nodata,\nodata \cr
GALEX1420+5250    & 215.186218   &  52.835056   &   0.2536 &    43.39 &     41.58 &      17$\pm$2 &          0.2519 &     66$\pm$2. &  0.168$\pm$0.038 &   2.38$\pm$ 0.11  &          \nodata  &    -14.8,  -15.1 \cr
GALEX0331-2742    &  52.755127   & -27.715334   &   0.3141 &    43.58 &     41.76 &      17$\pm$2 &          0.3107 &     75$\pm$1. &  0.135$\pm$0.009 &   2.65$\pm$0.028  &  0.091$\pm$0.045  &  \nodata,\nodata \cr
GALEX0333-2821    &  53.258461   & -28.357668   &   0.2507 &    43.62 &     41.71 &      17$\pm$2 &          0.2471 &     76$\pm$1. &  0.126$\pm$0.009 &   2.50$\pm$0.052  &          \nodata  &  \nodata,\nodata \cr
GALEX1422+5310    & 215.500092   &  53.168472   &   0.2392 &    43.10 &     41.17 &      16$\pm$5 &          0.2342(A) &     16$\pm$1. &  0.410$\pm$0.043 &   2.10$\pm$ 0.69  &          \nodata  &    -15.2,\nodata \cr
GALEX0333-2814    &  53.471249   & -28.248138   &   0.2018 &    43.36 &     41.45 &      16$\pm$3 &         \nodata &       \nodata &          \nodata &          \nodata  &          \nodata  &  \nodata,\nodata \cr
GALEX1419+5230    & 214.907211   &  52.507057   &   0.2853 &    43.41 &     41.55 &      15$\pm$3 &          0.2814(A) &     10$\pm$1. &  0.490$\pm$0.088 &          \nodata  &          \nodata  &    -15.3,  -15.8 \cr
GALEX0331-2748    &  52.800335   & -27.800335   &   0.2594 &    43.19 &     41.30 &      15$\pm$3 &          0.2581 &     98$\pm$1. &  0.133$\pm$0.010 &   2.32$\pm$0.035  &          \nodata  &  \nodata,\nodata \cr
GALEX0331-2753    &  52.850460   & -27.894693   &   0.2594 &    43.30 &     41.35 &      15$\pm$3 &          (STAR) &       \nodata &          \nodata &          \nodata  &          \nodata  &  \nodata,\nodata \cr
GALEX1436+3535    & 219.243362   &  35.597694   &   0.2077 &    43.36 &     41.40 &      15$\pm$3 &          0.2055 &     14$\pm$1. &  0.453$\pm$0.029 &  0.531$\pm$ 0.16  &          \nodata  &    -14.4,  -15.1 \cr
\enddata
\tablenotetext{1}{The optical redshift is shown in parentheses when it does not agree with the {\em GALEX\/} 
redshift. Intermediate-type Seyferts are labeled as Sy1.5 or Sy1.8, \\while AGNs identified in the BPT diagram 
are labeled as B. Galaxies with weak optical emission lines are labeled as A.\\}
\label{tab1b}
\end{deluxetable}

%
%
\begin{deluxetable}{lccccccccccc}
\renewcommand\baselinestretch{1.0}
\small\addtolength{\tabcolsep}{-1pt}
\tablenum{2}
\tablewidth{0pt}
\tablecaption{z$=0.65-1.25$ LAE SAMPLE}
\scriptsize
\tablehead{NAME & R.A. & Decl. & $z_{\rm galex}$ & $\log L$ & $\log L$ & EW & $z_{\rm opt}$\tablenotemark{1} & EW & [NII]/H$\alpha$ & [OIII]/H$\beta$ & [OIII]/H$\gamma$ \\ 
& (J2000.0) & (J2000.0) & & $L_{\nu}\nu$ & $L_\alpha$ & Ly$\alpha$ & & H$\beta$ & 6584/6563 & 5007/4861 & 4363/4341 \\ 
& & & & (erg/s) & (erg/s) & (\AA) & & (\AA) & & & \\ 
(1) & (2) & (3) & (4)  & (5) & (6) & (7) & (8) & (9) & (10) & (11) & (12)
}
\startdata
GALEX1421+5221    & 215.421539   &  52.359886   &   0.9475 &    44.60 &     43.21 &     52$\pm$ 4 &    0.9454 &   \nodata &  \nodata &  \nodata  &  0.26$\pm$0.06 \cr
GALEX0334-2753    &  53.623749   & -27.893917   &  1.0396  &  44.47 &  43.08 &     48$\pm$ 5 &   1.0362 & \nodata & \nodata & \nodata & 0.08$\pm0.04$ \cr
GALEX1712+6001    & 258.197723   &  60.022778   &   0.7830 &    44.76 &     43.19 &     37$\pm$ 4 &    0.7780 &    62 &          \nodata &   5.58$\pm$0.021  &  0.10$\pm$0.05 \cr
GALEX1437+3541    & 219.281967   &  35.692280   &   0.6630 &    44.54 &     42.91 &     29$\pm$ 5 &    0.6518 &    64 &          \nodata &   3.50$\pm$0.031  &  0.05$\pm$0.02 \cr
GALEX1420+5313    & 215.237045   &  53.218750   &   0.7488 &    44.53 &     42.84 &     27$\pm$ 3 &    0.7442(A) &     6 &          \nodata &  0.484$\pm$ 0.16  &   \nodata \cr
GALEX0331-2737    &  52.765331   & -27.621777   &  0.6797  &  44.30 &  42.53 &     22$\pm$ 4 &   0.6740 & 102 & \nodata & \nodata & 0.03$\pm$0.01 \cr
GALEX0332-2734    &  53.084042   & -27.573084   &  0.9705  &  44.72 &  42.82 &     18$\pm$ 2 &   0.9650 & \nodata & \nodata & \nodata & 0.10$\pm$0.03 \cr
GALEX0331-2754    &  52.854668   & -27.916639   &  0.8945  &  44.93 &  42.77 &     12$\pm$ 2 &    0.8425 & 63 & \nodata & \nodata  & 0.09$\pm$0.06 \cr
\enddata
\tablenotetext{1}{
Galaxies with weak optical emission lines are labeled as A.\\}
\label{tab2}
\end{deluxetable}

\section{AGN-Galaxy Discrimination}
\label{secagngal}

We might expect that sources classified as AGNs based on 
the presence of high-excitation lines or very wide lines in the UV
spectra are truly AGNs. However, the converse
is not true, which means our candidate LAE galaxies (Tables~\ref{tab1} 
and \ref{tab2}) may still have some degree of AGN contamination. 
In some cases the high-excitation lines fall at problematic 
wavelengths in the UV spectra, and in other cases these lines may be 
intrinsically weak.  Optical spectroscopic data support these points:
follow-ups of {\em GALEX\/} candidate LAE galaxy samples have shown 
that a significant fraction ($\sim 20\%$) have substantial
AGN contributions (Finkelstein et al.\ 2009b; Scarlata et al.\
2009; Cowie et al.\ 2010), though there is wide variation
in the estimates of the precise degree of contamination
reflecting the relatively small sample sizes and
the differences in the classification procedures.
(Finkelstein et al.\ 2009b find a higher AGN fraction in their sample
than other groups, so we make a detailed comparison with their results
in this section.)

We first searched our samples for optical spectra that showed 
broad emission lines. None of the spectra are quasars or Seyfert~1s,
a consequence of this class of source being easily picked out 
with the UV spectra (Barger \& Cowie 2010). However, three of the sources 
in the $z=0.195-0.44$ sample are intermediate-type Seyferts. 
In the optical redshift column (col.~8) of Table~\ref{tab1} we have labeled 
these sources as Sy1.5 or Sy1.8. We eliminate these sources from further 
consideration. 

We next constructed the BPT diagram (Baldwin et al.\ 1981) 
for the $z=0.195-0.44$ sample. This diagram uses the [NII]$\lambda6584$/H$\alpha$ 
and the [OIII]$\lambda5007$/H$\beta$ ratios to separate AGNs from star-forming
galaxies. H$\alpha$ is off the spectra in the $z=0.65-1.25$ sample, and we 
cannot run this test there. In addition, only sources with strong optical 
emission lines can be classified in this way. We restricted to galaxies
that had either [NII]$\lambda6584$ or H$\alpha$ detected with a 
signal-to-noise above 5 and either [OIII]$\lambda5007$ or H$\beta$ 
detected with a signal-to-noise above 5.

The BPT diagrams for the (a) LAE and (b) UV-continuum
samples having strong optical emission lines defined in this way 
are shown in Figures~\ref{bpt}(a) and \ref{bpt}(b), respectively.
The dotted curve traces the Kewley et al.\ (2001) division between
galaxies whose extreme UV ionizing radiation field is dominated by
an AGN ($>50\%$) and those dominated by star formation. It is a
theoretical curve based on photoionization models for giant HII regions
and a range of stellar population synthesis codes. The dashed
curve traces the Kauffmann et al.\ (2003) division between AGNs and
star formers based on an empirical separation of SDSS galaxies.
As mentioned above, we have eliminated the intermediate-type Seyferts 
in Table~\ref{tab1} from Figure~\ref{bpt}(a), but the BPT diagram 
suggests that a further four sources are also AGNs. 
We mark these with open black squares in Figure~\ref{bpt}(a) and
label them as ``B'' in the optical redshift column (col.~8) of Table~\ref{tab1}. 
Another four sources have weak [OIII] and H$\beta$
and do not appear on the diagram, but they
have strong $\log$([NII]$\lambda6584$/H$\alpha)>-0.25$ ratios,
which suggests that they are also AGNs.
We also label these sources as ``B'' in column~8 of Table~\ref{tab1}. 
Most of the remaining sources in Figure~\ref{bpt}(a) clearly lie along 
the star-forming galaxy track, as do all the UV-continuum 
sources in Figure~\ref{bpt}(b), though there are a small number
of sources that lie in positions where they may 
have mixed star formation and AGN contributions.
None of the results in the paper are changed if we
exclude these objects.

\begin{inlinefigure}
\includegraphics[width=4.0in]{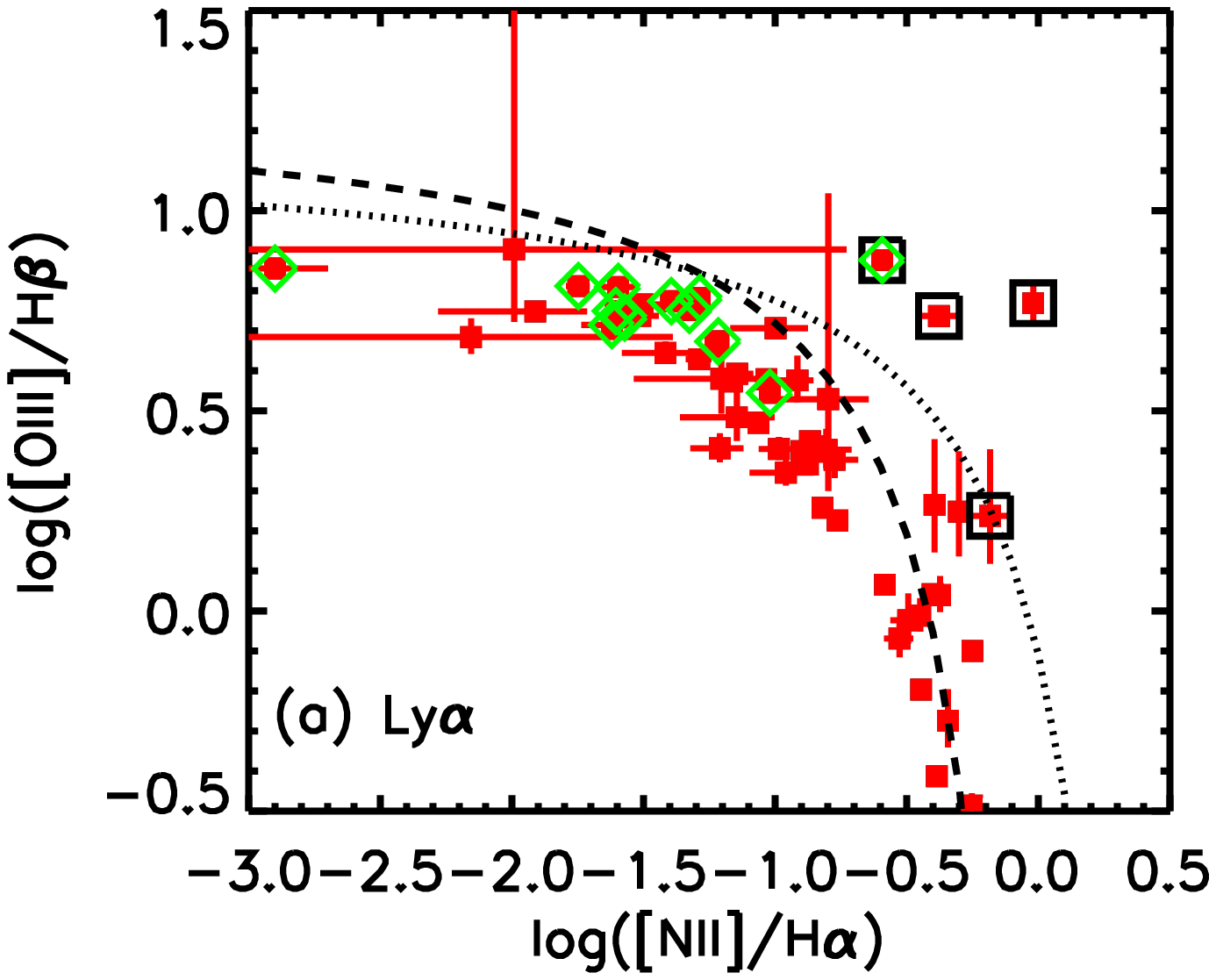}
\includegraphics[width=4.0in]{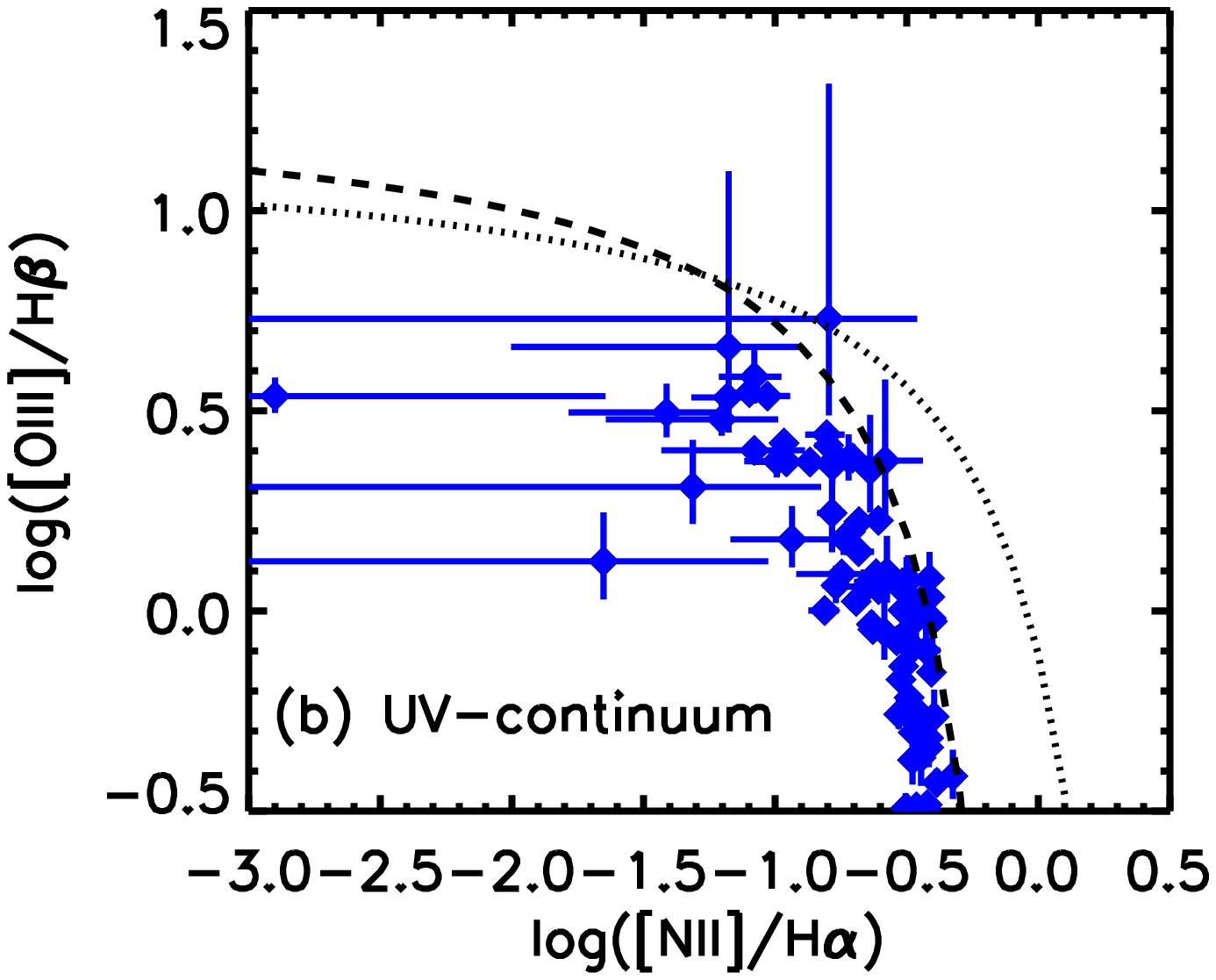}
\caption{(a) BPT diagram for the candidate LAE galaxy sample 
(Table~\ref{tab1}; red solid squares). Sources from both the main table 
and the supplement (the latter contains sources with rest-frame EW(Ly$\alpha$)
between 15 and 20~\AA) are included. Only the 58 sources with strong optical 
emission lines (as defined in the text) are shown.
The intermediate-type Seyferts in Table~\ref{tab1} are excluded.
The black open squares enclose the sources that we classify as 
AGNs on the basis of this diagram.
(b) BPT diagram for the sources in the
UV-continuum sample (blue solid diamonds).  Again, only
the 82 sources with strong optical emission lines are shown.
In both panels the green diamonds show sources where the 
[OIII]$\lambda4363$ auroral line is detected above the $3\sigma$ level,
the dashed curve denotes the Kauffmann et al.\ (2003) empirical division 
between AGNs (upper right) and star-forming galaxies (lower left), and 
the dotted curve denotes the Kewley et al.\ (2001) theoretical division.
\label{bpt}
}
\end{inlinefigure}

There remain 8 sources with optical spectra in the LAE
sample where the [OIII]$\lambda5007$
and H$\beta$ lines are too weak to place them on the BPT diagram
and where the [NII]$\lambda6584$/H$\alpha$ ratios are not high.
These galaxies are labeled as ``A'' in the optical redshift
column (col.~8) of Table~\ref{tab1} for their weak optical emission lines. 
In one case the optical spectrum is poor, but in the remaining cases
the sources either have more normal [NII]$\lambda6584$/H$\alpha$ ratios or are 
absorbers.  It is possible that these also contain AGNs.
The percentage of AGNs in the sample is $16\pm4\%$ if we include only the
classified AGNs, but that percentage would rise to $27\pm6\%$ if the 8 
weak emission-line sources were also AGNs.

Finkelstein et al.\ (2009b) optically classified 23 sources from the 
Deharveng et al.\ (2008) sample in the Groth strip. Fourteen of 
these overlap with sources in our Table~\ref{tab1}. The remaining 9 do not, 
either because the source lies outside the region of the field we used or outside the 
redshift range we used, or because we found the Ly$\alpha$ identification to be 
dubious, or because we classified the source as an AGN based on the UV spectrum.
Of the 14 sources in common, the classifications agree for 11 (3 AGNs and
8 star formers). We could not classify GALEX1420+5306, the first of the remaining
3 overlapped sources, as it has very weak emission lines. 
Finkelstein et al.\ (2009b) found this source, which they called EGS1, 
to be an AGN based on its [NII]$\lambda6584$/H$\alpha$ 
ratio. As discussed in Cowie et al.\ (2010), we believe the second of
the remaining 3 overlapped sources, 
GALEX1417+5224 (EGS2), is a high-excitation, very low metallicity galaxy 
rather than an AGN. Finally, we find that the third source, GALEX1421+5239 (EGS14), 
lies on the star-forming galaxy track in the BPT diagram. We do not 
detect high-ionization lines. This is in contrast to Finkelstein et al.\ (2009b), 
who classified the source as an AGN on the basis of both its position on 
the BPT diagram and the presence of high-ionization lines in its spectrum.

All eight of the $z=0.65-1.25$ LAEs in Table~\ref{tab2} have been optically 
observed (Figures~\ref{uv_spectra} and \ref{optical_spectra}), 
and in all cases the Ly$\alpha$ identification is confirmed with the
optical redshift. Seven of these have strong emission
lines, and none of them shows [NeV] emission or broad MgII$\lambda2800$
emission that might classify the source as an AGN. The remaining source 
(GALEX1420+5313; see Figure~\ref{optical_spectra}(e)) has a post-starburst 
spectrum. (This galaxy is also identified in the DR3 release of the DEEP2 
survey; Davis et al.\ 2007.)  We label it as ``A'' in column~8 of 
Table~\ref{tab2} for its weak optical emission lines. 
It is possible that the Ly$\alpha$
emission is AGN-dominated in this galaxy. We therefore
exclude this source from the $z=0.65-1.25$ LAE sample.

Throughout our subsequent analysis we restrict to the LAEs that are confirmed 
as star formers from the optical spectra. We eliminate all sources with AGN 
signatures in either the optical or UV (including those sources identified as 
AGNs based on the BPT diagram), and we also eliminate the optically 
unobserved galaxies, the unconfirmed galaxies, and the galaxies
with weak optical emission lines.  This leaves a sample of 40 LAEs 
with rest-frane EW(Ly$\alpha)\ge20$~\AA\ in the $z=0.195-0.44$ interval and a sample of
5 LAEs with EW(Ly$\alpha)\ge20$~\AA\ in the $z=0.65-1.25$ interval.

\begin{inlinefigure}
\includegraphics[width=3.7in]{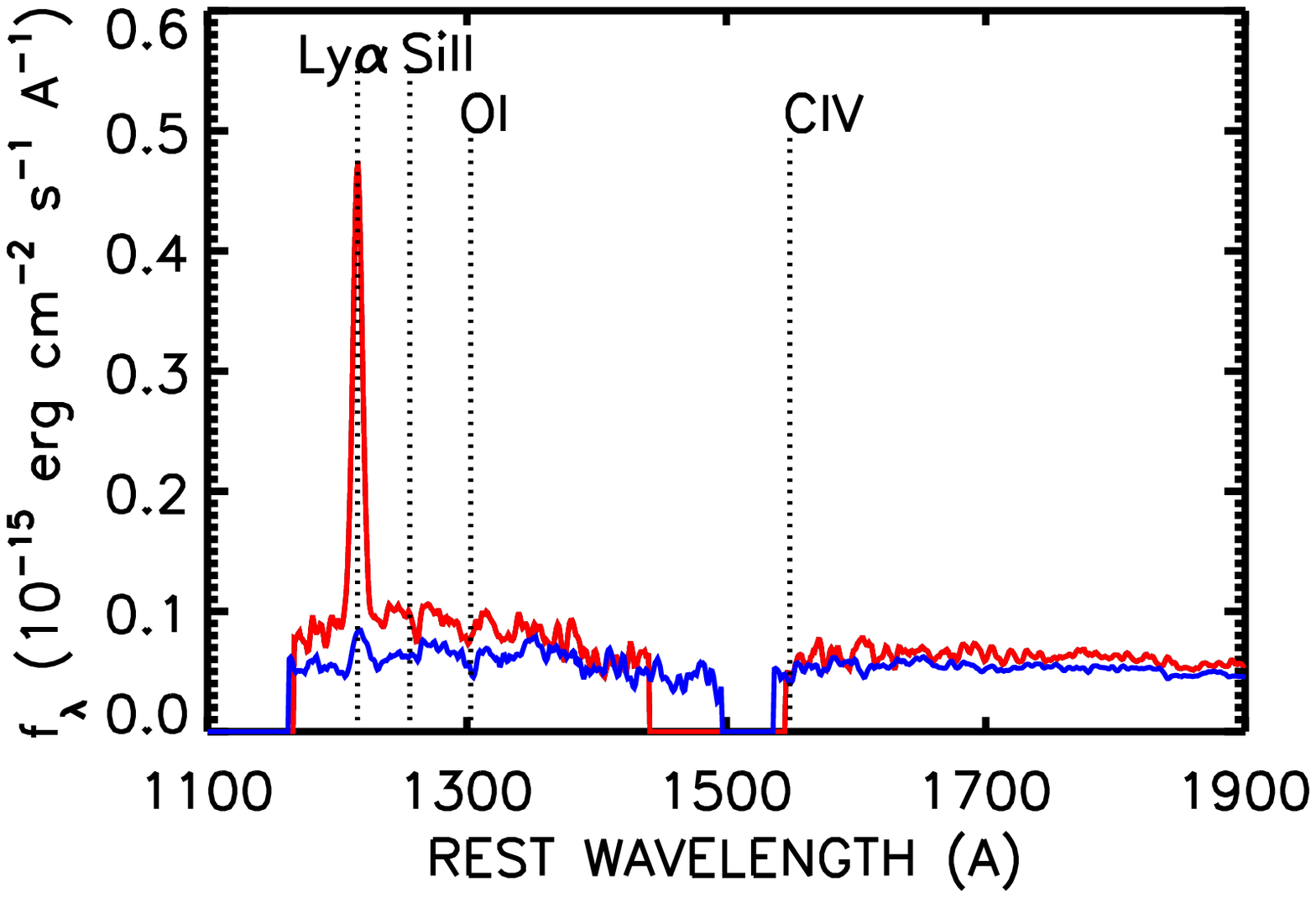}
\caption{
The average UV spectra of the 40 $z=0.195-0.44$ LAEs (red spectrum) 
and of the 101 UV-continuum sources with rest-frame 
EW(H$\alpha$)$>5$~\AA\ (blue spectrum). The gaps in the spectra 
correspond to the wavelength region between the FUV and NUV 
{\em GALEX\/} grisms. In addition to Ly$\alpha$, we mark
the positions of UV absorption features that are only weakly 
seen in these low-resolution spectra.
\label{compare_stacks}
}
\end{inlinefigure}

In Figure~\ref{compare_stacks} we compare the average UV
spectrum of the 40 $z=0.195-0.44$ LAEs (red spectrum)
with that of the UV-continuum sample
in the same redshift interval (blue spectrum).
In order to make the most direct comparison, we 
eliminate weak emission-line galaxies with rest-frame 
EW(H$\alpha$)$<5$~\AA\ in the UV-continuum
sample, since we have eliminated these sources
from the LAE sample. We also eliminate two objects
which are classified as BPT AGN in the UV-continuum sample.
This leaves 101 UV-continuum
galaxies. However, including the remaining
galaxies makes no difference to the results. 

The NUV normalizations are almost identical since the samples are 
chosen with the same distribution of NUV magnitudes. The shapes
are also nearly indistinguishable above $1400$~\AA. By selection
the LAE sample has strong emission, and the average spectrum for
this sample has a rest-frame EW(Ly$\alpha)=36$~\AA. While a
a stacking procedure is not the best way to analyze objects
with a mixture of Ly$\alpha$ emission and absorption, it
at least allows us to see that the UV-continuum sample has 
Ly$\alpha$ emission and that its stacked spectrum is
flatter than that of the LAE sample below $1400$~\AA.
Fitting the Ly$\alpha$ emission, we find a rest-frame 
EW(Ly$\alpha)=7$~\AA\ for the emission
feature. This is comparable to, but slightly higher than,
the median value of EW(Ly$\alpha)=4$~\AA\ found for
the $z\sim 3$ LBG population by Kornei et al.\ (2010). The $z\sim 3$ 
LBG sample includes the LAEs as a subsample; thus, if anything, our 
sample, which excludes the LAEs, should be lower.

\section{LAE Luminosity evolution}
\label{lae_lum_evol}

A rest-frame EW(Ly$\alpha)\ge20$~\AA\ is normally used to define 
the high-redshift LAE population (e.g., Hu et al.\ 1998), so in 
this section we apply that criterion. 
In Figure~\ref{lalum_evol} we plot Ly$\alpha$ luminosity versus 
redshift and compare it with the measured luminosities of high-redshift LAE samples 
taken from Shimasaku et al.\ (2006), Ouchi et al.\ (2008), and Hu et al.\ (2010),
all of which were chosen with the same EW selection criterion.

\begin{inlinefigure}
\includegraphics[width=4.0in]{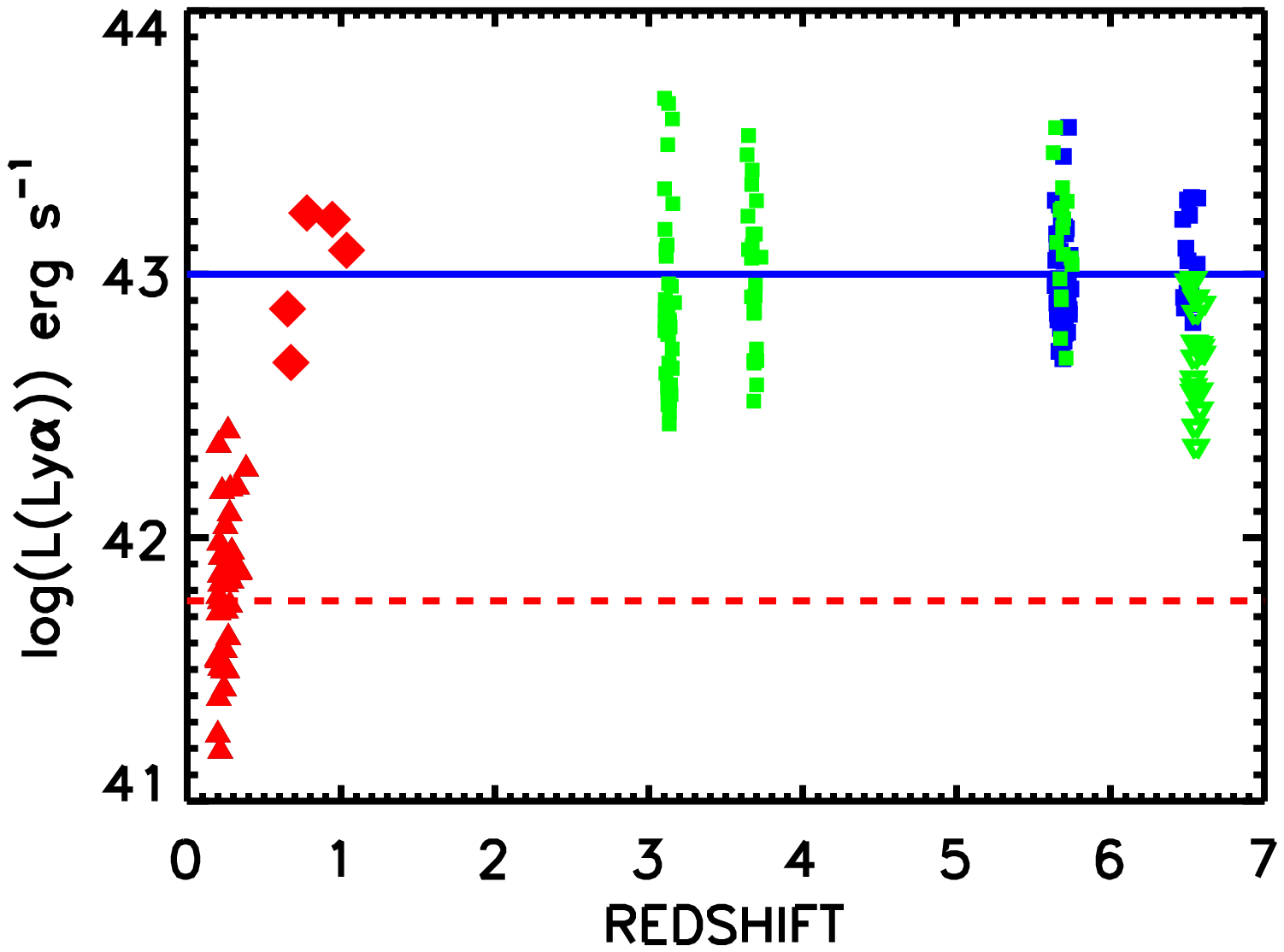}
\caption{
Ly$\alpha$ luminosity vs. redshift for the LAEs ($z=0.195-0.44$ --- 
small red triangles; $z=0.65-1.25$ --- larger red diamonds).
Only sources with rest-frame EW(Ly$\alpha)\ge20$~\AA\ are shown.
At higher redshifts we show the Ly$\alpha$ luminosities from 
Ouchi et al.\ (2008) (green squares),
Shimasaku et al.\ (2006) (green downward pointing triangles),
and Hu et al.\ (2010) (blue squares). 
The red dashed line shows the $L_\star$ derived by Cowie et al.\ (2010) 
from a Schechter (1976) function fit to the $z=0.195-0.44$ luminosity
function with a slope of $\alpha=-1.36$. The blue solid line
shows the $L_\star$ at $z=5.7$ for $\alpha=-1.36$ (Hu et al.\ 2010).
\label{lalum_evol}
}
\end{inlinefigure}

The primary conclusion that we draw from Figure~\ref{lalum_evol}
is that there is a dramatic change in the properties of
the LAEs between $z=1$ and $z=0$. There are
no sources in the $z=0.195-0.44$ sample with luminosities
close to those of the brightest LAEs at high redshifts,
but such sources are present by $z\sim1$.
This is not simply a volume effect: Cowie et al.\ (2010)
determined the LAE luminosity function in the $z=0.195-0.44$ redshift
interval and showed that the $L_\star$ for a Schechter (1976)
function fit with $\alpha=-1.36$ (red dashed line) is a factor of 
8 fainter than that measured by Gronwall et al.\ (2007) at $z\sim3$ 
and more than an order of magnitude fainter than that
measured by Hu et al.\ (2010) at $z=5.7$ (blue solid line).

We have computed the $z=0.65-1.25$ Ly$\alpha$ luminosity
function for the observed sample using the $1/V$ technique
(Felten 1976) with the accessible volumes calculated from
the areas versus NUV magnitudes in the sample. We show
the luminosity function with the open black squares in 
Figure~\ref{high_la_lumfun}.  As discussed extensively in 
Deharveng et al.\ (2008) and Cowie et al.\ (2010),
this Ly$\alpha$ luminosity function corresponds to sources
with NUV$<22$ and must be corrected for LAEs
at fainter continuum magnitudes. In the redshift interval
$z=0.195-0.44$ most sources with
Ly$\alpha$ luminosities in the $10^{42}-10^{43}$~erg~s$^{-1}$ 
range lie at NUV$<22$ and the corrections are
small (Cowie et al.\ 2010). In contrast, many of the LAEs
wth Ly$\alpha$ luminosities
$\sim 10^{43}$~erg~s$^{-1}$ at $z=0.65-1.25$ 
are fainter than NUV$>22$ and the correction is substantial. 
Following the procedures outlined in Deharveng et al.\ (2008) 
and Cowie et al.\ (2010), we have computed the correction by
assuming the $z \sim 1$ NUV counts derived from
the luminosity functions of Arnouts et al.\ (2005)
and the shape of the equivalent width
distribution derived in Cowie et al.\ (2010)
for the $z=0.195-0.44$ sample. (This shape is
almost identical to that seen at $z \sim 3$
by Shapley et al.\ 2003.) We show the corrected points
with the solid squares in Figure~\ref{high_la_lumfun}.

We compare with the Ly$\alpha$ luminosity
functions at $z\sim0.3$ (blue curve), 
$z\sim3$ (red curve), $z\sim5.7$ (green solid curve),
and $z\sim6.5$ (green dashed curve) using the Schechter (1976)
function fits given in Cowie et al.\ (2010),
Gronwall et al.\ (2007), and Hu et al.\ (2010). Even allowing 
for the small numbers in the present sample and the substantial
corrections, we can see that
there is a dramatic evolution in the luminosity
function between $z \sim 0.3$ and $z \sim 0.95$.
The data do not justify a Schechter (1976) function fit,
but we show the Gronwall et al.\ (2007) luminosity function
with $L_\star$ reduced by a factor of 2.5 as
the black dashed curve. This function 
($\phi_\star =1.28\times10^{-3}$, $\alpha = -1.36$, and
$\log L_\star = 42.26$) provides a reasonable fit
to the data, though it is by no means unique. It
would imply a reduction by a factor of $2.5$ in the 
Ly$\alpha$ luminosity density between $z=3$ and $z=1$, 
which is much smaller than the dramatic factor of 50 reduction
that Cowie et al.\ (2010) estimate between $z=3$ and $z=0.3$. 
This suggests that much of the evolution occcurs 
in the $z=0-1$ range, as is seen in many other populations.

\begin{inlinefigure}
\includegraphics[width=3.5in]{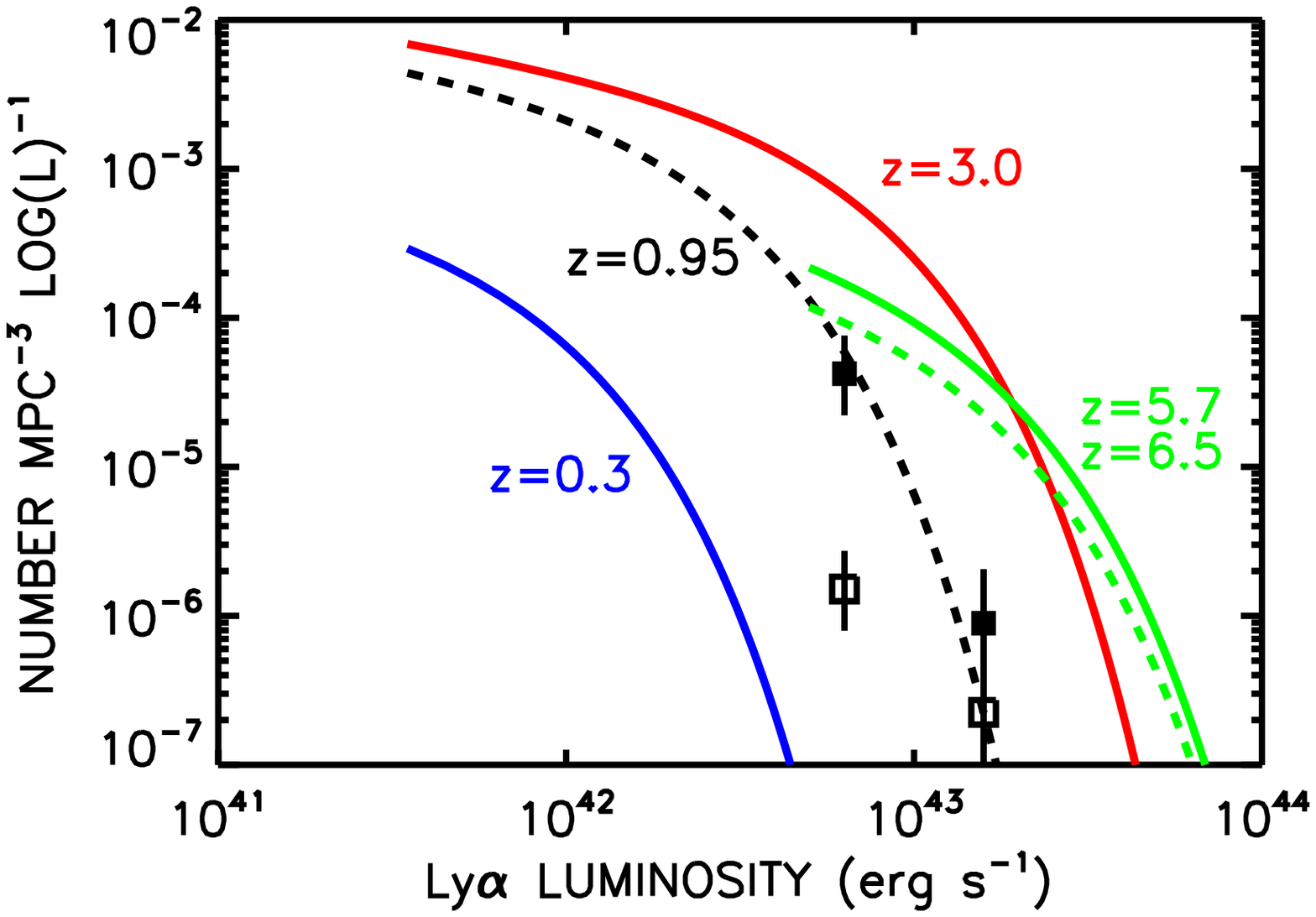}
\caption{The derived Ly$\alpha$ luminosity function at 
$z=0.65-1.25$ (black squares). The open squares show
the values calculated for the present sample,
which is drawn from galaxies with NUV$<22$.
The solid squares show the values corrected to
allow for sources with fainter continuum mangitudes
(see text). The error bars are $\pm1\sigma$ from the Poisson 
errors corresponding to the number of sources in the bin.
The blue curve shows the Schechter (1976) function
fit to the Ly$\alpha$ luminosity function at $z=0.194-0.44$
from Cowie et al.\ (2010), the red curve shows the fit
at $z=3$ from Gronwall et al.\ (2007), and the green curves
show the fits at $z=5.7$ (solid) and $z=6.5$ (dashed)
from Hu et al.\ (2010). The black dashed
curve shows the Gronwall et al.\ luminosity function with
$L_{\star}$ reduced by a factor of 2.5 
to fit the $z=0.65-1.25$ points.
\label{high_la_lumfun}
}
\end{inlinefigure}

\section{What are the LAEs?}
\label{la_metal}

\subsection{Overview}
\label{overview}

We now turn to the $z=0.195-0.44$ samples alone and what they can 
tell us about how LAE galaxies are drawn from the more general
UV-continuum selected galaxy population. In this section we will
occasionally include the LAEs with EW(Ly$\alpha)<20$~\AA\ from the 
Table~\ref{tab1} supplement when this is appropriate. We parameterize
the data with the rest-frame EW(H$\alpha$). Using the EW(H$\alpha$) 
allows us to intercompare easily the properties of the 
Ly$\alpha$ and UV-continuum samples. In addition, the EW(H$\alpha$)
is a rough measure of the age of a galaxy or, more
precisely, its specific SFR (SSFR), since the
conversion from SSFR to age is dependent on the star formation
history of the source (e.g., Leitherer et al.\ 1999). 
To zeroth order we may also expect the EW(H$\alpha$) to be
independent of extinction. In contrast, the complex escape path
of the Ly$\alpha$ photons relative to the continuum photons
combined with the effects of extinction
may result in substantial changes in the EW(Ly$\alpha$). 
These have been observed by some as decreases (Shapley et al.\ 2003;
Pentericci et al.\ 2009) but not by others (Atek et al.\ 2008;
Finkelstein et al.\ 2009c).
We note, however, that even the EW(H$\alpha$) 
may have some dependence on extinction, since the stars
producing the ionizing photons may lie in different spatial
regions of the galaxy than those producing the optical continuum 
photons.

\begin{inlinefigure}
\includegraphics[width=4.0in]{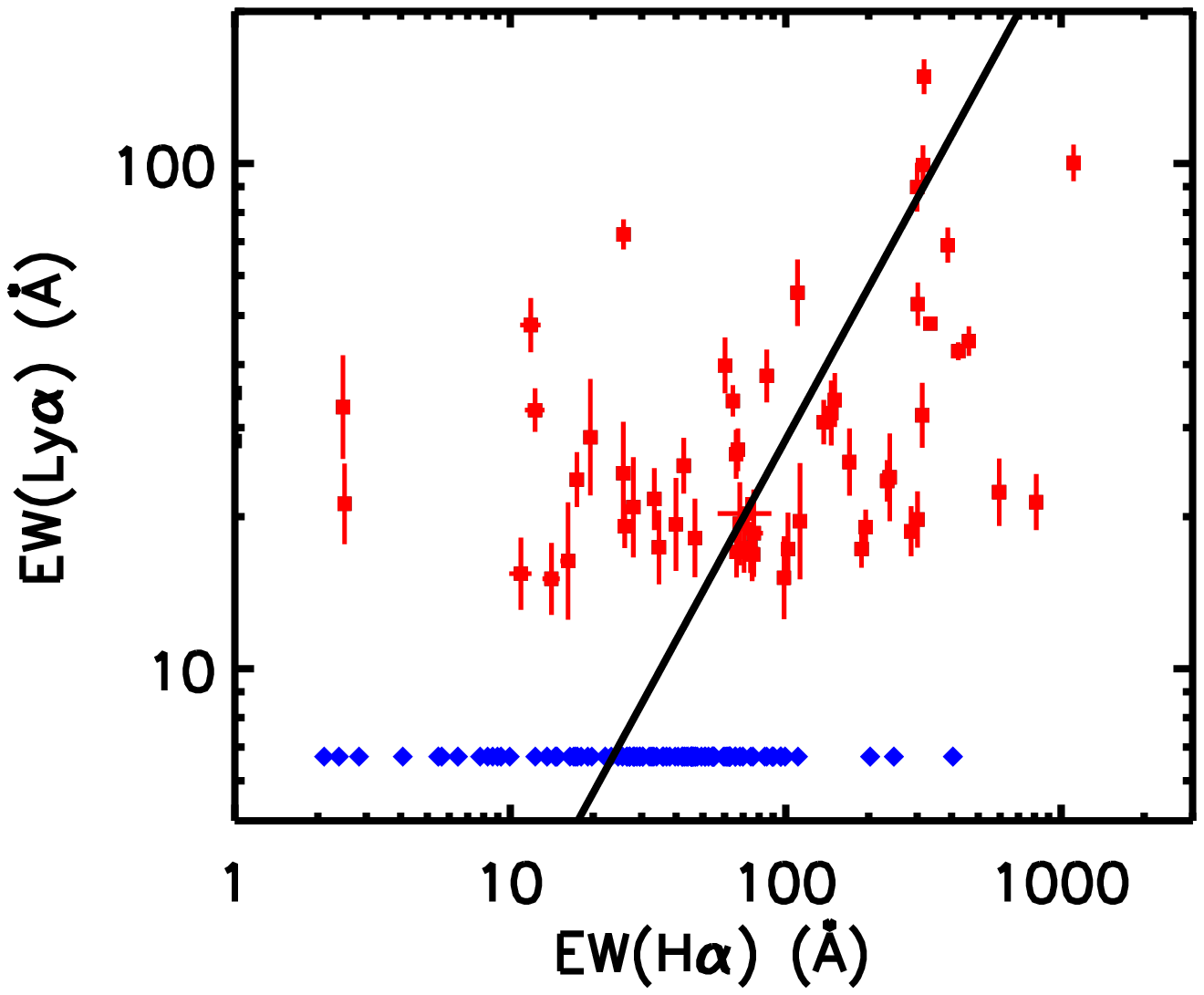}
\caption{The rest-frame EW(Ly$\alpha$) vs. the rest-frame EW(H$\alpha$). 
The red squares show the LAE galaxies (the error bars are $\pm1\sigma$),
including those with rest-frame EW(Ly$\alpha$) between 15 and 20~\AA\ 
from the supplement to Table~\ref{tab1}.
The blue diamonds show the UV-continuum sample placed at the average
EW(Ly$\alpha)=7$~\AA\ that was measured from the stacked spectrum. 
The positioning of these objects in $y$ is purely for display
purposes and is not used in the analysis.
The black line shows a linear relation corresponding to the median 
EW(H$\alpha$)/EW(Ly$\alpha)=3.5$ that was measured from the full LAE
sample with rest frame EW(Ly$\alpha)>15$\AA. 
\label{ew_ewha}
}
\end{inlinefigure}

\begin{figure*}
\includegraphics[width=3.6in]{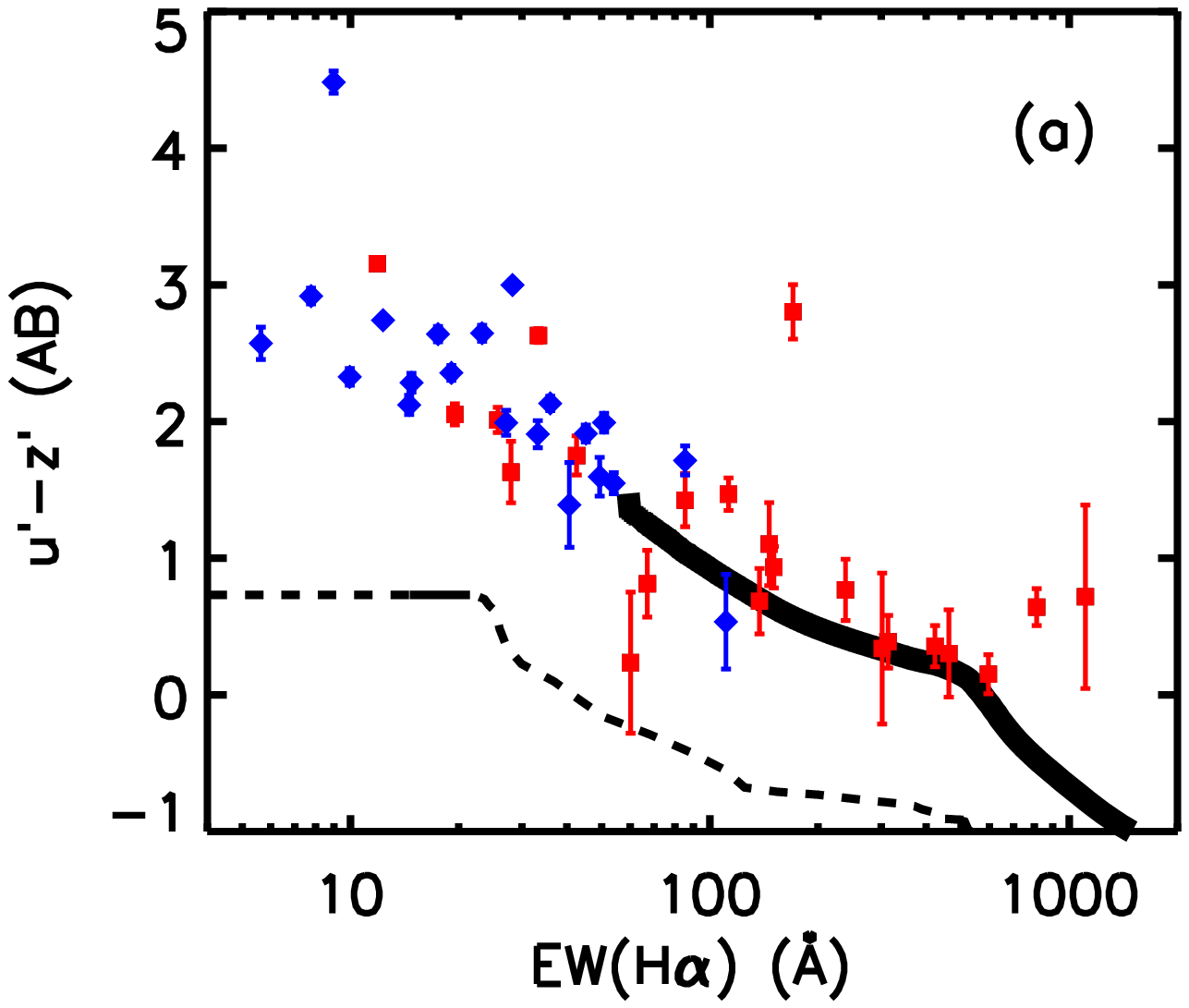}
\includegraphics[width=3.6in]{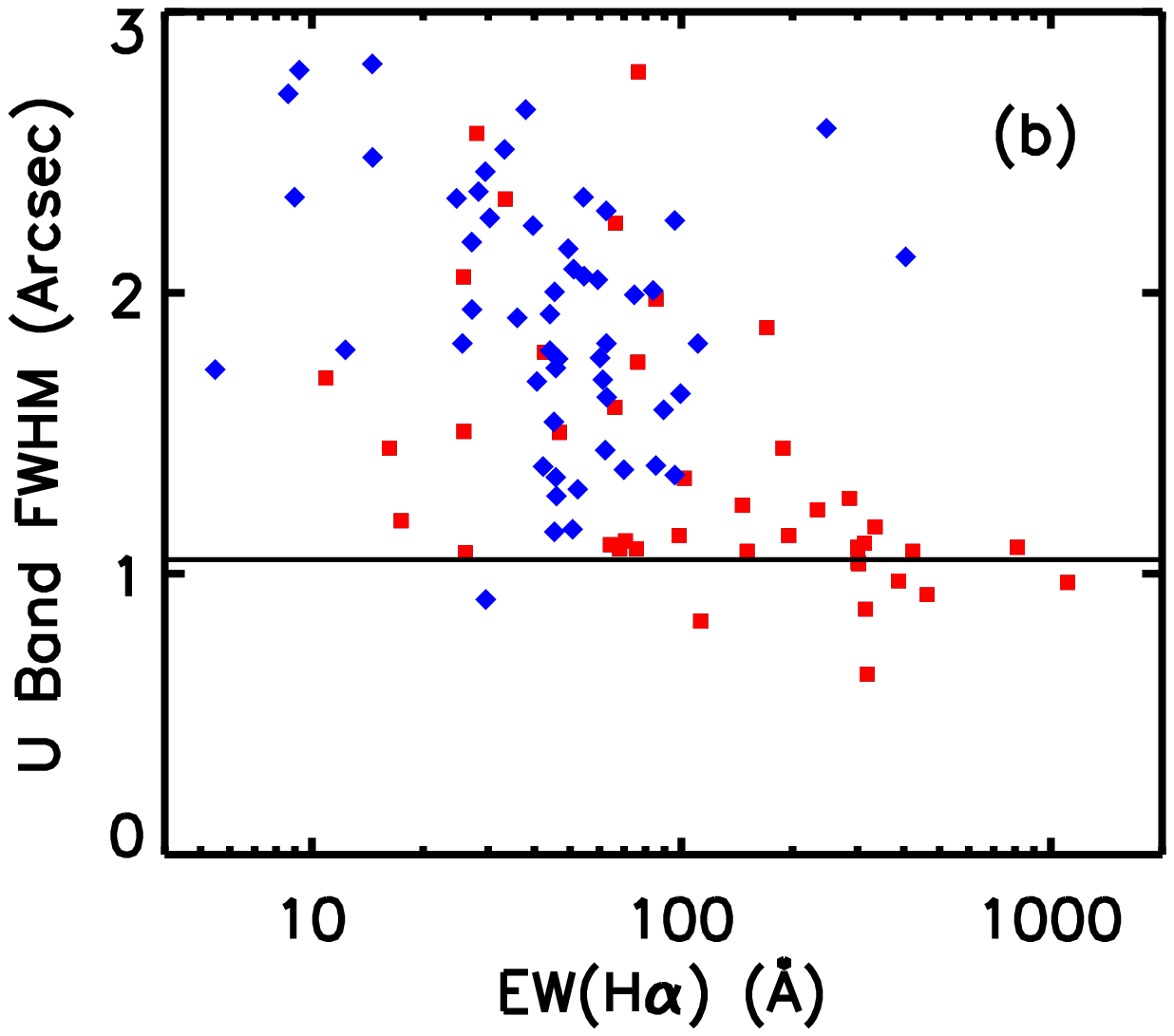}
\includegraphics[width=3.6in]{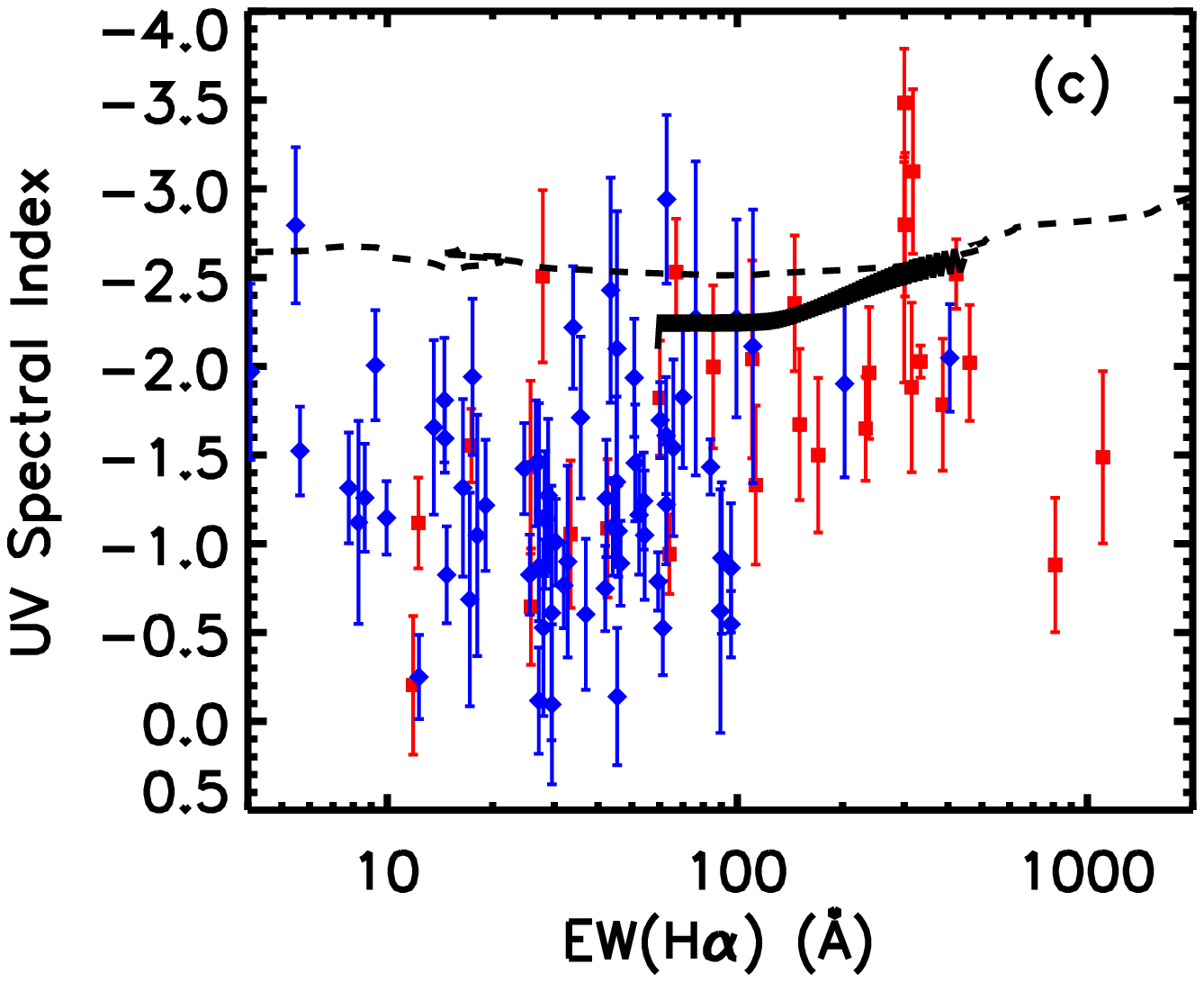}
\includegraphics[width=3.6in]{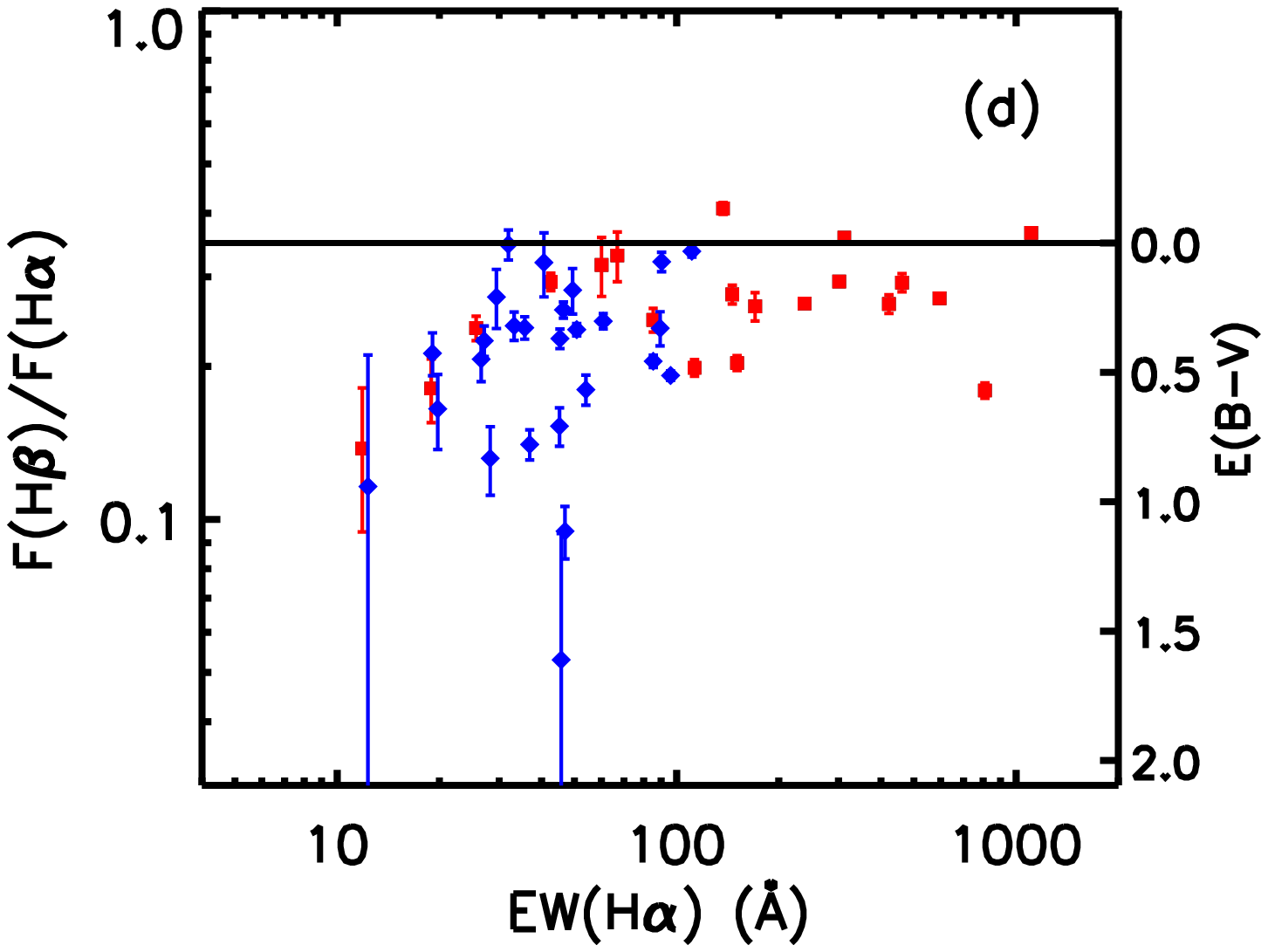}
\includegraphics[width=3.6in]{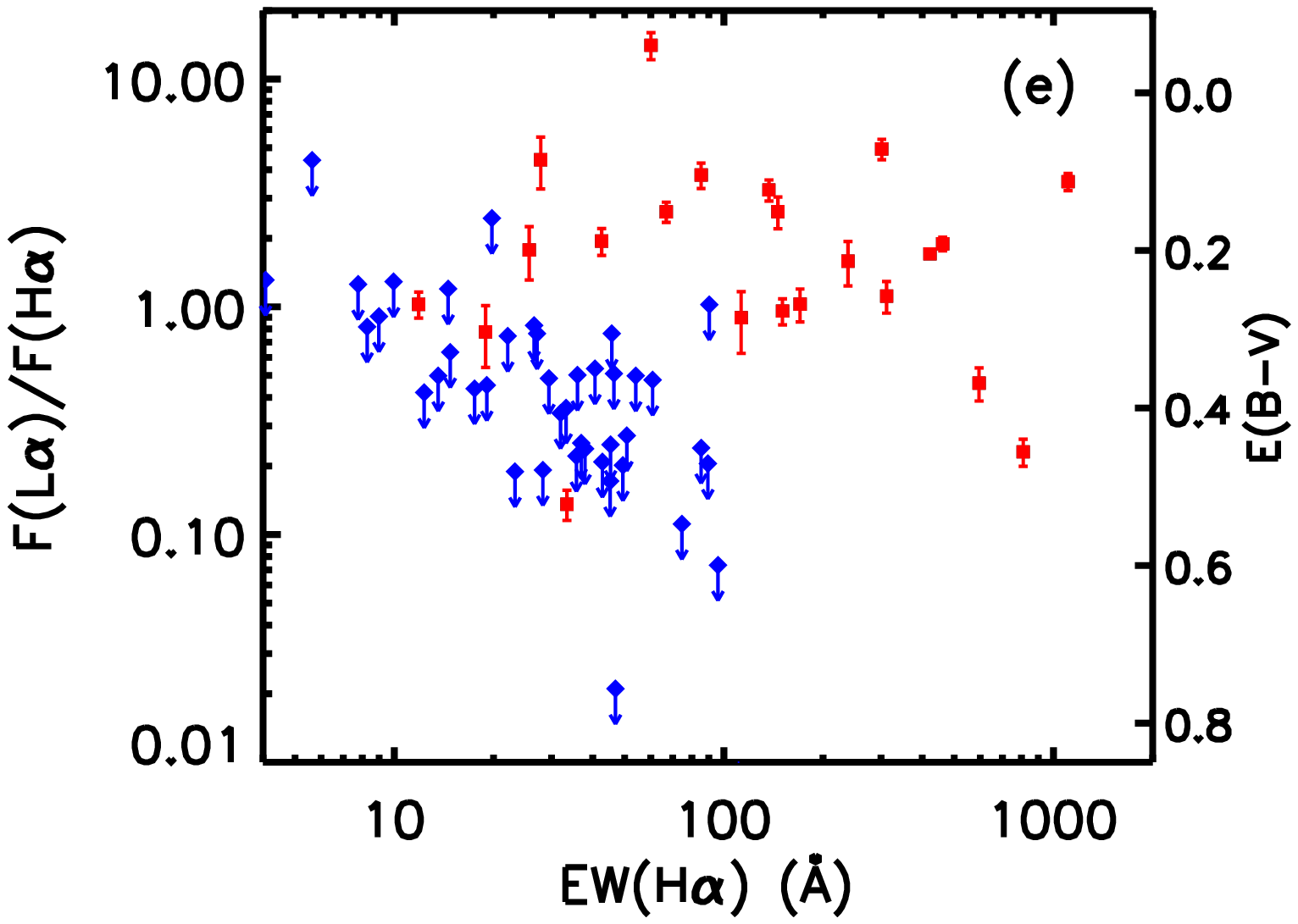}
\includegraphics[width=3.6in]{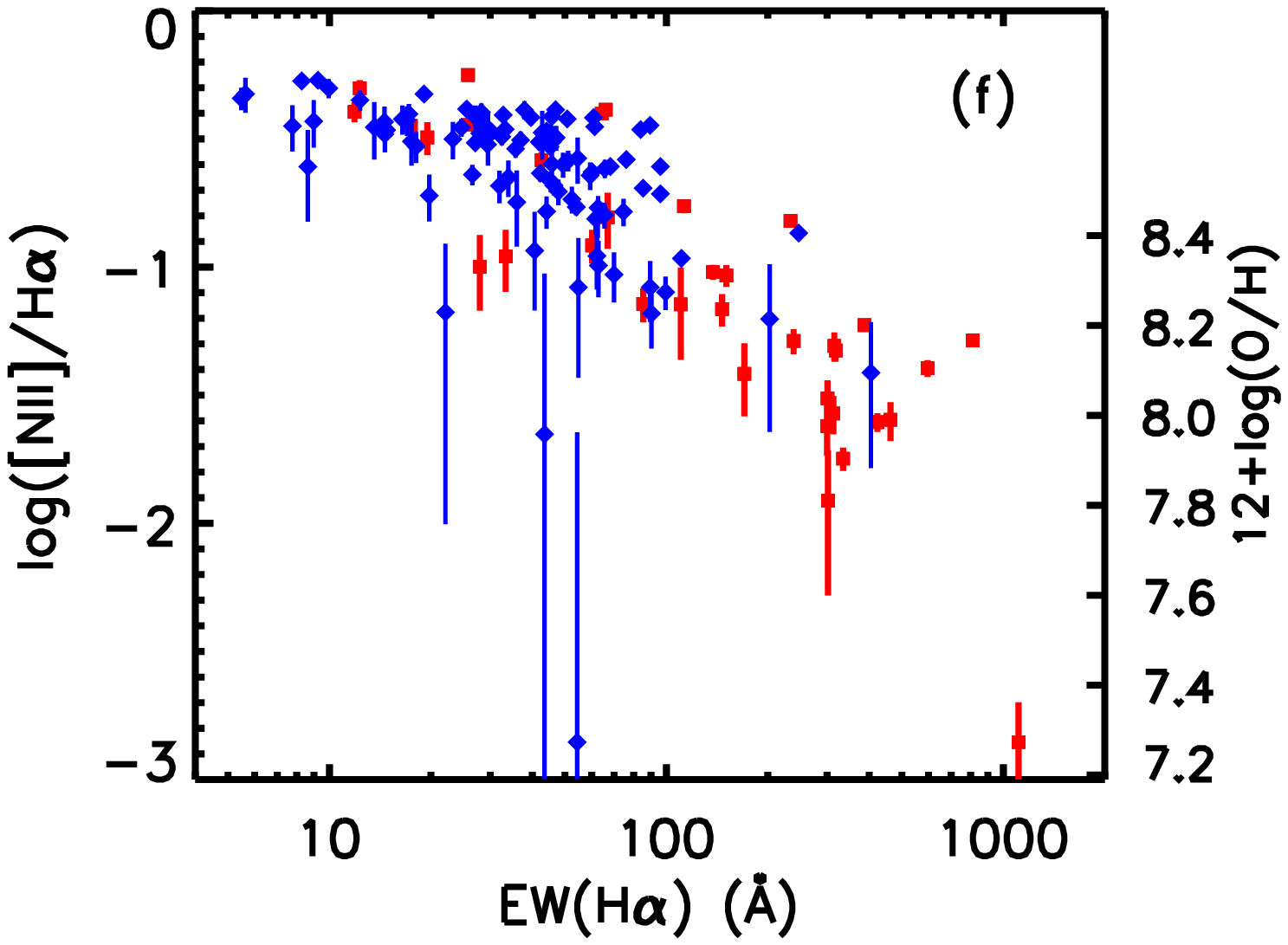}
\caption{The (a) SDSS $u'-z'$ color, (b) FWHM
in CFHT MegaPrime $U$-band images, (c) UV power law slope, 
(d) $f$(H$\beta$)/$f$(H$\alpha$),
(e) $f$(Ly$\alpha$)/$f$(H$\alpha$), and 
(f) $\log$($f$([NII]$\lambda6584$)/$f$(H$\alpha))$,
all plotted vs. the rest-frame EW(H$\alpha$). 
The red squares show the LAE galaxies with EW(Ly$\alpha)\ge20$~\AA. 
The blue diamonds show the UV-continuum sample. 
Error bars are $\pm1\sigma$.
In (a) we have used the spectra to correct for the emission-line
contributions to the $z'$ magnitude (see Section~\ref{ages}).
In (a) and (c) the black curves show solar metallicity models 
from STARBURST99 with no extinction: a constant SFR model
(thick solid) and a single instantaneous starburst model
(dashed). 
The right-hand axes in (d) and (e) show 
the E(B-V) values that would be derived from the Calzetti et al.\ (2000) 
reddening law.  The right-hand axis in (f) shows the O abundance 
that would be derived from $f$([NII]$\lambda6584$)/$f$(H$\alpha$)
using the Pettini \& Pagel (2004) conversion.
\label{ewha_plots}
}
\end{figure*}

While in the absence of dust we may expect the flux of Ly$\alpha$ to be related 
to that of H$\alpha$ by the case~B ratio (8.7 for a temperature of 10,000~K;
Brocklehurst 1971), 
the relation between the EW(H$\alpha$) and the EW(Ly$\alpha$) is more complex 
because of the relative evolution of the continuum fluxes.  The EW(Ly$\alpha$) 
declines rapidly after a single burst of star formation, as does
the EW(H$\alpha$).  However, for a constant SFR model, the EW(Ly$\alpha$)
instead becomes roughly constant, since the UV-continuum flux and the Ly$\alpha$ 
line strength are governed by ongoing massive star formation
(Charlot \& Fall 1993; Schaerer \& Verhamme 2008).
In contrast, the EW(H$\alpha$) continues to decline, even for a constant SFR model, 
as the older stars build up in the galaxy. For a galaxy with a large old population 
in place, the onset of a new burst of star formation will give a high EW(Ly$\alpha$) 
while leaving the EW(H$\alpha$) relatively weak. It is this ongoing build-up of the 
older stars that makes the EW(H$\alpha$) a function of the age of the source.

In Figure~\ref{ew_ewha} we show the rest-frame EW(Ly$\alpha$) versus the 
rest-frame EW(H$\alpha$) for the LAE galaxies (red squares), 
including those with rest-frame EW(Ly$\alpha$) between 15 and 20~\AA\ 
from the supplement to Table~\ref{tab1}.  
We also show the spread of the EW(H$\alpha$) for the 
UV-continuum sample (blue diamonds).  
While there is a large scatter, there is a 
broad general trend for the LAEs to have higher EW(H$\alpha$).  Specifically, we 
find that in the EW(Ly$\alpha)\ge20$~\AA\ and EW(Ly$\alpha$)=$15-20$~\AA\ 
samples, the median EW(H$\alpha$) is 113 $(66-233)$~\AA\ and 71 $(47-77)$~\AA,
respectively, while in the UV-continuum sample, it is 38 $(31-44)$~\AA,
where the quantities in parentheses are the $68\%$ confidence range.  
A Mann-Whitney rank-sum test gives less than a $3\times10^{-6}$ probability
that the distribution of EW(H$\alpha$) is the same in the LAE
sample as in the UV-continuum sample. For the 12 highest EW((Ly$\alpha$)
objects with rest-frame EW(Ly$\alpha)\ge40$~\AA\ the rank sum
test give less than a $8\times10^{-3}$ probability that the distribution of EW(H$\alpha$)  
is the same as that for those with with weaker EWs between 15 and 40~\AA.
This suggests that the LAEs are preferentially
drawn from the younger galaxies in the sample. 

These results differ from those of \"{O}stlin et al.\ (2009), who suggested, 
based on their observations of local blue compact galaxies, that there 
might be an anti-correlation between EW(Ly$\alpha$) and EW(H$\alpha$).
As \"{O}stlin et al.\ noted, their small sample is highly selected and their 
result is not statistical. The \"{O}stlin et al.\ sample does emphasize, however,
that there are galaxies such as SBS~0335-052 that
have a very high EW(H$\alpha$)=1434~\AA\ but
do not have strong Ly$\alpha$ emission. 
We do not detect any sources this extreme in 
our UV-continuum sample, where only four sources even have 
EW(H$\alpha$)$>100$~\AA, so they are clearly rare.

We can take this further by comparing other properties
of the galaxies with the EW(H$\alpha$). In Figure~\ref{ewha_plots}
we plot the following versus the rest-frame EW(H$\alpha$):
(a) the SDSS $u'-z'$ colors, (b) the FWHM of the
galaxies in the CFHT $U$-band images, (c) the UV
spectral index, (d) the flux ratio $f$(H$\beta$)/$f$(H$\alpha$),
(e) the flux ratio $f$(Ly$\alpha$)$/f$(H$\alpha$), and 
(f) the logarithm of the flux ratio $f$([NII]$\lambda6584$)/$f$(H$\alpha$). 
Panel (b) is restricted
to the subsample covered by the $U$-band CFHT images, while panels 
(a), (d), and (e) correspond to the subsample covered by the SDSS 
observations. Panels (c) and (f) contain the full sample. 
In panel (a) we have used the spectra to correct for the contributions 
of emission lines to the $z'$ magnitude (see Section~\ref{ages}), since 
these can make significant contributions that are a function of the 
EW(H$\alpha$).  We now consider each of these relations in detail.

\subsection{Colors and Ages}
\label{ages}

The $u'-z'$ colors of Figure~\ref{ewha_plots}(a) follow a single 
well-defined track (see also Overzier et al.\ 2008, who show a 
similar figure for their low-redshift analogs to the LBGs).  
We compare the data with a constant star formation model (solid curve) 
computed with the STARBURST99 code (Leitherer et al.\ 1999). 
Here the EW(H$\alpha$) decreases as the age increases so time
increases to the left in the plot.
The predicted evolution of the $u'-z'$ color versus the EW(H$\alpha$) 
up to an age of $10^{9}$~yr matches to the emission-line corrected 
colors. By contrast, a single starburst (dashed curve) does not match, 
 since the EW(H$\alpha$) drops rapidly with time, while the $u'-z'$ 
colors remain blue. 

We can use the full color information to infer ages and extinctions 
by spectral energy distribution (SED) fitting.  
For sources with strong optical emission lines
we emphasize that it is critical to remove the substantial 
line contributions.  The correction is a function of the EW(H$\alpha$)
and hence introduces a systematic bias into any comparison of LAEs with 
UV-continuum selected galaxies. For strong-line sources SED fitting to 
the uncorrected data substantially overestimates the ages. This then 
causes the extinctions to be underestimated and the masses to be 
overestimated.

We illustrate the problem in Figure~\ref{sed_fit}, where we show
the broadband fluxes before (black open squares) and after (red diamonds)
line correction for three LAE galaxies spanning a wide 
range in the EW(H$\alpha$). The strong [OIII]$\lambda5007$ and H$\alpha$ line 
contributions raise the SED at wavelengths above $4000$~\AA. In the fitting
procedure this is spuriously interpreted as a strong $4000$~\AA\ break,
which results in a substantial age overestimate. The strongest 
emitters (Figures~\ref{sed_fit}(a) and \ref{sed_fit}(b)) are, in fact, 
very young but are misinterpreted as having ages of 0.45 and 0.9~Gyr 
when fits to the uncorrected data are used.

We have fitted ages and extinctions for all of
the galaxies covered by the SDSS in both the LAE
and UV-continuum samples. We used the Bruzual \& Charlot
(2003) models with solar metallicities and a range
of exponentially declining SFR models with e-folding
times from 1~Gyr to 20~Gyr. We combined these with a Calzetti
et al.\ (2000) extinction law. For each star formation 
history we determined the age
and extinction that produced a $\chi^2$ minimized
fit using the measured errors in the broadband fluxes,
together with a $10\%$ error to allow for systematic
uncertainties. 

It is well known that the EW(H$\alpha$) can be used as an age 
indicator (e.g., Leitherer et al.\ 1999; Bruzual \& Charlot 2003).
We plot the galaxy ages determined from the SED fits versus
the EW(H$\alpha$) in Figure~\ref{ewha_time}. 
In Figure~\ref{ewha_time}(a) we show the spurious ages
that would be derived from a simple fit to the
data without the emission-line corrections. 
In Figure~\ref{ewha_time}(b) we show the ages derived from 
the corrected broadband fluxes.
The inferred ages are somewhat degenerate with
the assumed exponential star formation history, so
we also show the range of ages 
corresponding to the range of SFR models to give a measure 
of the modeling uncertainties.
We can see from Figure~\ref{ewha_time}(a) that
nearly all of the sources with strong EW(H$\alpha$)
are assigned ages of about a Gyr. Thus, without
the line correction, we would consider there to
be no difference in age for a wide range of
EW(H$\alpha$). However, we can see from Figure~\ref{ewha_time}(b) 
that when the line correction is made, we obtain a smooth 
evolution in age with the strongest EW(H$\alpha$) sources
having the lowest ages of about 10~Myr. The high EW(H$\alpha$)
sources are quite well described by a constant
SFR model (black solid curve in Figure~\ref{ewha_time}(b)),
but at lower EW(H$\alpha$), the sources tend to fall below this
curve, presumably as the SFR begins to decline. The data are quite 
well represented by a single power law fit 
\begin{equation}
\log {\rm T} = 5.48\pm0.23 + (-1.53\pm0.12) \times {\rm log}({\rm EW(H}\alpha)) \,,
\label{ageha}
\end{equation}
where the age T is in Myr and the EW(H$\alpha$) is in \AA.
This is shown by the green dashed line in Figure~\ref{ewha_time}(b).

\begin{inlinefigure}
\vskip 0.2cm
\includegraphics[width=3.23in]{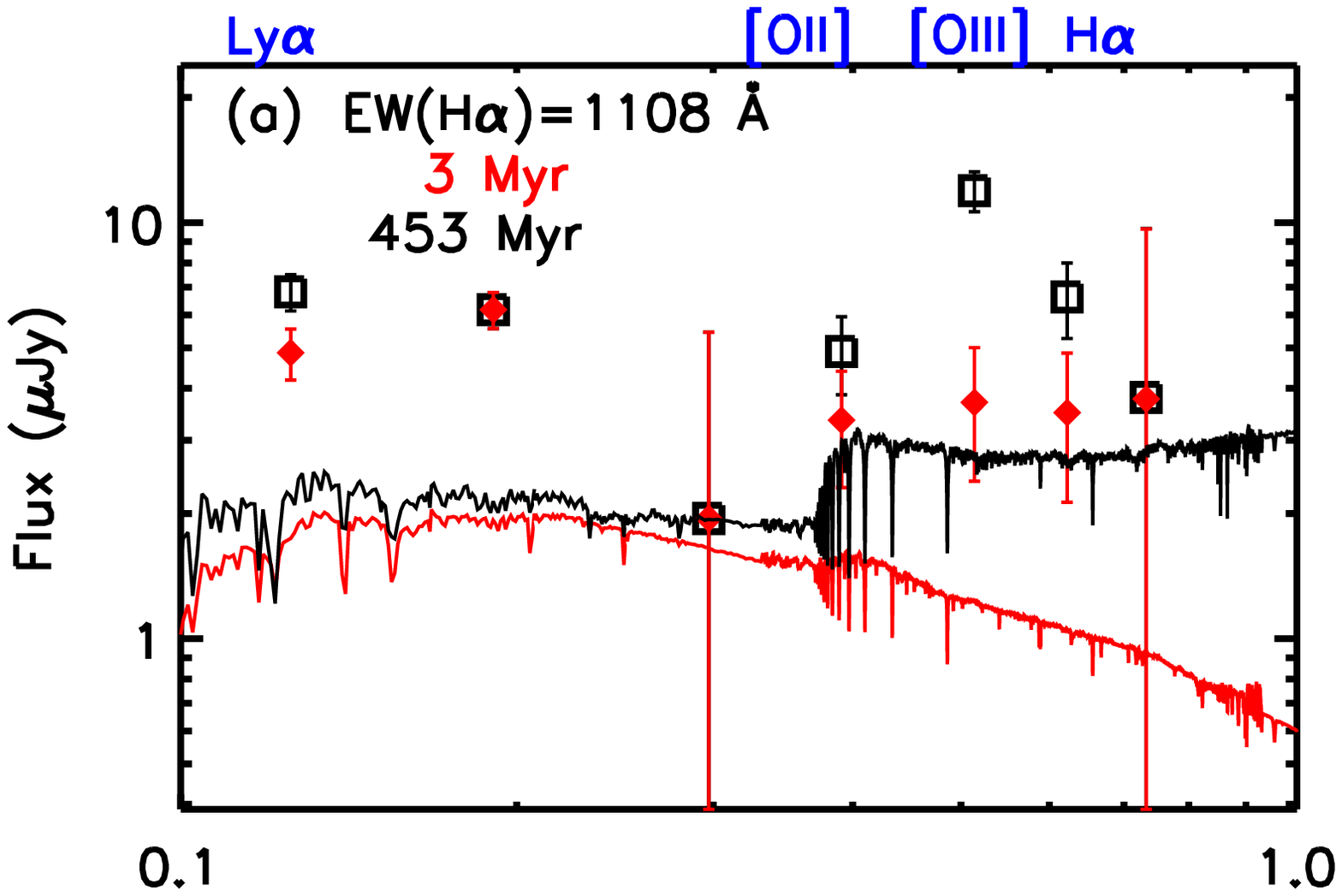}
\vskip -1.2cm
\includegraphics[width=3.23in]{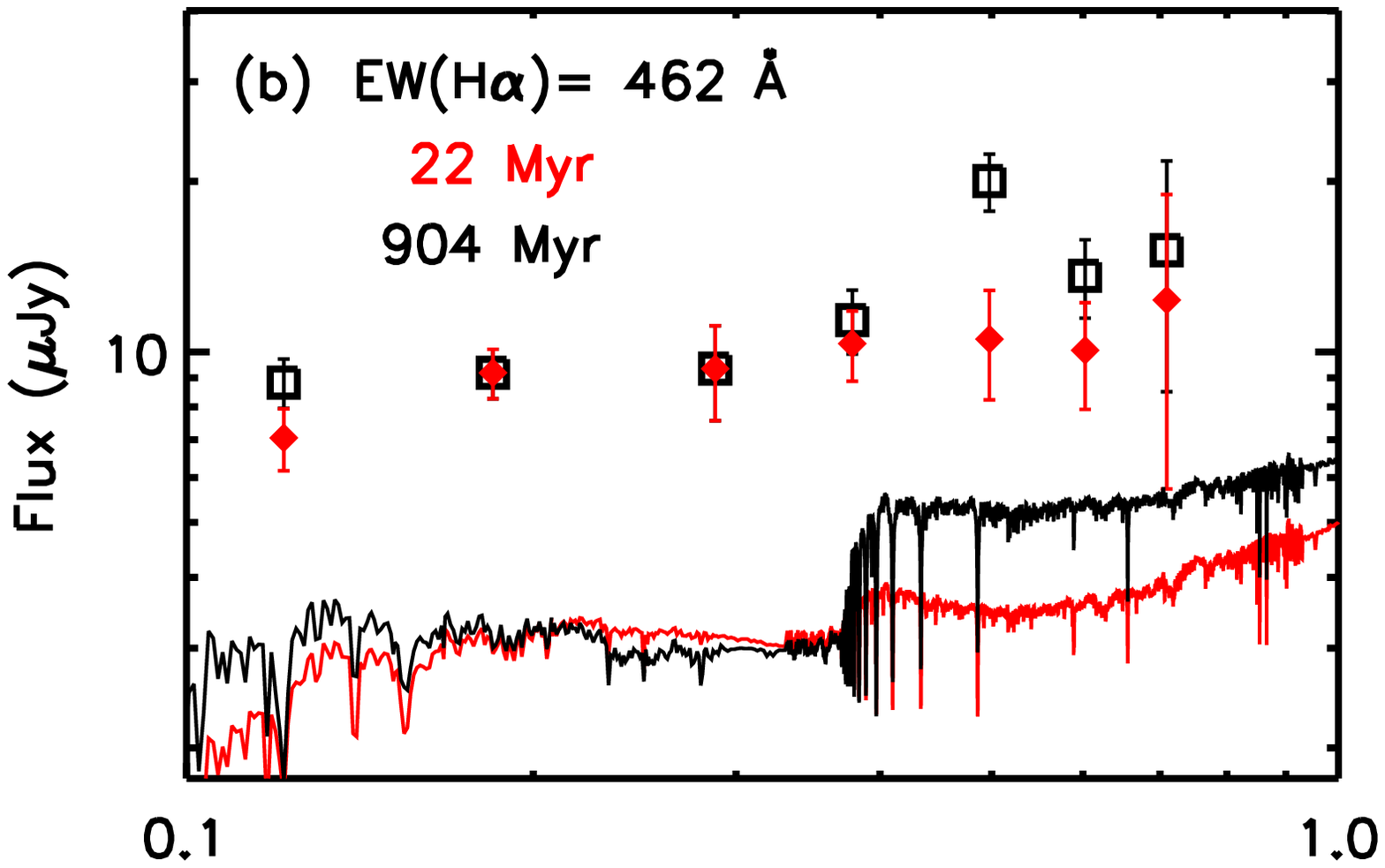}
\vskip -1.2cm
\includegraphics[width=3.23in]{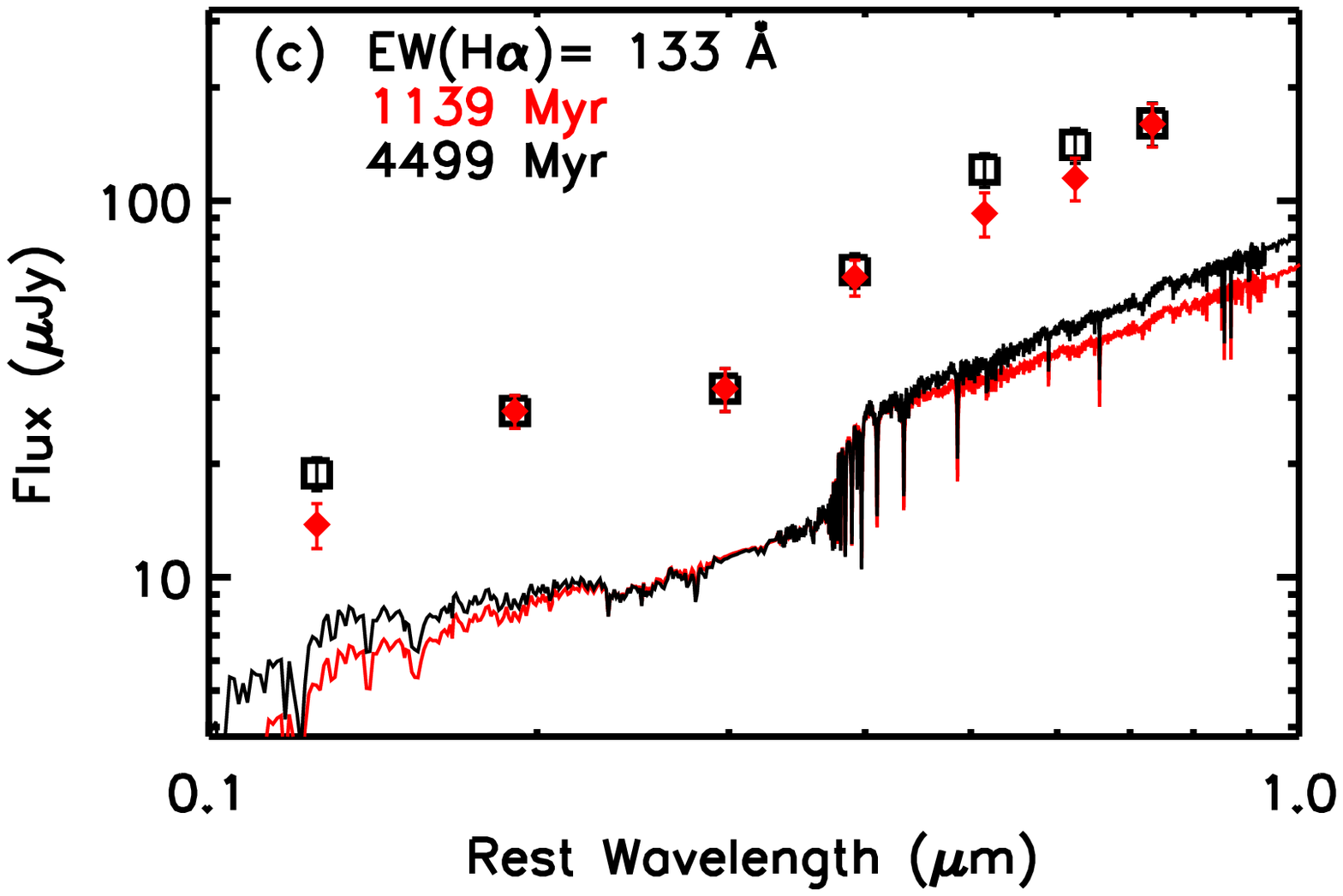}
\caption{
\label{sed_fit}
SEDs before (black open squares) and after (red solid diamonds) removal 
of the emission-line contributions. Each panel shows a LAE
galaxy with the rest-frame EW(H$\alpha$) given in the upper-left corner.
For each spectrum we removed the Ly$\alpha$, [OII]$\lambda3727$, H$\gamma$, 
H$\beta$, [OIII]$\lambda4959$ and $\lambda5007$, H$\alpha$, [NII]$\lambda6548$ and
$\lambda6584$, and [SII]$\lambda6716$ and $\lambda6738$ emission lines using the 
Gaussian fits. We then integrated the spectrum (with and without the lines) through 
the filter response to determine the fraction of the light that is produced 
by the continuum and used this to correct the magnitudes.  The principal 
contributing lines are marked at the top of the figure. The error bars are 
$\pm 1\sigma$, including a $10\%$ flux error to allow for possible
systematic errors in the determination of the total magnitudes.  The line 
effects are the largest for the higher EW(H$\alpha$) galaxies. The 
corrections act to smooth the SEDs. For each galaxy we show the constant
SFR model fits from Bruzual \& Charlot (2003). The black (red) curve 
shows the fit to the raw (corrected) data.
Both are reduced by a factor of three so the data points can be clearly 
seen. The fitted ages are shown in the upper left corner for the raw fit 
(black) and for the line-corrected fit (red).}
\end{inlinefigure}

\begin{inlinefigure}
\includegraphics[width=3.8in]{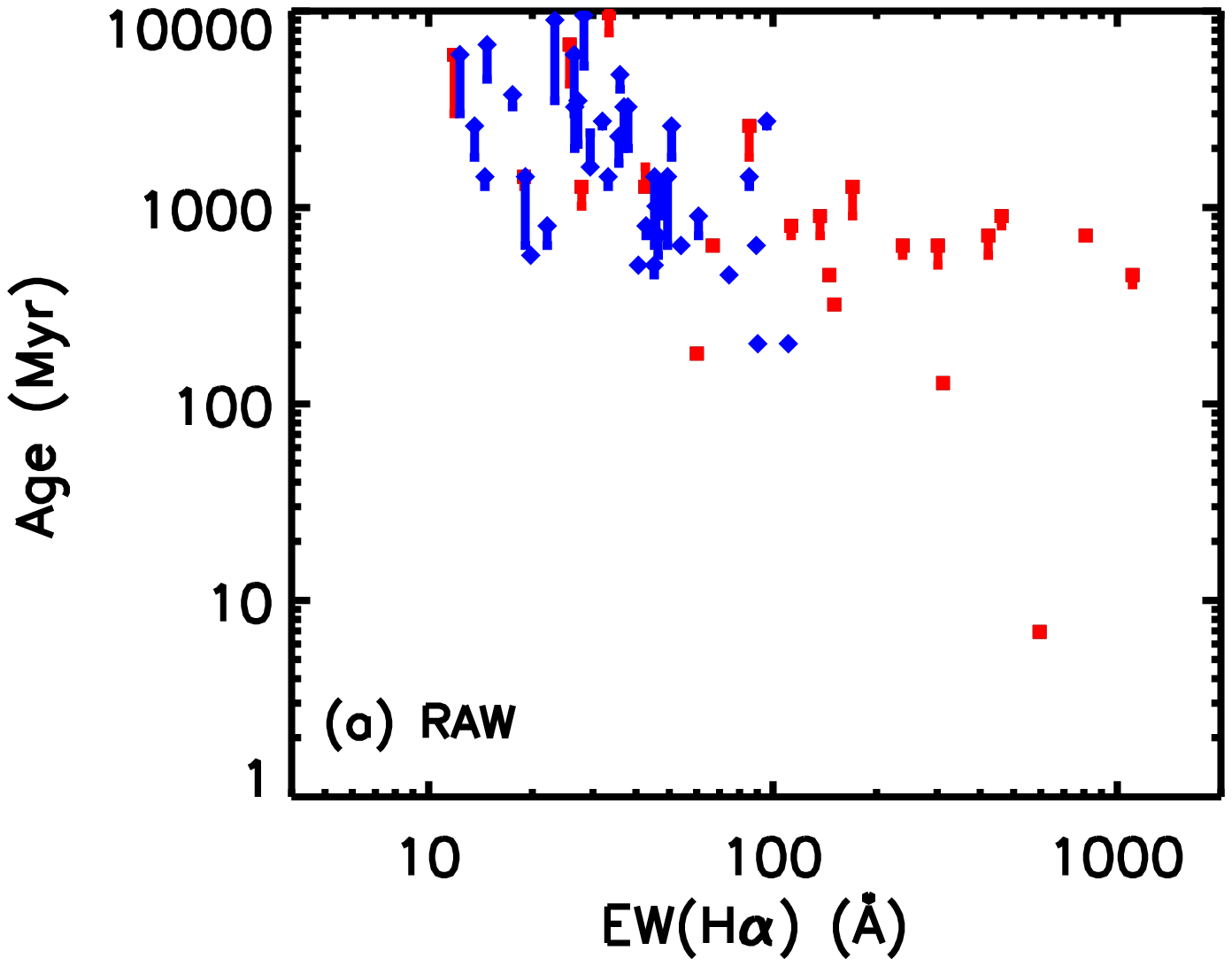}
\includegraphics[width=3.8in]{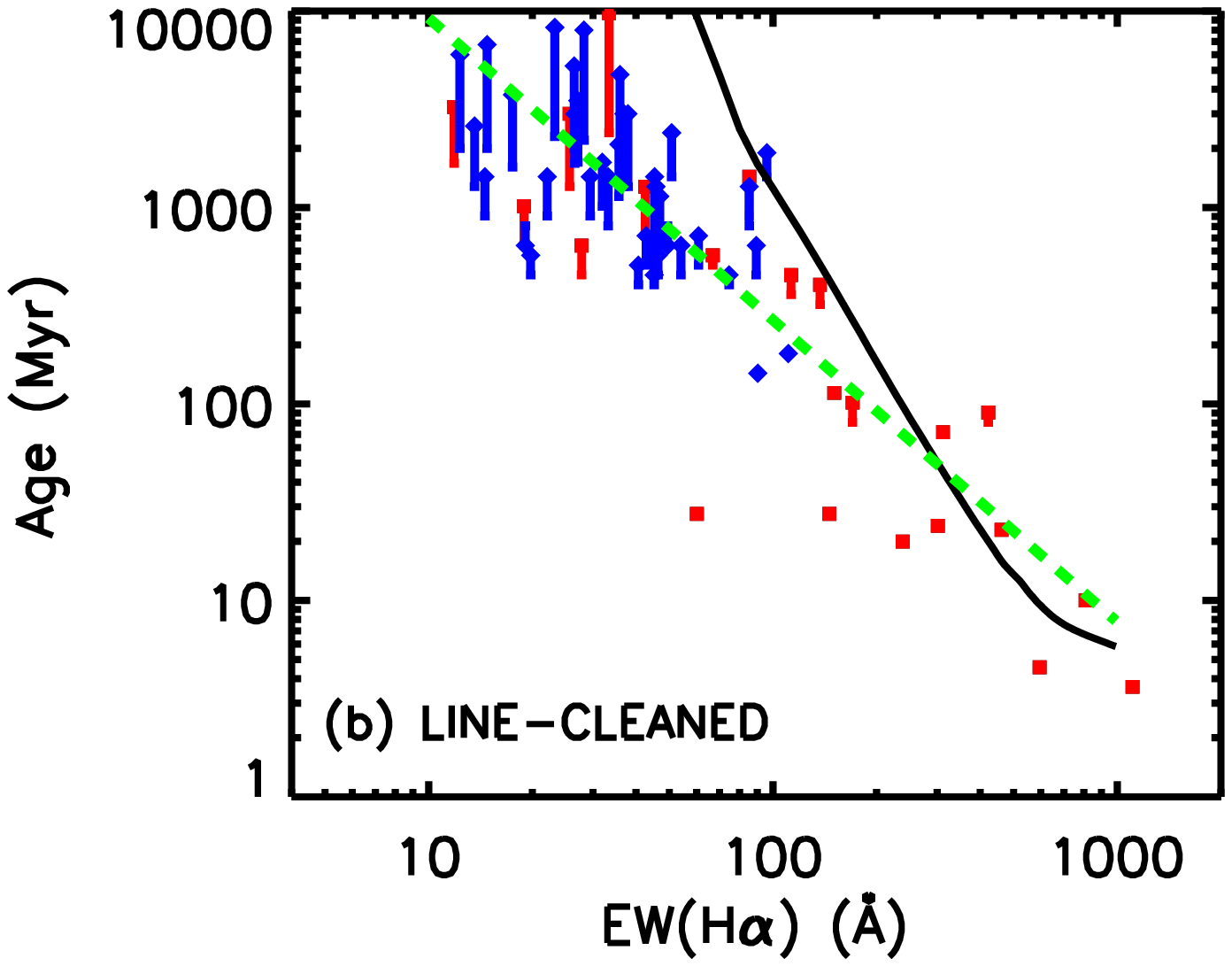}
\caption{The ages determined from spectral synthesis fitting to the 
Bruzual \& Charlot (2003) models vs. the EW(H$\alpha$). In (a) we show
the ages derived when no correction is applied for the emission lines. 
In (b) we show the ages derived when the broadband colors are properly 
corrected. For each object 
(red squares---LAE galaxies with EW(Ly$\alpha)\ge20$~\AA;
blue diamonds---UV-continuum sample) 
we show the range of ages corresponding 
to the models with exponentially declining SFRs with e-folding times
from 1~Gyr to 20~Gyr (vertical bars). The data point 
corresponds to the 20~Gyr case.
In (b) the black solid curve shows the age vs. the 
EW(H$\alpha$) relation for a constant SFR model computed with 
STARBURST99, and the green dashed line shows the optimal power law 
fit given in the text.
}
\label{ewha_time}
\end{inlinefigure}

\begin{inlinefigure}
\includegraphics[width=4.0in]{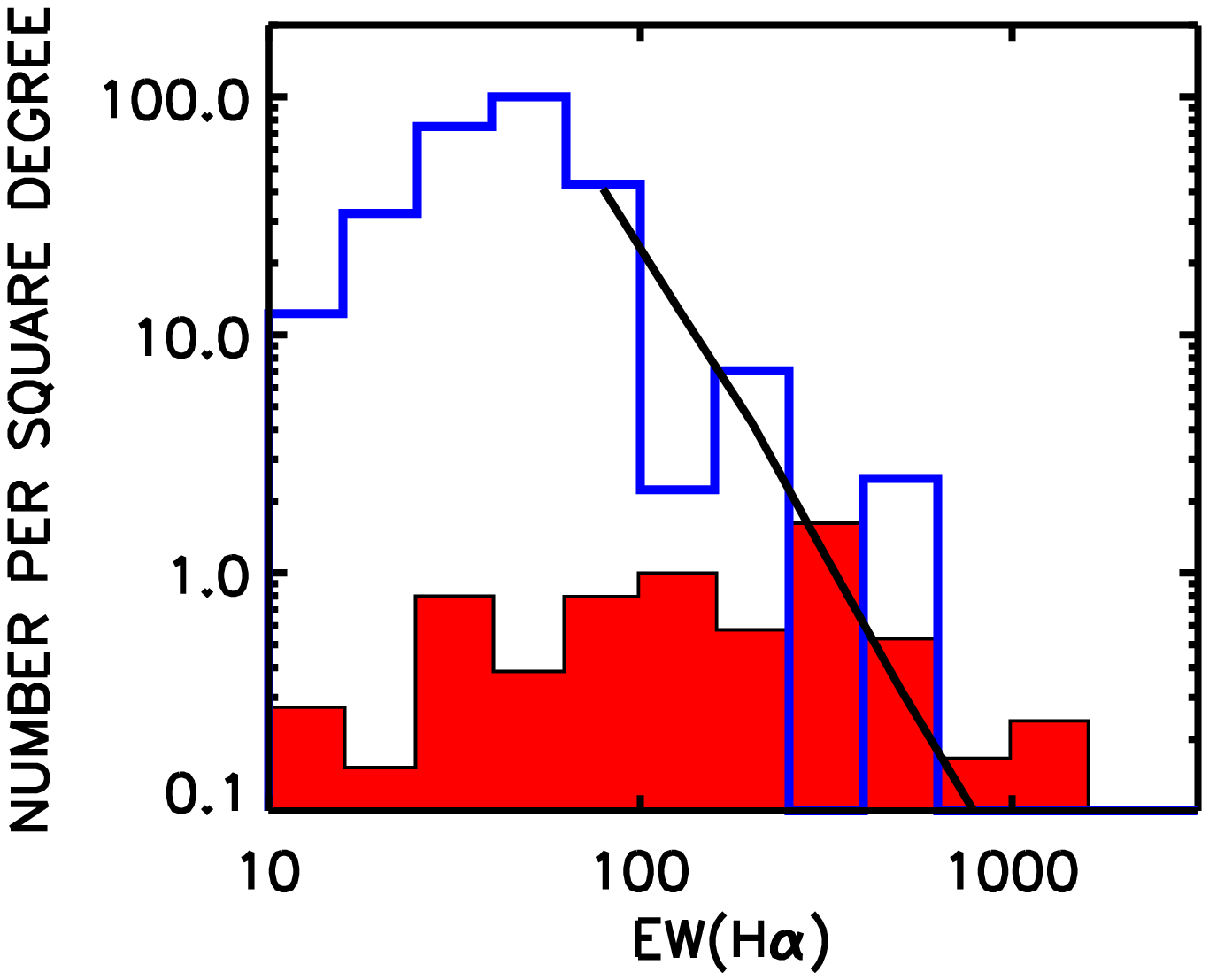}
\caption{The surface density of sources vs. the EW(H$\alpha$). The red 
shaded histogram shows the LAE galaxies with rest-frame 
EW(Ly$\alpha)\ge20$~\AA.  The blue histogram shows the UV-continuum sample,
which is much more heavily weighted to low values of the EW(H$\alpha$). 
The black solid curve shows the shape computed from a
model where we assume a constant rate of formation of galaxies and
that the time spent in each equivalent width bin is determined
from Equation~1. The normalization is matched to the observed surface 
density. 
\label{ewha_dist}
}
\end{inlinefigure}

In Figure~\ref{ewha_dist} we plot the surface
densities of galaxies in our LAE and UV-continuum
samples versus the EW(H$\alpha$). It is worth
emphasizing that in both the LAE and UV-continuum samples the galaxies 
are chosen solely on the basis of their NUV 
magnitudes; that is, they have NUV$<22.1$, lie in the redshift 
interval $z=0.195-0.44$, and have the same NUV magnitude distribution. 
The only difference between the two samples is the presence or absence
of strong Ly$\alpha$ emission. Thus, while there
might be a relation between the NUV magnitude and the EW(H$\alpha$),
this will operate in the same way in both samples and will not
affect any comparison. 

For the LAE sample (red shaded histogram) we plot the sum of the inverse
areas over which a galaxy with a given NUV magnitude could be observed 
in the {\em GALEX\/} fields (Cowie et al.\ 2010). 
(We also corrected the LAE density for the $10\%$ of sources that were 
not spectroscopically observed.) For the UV-continuum
sample (blue histogram) we also weight the areas with the fraction of 
sources that we observed as a function of NUV magnitude. 
The black solid curve shows the shape computed from a 
model where we assume a constant rate of formation of galaxies and
that the time spent in each equivalent width bin is determined
from Equation~1. The normalization is proportional to the
birthrate of galaxies, and we have matched it to the observed surface 
density. We can see that this model matches well to the shape 
at high EW(H$\alpha$), so the surface density of galaxies versus the 
EW(H$\alpha$) is also consistent with a constant production rate of 
new galaxies that then evolve with a constant SFR. 

We can also see from the figure that Ly$\alpha$ emission is rare
in sources with EW(H$\alpha$)$<100$~\AA\ (about $0.7\%$ of galaxies) but 
common in higher EW(H$\alpha$) sources where the blue and red histograms
become comparable. $31\pm13\%$ of the EW(H$\alpha$)$>100$~\AA\ galaxies 
and $57\pm30\%$ of the EW(H$\alpha$)$>250$~\AA\ galaxies have Ly$\alpha$ 
emission with the EW(Ly$\alpha)\ge20$~\AA. The uncertainties reflect the 
small number of UV-continuum sources with high EW(H$\alpha$). Finally,
we can see that a large fraction of the LAEs ($75\pm12\%$) 
are drawn from the high EW(H$\alpha$)$>100$~\AA\ population. This again 
shows that the LAEs are drawn primarily from the youngest galaxies.
The remaining $25\%$ of LAEs with EW(H$\alpha$)$<100$~\AA\
may have geometries, kinematics, or orientations that are unusually 
conducive to Ly$\alpha$ escape. However, some may be objects where the 
Ly$\alpha$ emission is produced by AGN activity, but the AGN signatures 
in the optical are swamped by the galaxy contributions.

In the following sections we will use the EW(H$\alpha)=100$~\AA\ as a 
rough cut above which Ly$\alpha$ emission is more common.

\subsection{Sizes}
\label{Sizes}

Figure~\ref{ewha_plots}(b) shows that there
is also a size evolution as a function of EW(H$\alpha$). 
The higher EW(H$\alpha$) sources are generally unresolved
at the $\sim1''$ resolution of the CFHT MegaPrime $U$-band images, 
while the lower EW(H$\alpha$) sources are generally extended. 
However, unlike in the other correlations, there is a significant 
difference between the LAEs and the UV-continuum sources at the same
EW(H$\alpha$). From Figure~\ref{ewha_plots}(b) we can see that at the 
same EW(H$\alpha$), sources with Ly$\alpha$ emission are significantly 
smaller than those without; that is, where Ly$\alpha$ is strong is in the 
most compact sources at a given EW(H$\alpha$). A rank-sum test shows that 
there is a $<3.6\times10^{-5}$ probability that the LAEs
at EW(H$\alpha)>80$~\AA\ are physically as large as the UV-continuum
sources with the same EW(H$\alpha$) selection. (We have chosen the
EW(H$\alpha$) to provide a significant number of UV-continuum sources
since there are only 4 UV-continuum sources with EW(H$\alpha)>100$~\AA\
as can be seen in Figure~\ref{ewha_plots}(b)). The
median FWHM for these LAEs is $1.1''$, comparable to
the resolution in the CFHT MegaPrime images, while that for the 
UV-continuum sources with the same EW(H$\alpha$) selection is $1.6''$.

\begin{inlinefigure}
\includegraphics[width=4.0in]{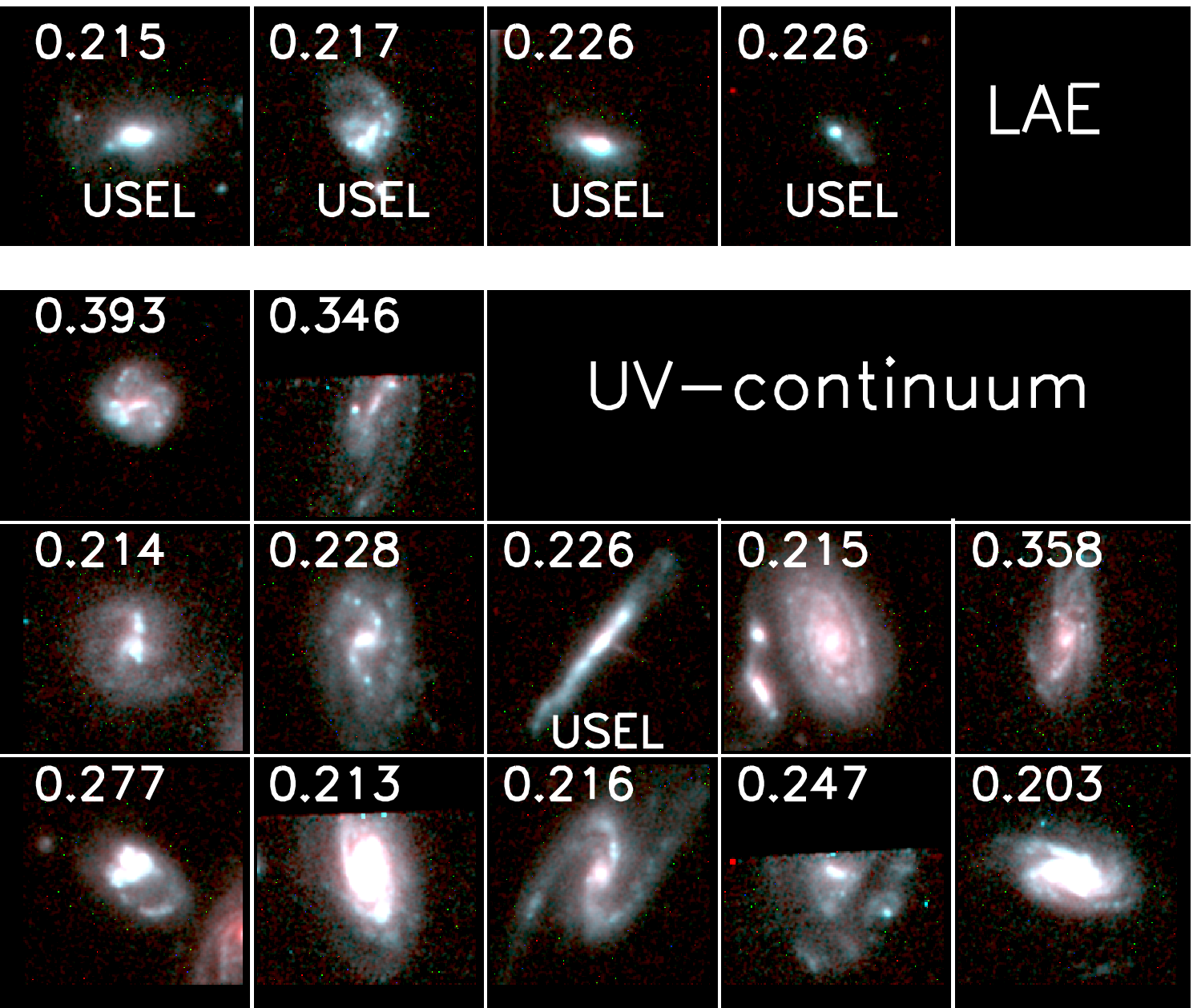}
\caption{Images from the GEMS survey of the CDF-S.
The lower 12 sources 
are UV-continuum sources in the region, and 
the top 4 sources are LAEs.  Each thumbnail is 
$6''$ on a side. The blue and green colors correspond to the 
F606W GEMS image, and the red color corresponds to the F850W 
GEMS image.
\label{gems_thumbs}
}
\end{inlinefigure}

\begin{inlinefigure}
\includegraphics[width=4.0in]{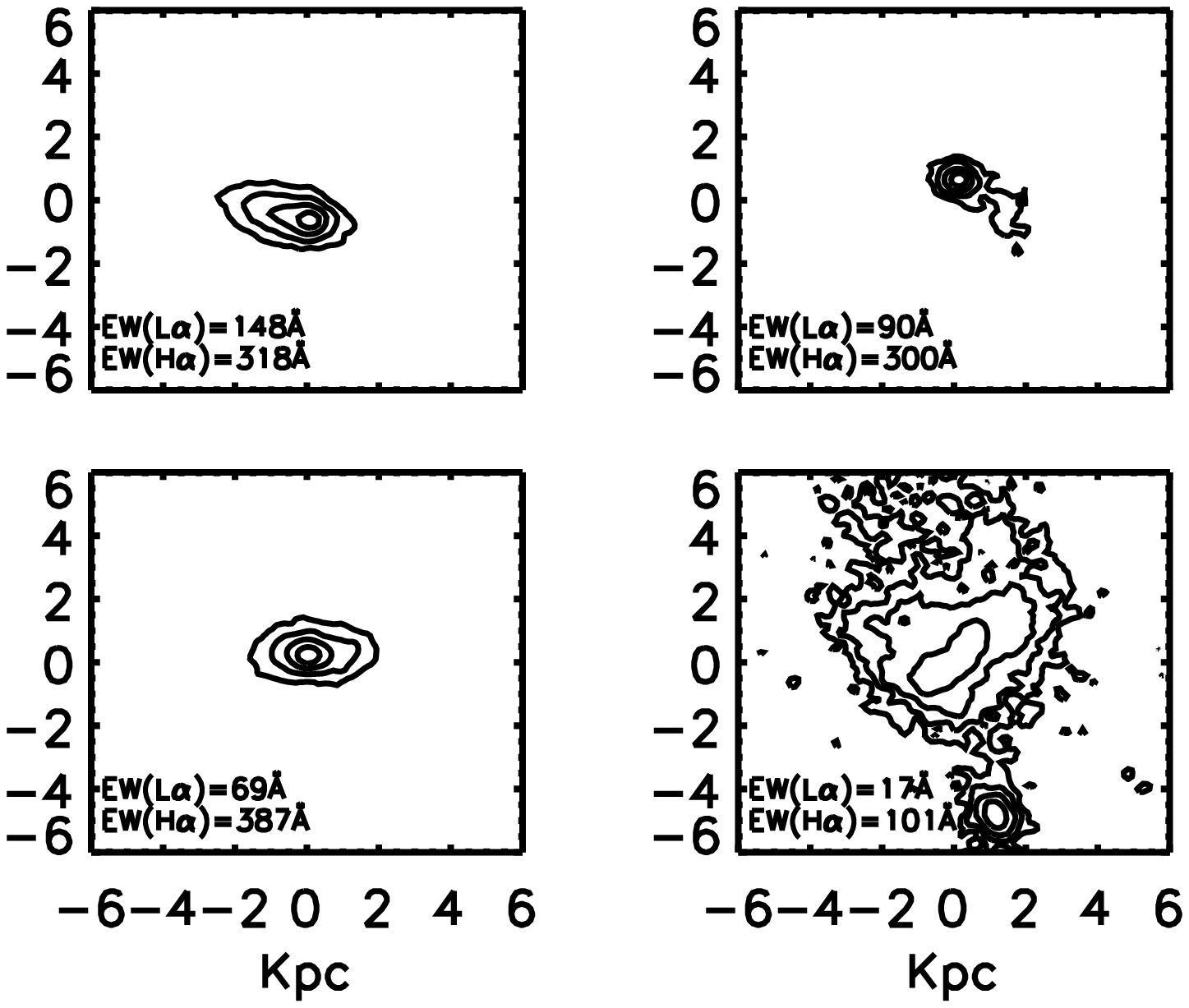}
\caption{Contour plots for the 4 LAEs in the GEMS survey of the 
CDF-S. Each contour in the combined F606W and F850W image
rises by a factor of two. The rest-frame EW(Ly$\alpha$) and the 
rest-frame EW(H$\alpha$) are given in the lower-left corner.
\label{gems_sizes}
}
\end{inlinefigure}

The ground-based $U$-band imaging is inadequate to resolve the
sizes for most of the LAEs. Thus, although the optical
morphologies may differ from those in the UV, here 
we use {\em HST\/}-based images to obtain rest-frame optical sizes. 
In Figure~\ref{gems_thumbs}
we show the combined F606W and F850W {\em HST\/} ACS images for the 
sources that lie in the GEMS survey field. In Figure~\ref{gems_sizes}
we show the contours from the combined F606W and F850W images 
for the four LAEs.  The FWHM in these higher resolution images 
is 0.5~kpc for the three sources with EW(Ly$\alpha)\ge20$~\AA, though
the one source with EW(Ly$\alpha)<20$~\AA\ is more extended
with a FWHM of about 2~kpc. These sizes are very similar
to the continuum sizes of LAEs in the rest-frame UV at 
$3<z<6.5$ (Bond et al.\ 2009, 2010; Cowie et al.\ 2011)
though again we wmphasize that it would be desirable to make the comparison
at the same rest-frame wavelength.

\subsection{Extinction}
\label{Extinction}

We can measure the dust extinction from the galaxies in several 
ways: from the UV spectral slopes, from the SED fits, and from 
the Balmer ratios. We may intercompare these measurements
and also use each of them to test the effects of
extinction on the Ly$\alpha$ line.

In Figure~\ref{ewha_plots}(c) the UV spectral index,
which is often used as a measure of the stellar extinction,
shows a general trend towards less negative values as
we move to lower EW(H$\alpha$), which would be expected
if the extinction is rising as the galaxies age. However,
there is a good deal of scatter, and at least part
of the change is caused by evolution in the intrinsic spectrum. 
The evolution of the reddening may be more clearly
seen in Figure~\ref{ewha_plots}(d):
the Balmer ratio shows a systematic decrease
as the EW(H$\alpha$) decreases, indicating an increase in
the nebular extinction. The derived E(B-V) is similar whether
we use the Cardelli et al.\ (1989) extinction law or 
the Calzetti et al.\ (2000) extinction law. With the Calzetti et al.\ 
reddening we obtain a median E(B-V) of 0.23 (0.15, 0.32)
from the Balmer ratio for sources with EW(H$\alpha)>100$~\AA\ 
and 0.33 (0.30, 0.38) for sources with EW(H$\alpha)=20-100$~\AA, 
where the quantities in parentheses are the $\pm1\sigma$ range. 
Interestingly, in both Figures~\ref{ewha_plots}(c) [UV spectral index versus 
EW(H$\alpha$)] and \ref{ewha_plots}(d) [Balmer ratio versus EW(H$\alpha$)]
there appears to be little differentiation between LAEs and UV-continuum 
galaxies at the same EW(H$\alpha$), suggesting that extinction 
is not the primary reason for Ly$\alpha$ being suppressed at a 
given EW(H$\alpha$).  This is a well-known result
which has been found by many authors (e.g., Giavalisco et al.\ 1996; Mas-Hesse et al.\ 2003; 
Atek et al.\ 2009b; Finkelstein et al.\ 2009a).

\begin{inlinefigure}
\includegraphics[width=4.0in]{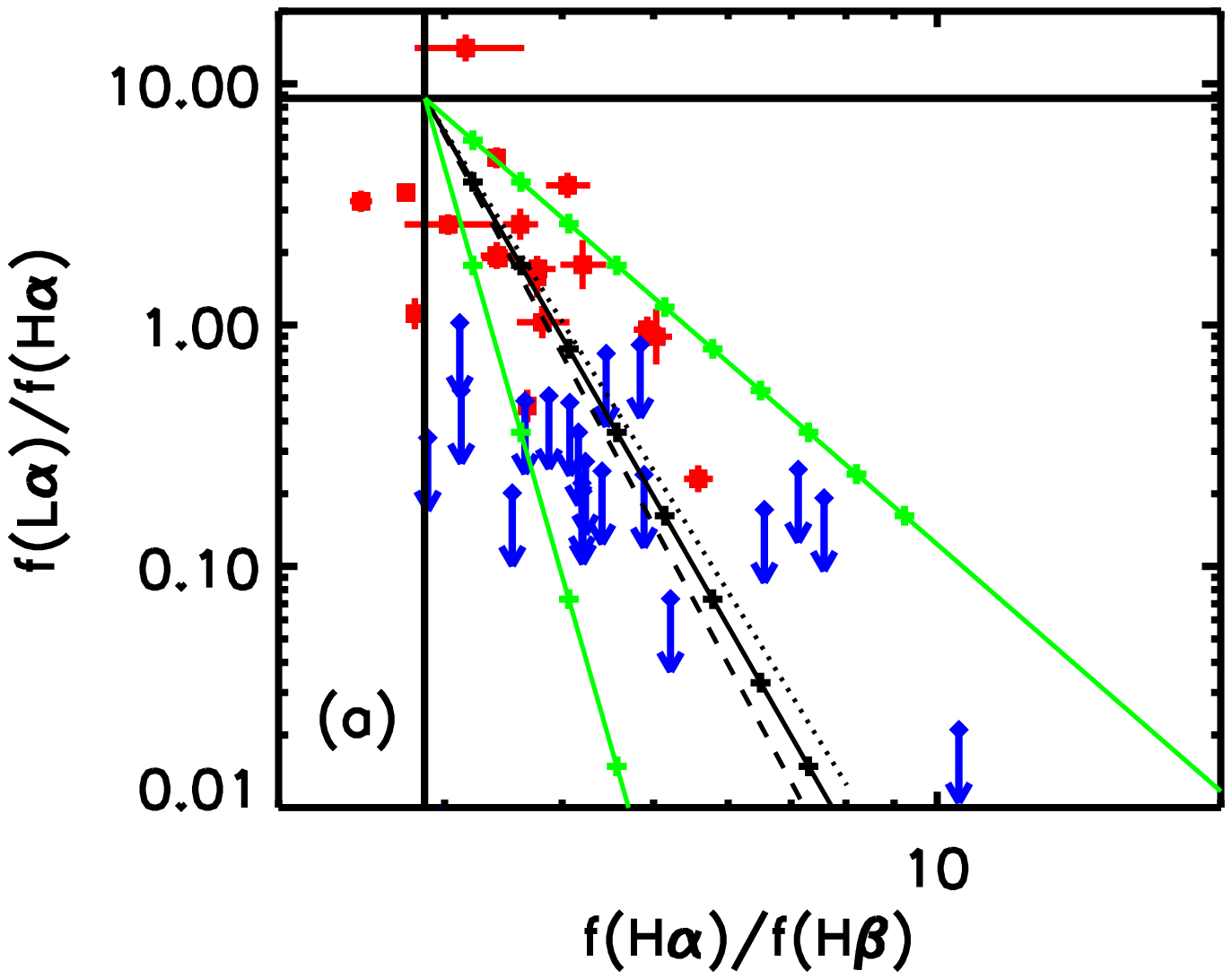}
\includegraphics[width=4.0in]{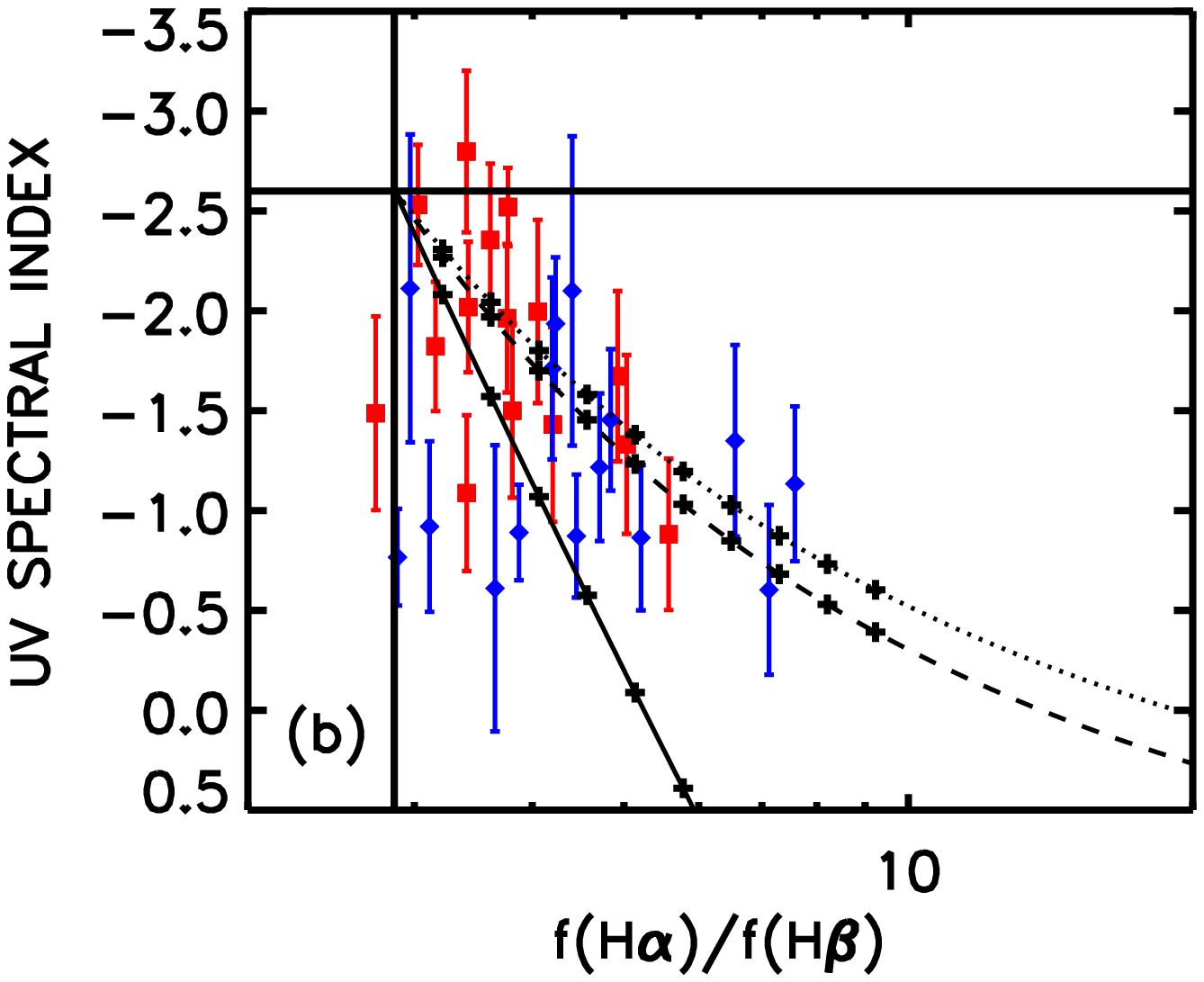}
\caption{The evolution of (a) $f$(Ly$\alpha$)/$f$(H$\alpha$) and 
(b) the UV spectral index vs. the Balmer ratio 
$f$(H$\alpha$)/$f$(H$\beta$). 
All sources with rest-frame EW(H$\alpha)>20$~\AA\
are shown (red squares --- LAEs; blue diamonds --- UV-continuum 
sources). The EW(H$\alpha$) limit is chosen to include
nearly all of the LAEs (see Figure~\ref{ewha_dist}) and to provide 
a sample of UV-continuum objects with strong H$\alpha$. 
The two figures contain slightly different sets of objects
since the UV spectral indices are only measured for a limited
redshift range.
The upper limits on the Ly$\alpha$ fluxes for the UV-continuum 
galaxies are computed for an observed-frame EW(Ly$\alpha)=10$~\AA.
The expected ratios are shown for several reddening laws:
Calzetti et al.\ (2000; solid), Cardelli et al.\ (1989; dashed), 
and Fitzpatrick et al.\ (1999; dotted).
The crosses on the curves show increments of 0.1 in E(B-V) from the
adopted intrinsic values, which are shown by the solid vertical
and horizontal lines. In (a) the lower (upper) green line shows
the expected relationship for the Calzetti et al.\ reddening law 
if the E(B-V) is twice (half) as large for Ly$\alpha$ as for the
Balmer lines.
\label{red_plot}
}
\end{inlinefigure}

As has been pointed out by Atek et al.\ (2009a) and
Scarlata et al.\ (2009) based on the GALEX LAE samples
(see also Hayes et al.\ (2010) for a discussion of a higher
redshift sample), there is a relation between 
$f$(Ly$\alpha)/f$(H$\alpha$) and $f$(H$\alpha)/f$(H$\beta$)
for LAE galaxies but with a wide dispersion.
In Figure~\ref{red_plot}(a) we show such a plot for our LAE galaxies 
(red squares) and UV-continuum galaxies (blue diamonds) with 
EW(H$\alpha)>20$~\AA, where the H$\alpha$ limit is chosen
to include nearly all the LAEs (see Figure~\ref{ewha_dist}). 
The upper limits on the Ly$\alpha$ fluxes for the UV-continuum
galaxies are computed for an observed-frame EW(Ly$\alpha)=10$~\AA.
We emphasize again that, unlike Atek et al.\ (2009a), our
H$\alpha$ measurements are not matched to the {\em GALEX\/} aperture,
so our $f$(Ly$\alpha)/f$(H$\alpha$) ratio may have additional
systematic scatter. 
In addition, the relative flux calibration
and uncertainties in the underlying absorption correction
for the H$\beta$ line may result in scatter in $f$(H$\alpha)/f$(H$\beta$)
and occasional unphysical values (negative extinction) in this ratio.

The inclusion of the UV-continuum sample shows that the 
LAEs correspond to the maximum values of $f$(Ly$\alpha)/f$(H$\alpha$)
at a given  $f$(H$\alpha)/f$(H$\beta$), as would
be expected from the initial selection in the Ly$\alpha$
line. At any given $f$(H$\alpha)/f$(H$\beta$) there is
a range in $f$(Ly$\alpha)/f$(H$\alpha$) stretching up to
roughly the value set by simple extinction of the
Ly$\alpha$ line if it followed the same escape path as
the H$\alpha$ line.  The LAEs primarily lie along this
relation. The line followed by the LAEs is consistent either with the
Calzetti et al.\ (2000) reddening law (black solid line)
or with uniform screen models, such as those of 
Cardelli et al.\ (1989; black dashed) or 
Fitzpatrick et al.\ (1999; black dotted), all of which give 
nearly identical predictions. There is considerable scatter
about the relation, but we suspect that this is a consequence
of systematic errors in the flux calibration discussed above, rather
than an indication of the need for more complex dust models, 
as suggested by Scarlata et al.\ (2009). We can also see that 
the data are not consistent with the Ly$\alpha$ photons in
the LAE galaxies having
a widely different E(B-V) path from the H$\alpha$ photons.
Following Scarlata et al.,
we have plotted the relation when the Ly$\alpha$
path is twice that of the Balmer photons (lower green line)
and also when it is half that of the Balmer photons (upper
green line). It is clear that these are a much poorer fit to 
the LAE observations.

In contrast, if we plot the UV spectral index for the same 
sample versus $f$(H$\beta$)/$f$(H$\alpha$) (Figure~\ref{red_plot}(b)), 
then we see that 
the Calzetti et al.\ (2000) reddening law gives a substantially 
different result than the Cardelli et al.\ (1989) or 
Fitzpatrick et al.\ (1999) uniform screen models, reflecting their 
different shapes at UV wavelengths. Most of the data appear to 
follow the  models of Cardelli et al.\
or Fitzpatrick et al.\ rather than the reddening law of 
Calzetti et al., though there are a handful of outlying points.  
This suggests that for these galaxies the stellar extinction is better
reperesented as a uniform screen rather than a patchy distribution.
This is the case for both the LAEs and the UV-continuum sources.

\subsection{Metallicity}
\label{Metallicity}

Perhaps the most interesting plot in Figure~\ref{ewha_plots} is 
that of N2=$\log(f$([NII]$\lambda6584$)/$f$(H$\alpha$)) versus 
EW(H$\alpha$) (panel f). N2 is a widely used metallicity indicator
with low values corresponding to low metallicities. The fact that 
both the LAE and UV-continuum galaxies follow a surprisingly tight 
track in this diagram with N2 rising 
as EW(H$\alpha$) decreases appears to show that we are seeing a 
smooth metal build-up with age in both classes of galaxies.

In order to translate N2 to
a metallicity in the galaxies we use the Pettini \& Pagel (2004) local 
relationship, $12+\log$(O/H)$=8.90+0.57$N2, which was determined 
by comparing to direct measures of the O abundance over the range 
N2=$-2.5$ to $-0.5$. Extrapolating the Pettini \& Pagel relation to these 
galaxies requires assuming that there is no change in the ionization 
parameter, which may well be incorrect. However, at $z=0.195-0.44$ 
Cowie \& Barger (2008) found a narrow range of ionization parameters 
($q\sim 2\times10^7$) and a similar relation to that of Pettini \& Pagel. 
Cowie \& Barger (2008) also showed that other line diagnostics 
gave similar metal-luminosity relations to that derived from N2 in the 
redshift interval $z=0.195-0.44$.
\begin{inlinefigure}
\includegraphics[width=3.8in]{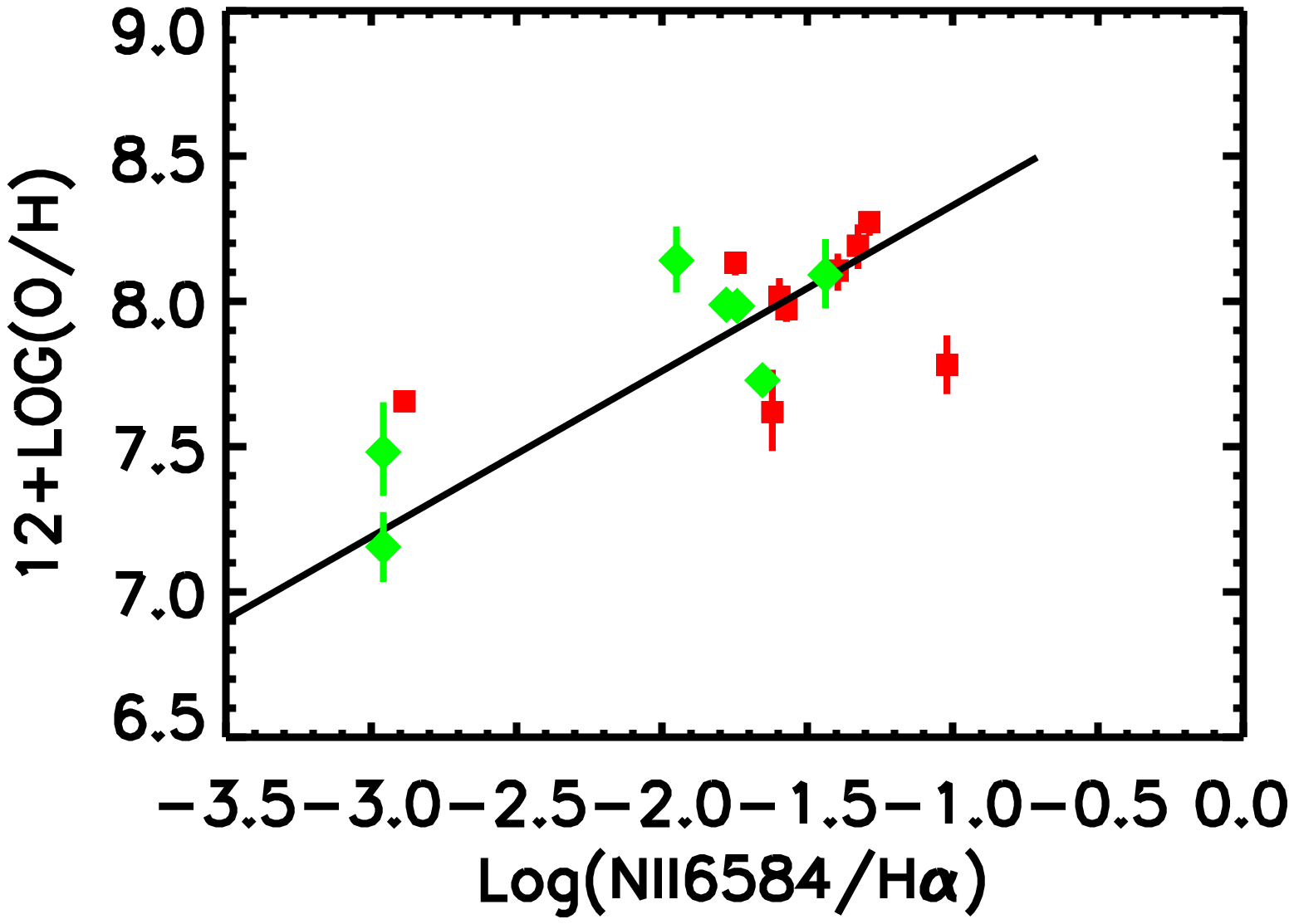}
\caption{$12+\log$(O/H) derived from the direct method vs. N2.
The {\em GALEX\/} sources in the redshift interval $z=0.195-0.44$ 
are shown with red squares with $1\sigma$ error bars. 
A sample of ultra-strong emission-line galaxies in the 
same redshift interval with direct O measurements from Hu et al.\ (2010)
are shown with green diamonds. Both are broadly consistent
with the Pettini \& Pagel (2004) relation (black line), though 
with substantial scatter.
\label{n2ha_test}
}
\end{inlinefigure}

We can also test the Pettini \& Pagel (2004) relation with the present 
data. Ten of the $z=0.195-0.44$ sources have [OIII]$\lambda4363$ 
detected at above the $3\sigma$ level.  These are marked with
green diamonds in Figure~\ref{bpt}(a). None of the UV-continuum sample 
at $z=0.195-0.44$ has detected [OIII]$\lambda4363$ at the $3\sigma$ level.
All of the [OIII]$\lambda4363$ galaxies have low values of N2, except for 
{\em GALEX\/} 1240+6233, which is classified as an AGN based on the BPT 
diagram. Conversely, most of the low-N2 galaxies are detected in 
[OIII]$\lambda4363$.  For the sources that are not AGNs and where we 
have detected [OIII]$\lambda4363$ we used the `direct' or $T_e$ method 
to determine the metallicity
(e.g., Seaton 1975; Pagel et al.\ 1992; Pilyugin \& Thuan 2005; 
Izotov et al.\ 2006).
To derive $T_e$\oiii\ and the oxygen abundances, we used the
Izotov et al.\ (2006) formulae, which were
developed with the latest atomic data and photoionization models.
We compare the derived O abundances with N2 in Figure~\ref{n2ha_test} 
(red squares), where we also show similar measurements from a sample of 
ultra-strong emission-line galaxies taken from Hu et al.\ (2009;
green diamonds). The O abundance versus N2 is broadly consistent 
with the Pettini \& Pagel relation (black line), though there is a 
good deal of scatter, probably reflecting the variation in the 
ionization parameter.

We can see immediately from Figure~\ref{ewha_plots}(f) that the
LAEs in the redshift interval $z=0.195-0.44$ have lower values 
of N2 than the UV-continuum sources
(see also Figure~\ref{bpt}). This effect was previously
noted by Cowie et al.\ (2010), though the present larger sample 
substantially increases the statistical significance of the result. 
This result may be more clearly seen in Figure~\ref{n2ha_dist}, 
where we show the N2 distributions for the LAE (red shaded histogram) 
and UV-continuum (blue histogram) samples. While the N2 distributions 
overlap, the distribution for the LAEs clearly extends 
to lower values, and the median N2 is lower. A rank-sum test 
gives only a $10^{-6}$ probability that the two {\em GALEX\/} 
samples are similar. Using the Pettini \& Pagel (2004)
conversion gives a median $12+\log$(O/H)=8.24 (8.17, 8.35) for the 
LAEs, where the quantities in parentheses are the $\pm1\sigma$ range. 
This is about 0.4~dex lower than that of the UV-continuum sample, though 
the median value for the UV-continuum sample is more poorly determined, 
since at the higher values of N2 the conversion breaks down 
as the N2 value saturates.

We can now see from Figure~\ref{ewha_plots}(f)
that the difference in N2 distributions is primarily
a reflection of the evolutionary state of the galaxy. That is, the 
metallicity is building up with age, and the LAEs are preferentially 
drawn from the younger galaxies. UV-continuum galaxies have 
similar metallicities to LAEs with the same EW(H$\alpha$). However,  
the LAE sample is much more weighted to high EW(H$\alpha$) galaxies,
which are younger and more metal poor.

\begin{inlinefigure}
\includegraphics[width=4.0in]{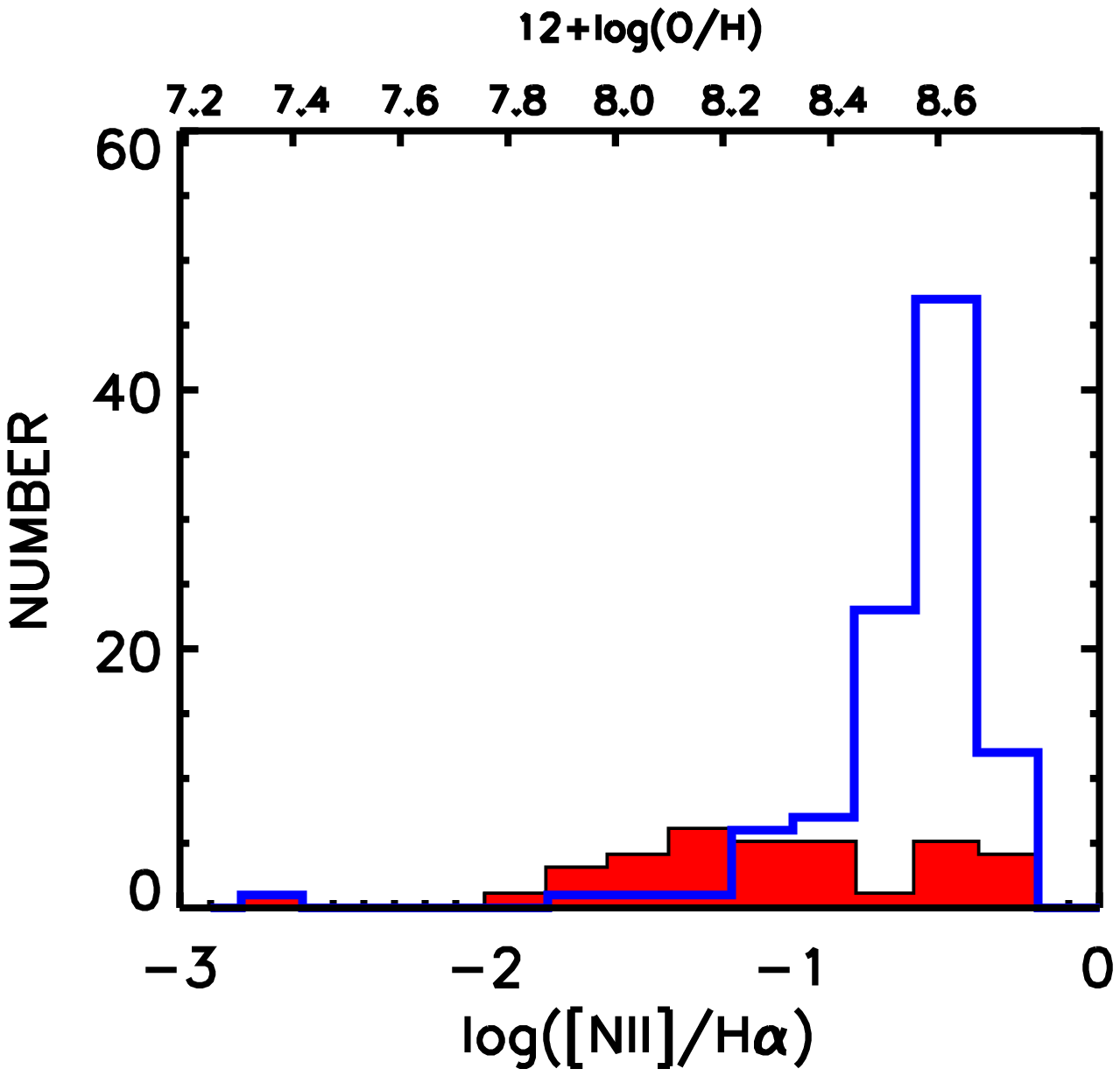}
\caption{The number distributions of the LAE galaxies
with EW(Ly$\alpha)\ge20$~\AA\ (red shaded
histogram) and the UV-continuum 
sample (blue histogram) as a function of N2.
The UV-continuum sample is much more heavily weighted
to high values of N2. The top axis shows the O abundance that
would be derived from the N2 ratio using the Pettini
\& Pagel (2004) conversion.
\label{n2ha_dist}
}
\end{inlinefigure}


\section{The Physical Properties of the Galaxy Samples}
\label{secint}

In order to convert the observed quantities into physical
properties of the galaxies, we will assume a constant SFR model 
with a Salpeter (1955) initial mass function (IMF). As we have
seen, this appears to provide a self-consistent fit to most of 
the observed properties. For simplicity, we assume that the IMF 
extends smoothly from 0.1~M$_{\sun}$ to 100~M$_{\sun}$, though 
this can be easily converted into other favored IMFs with flatter 
slopes below M$_{\sun}=1$ by a simple renormalization of the SFR.

In Figure~\ref{ewha_halum} we show the H$\alpha$ luminosities
of the LAE and UV-continuum samples corrected for extinction 
using the Balmer ratio and a Fitzpatrick et al.\ (1999) extinction 
law. The median corrections for the LAE and UV-continuum
samples are 1.7 and 2.1, respectively.

For our assumed constant SFR model the conversion from H$\alpha$ 
luminosity to SFR is roughly invariant over the range of EW(H$\alpha$) 
and is given by the Kennicutt (1998) relation:
\begin{equation}
\log {\rm SFR} = -41.1 + \log L({\rm H}\alpha) \,.
\label{eqhaSFR}
\end{equation}
On the right-hand axis of Figure~\ref{ewha_halum} we give the SFR in 
M$_{\sun}$~yr$^{-1}$ corresponding to a given H$\alpha$ luminosity.
Individual galaxies have SFRs that range from about 1~M$_{\sun}$~yr$^{-1}$ 
up to about 30~M$_{\sun}$ yr$^{-1}$. The lower bound is roughly set by 
the NUV magnitude limit for the selection, but the upper bound on the SFR 
is a measure of the maximum SFRs seen in UV-continuum selected galaxies at 
these redshifts. On the top axis we give the logarithm of
the age in years corresponding to a given EW(H$\alpha$). The LAEs 
are primarily drawn from sources younger than $1-2$~Gyr.

We can see from  Figure~\ref{ewha_halum} that there is
no evolution in the distribution of the SFRs as a function of 
the EW(H$\alpha$). This is consistent with our constant SFR
assumption. There is also no difference in the distribution 
of the SFRs for the LAE and UV-continuum galaxies. If the
highest SFR galaxies maintain their rates for a subtantial 
fraction of the local age of the universe, then the final 
stages will have stellar masses in excess of $10^{11}$~M$\sun$.

\vskip 0.25cm
\begin{inlinefigure}
\includegraphics[width=3.8in]{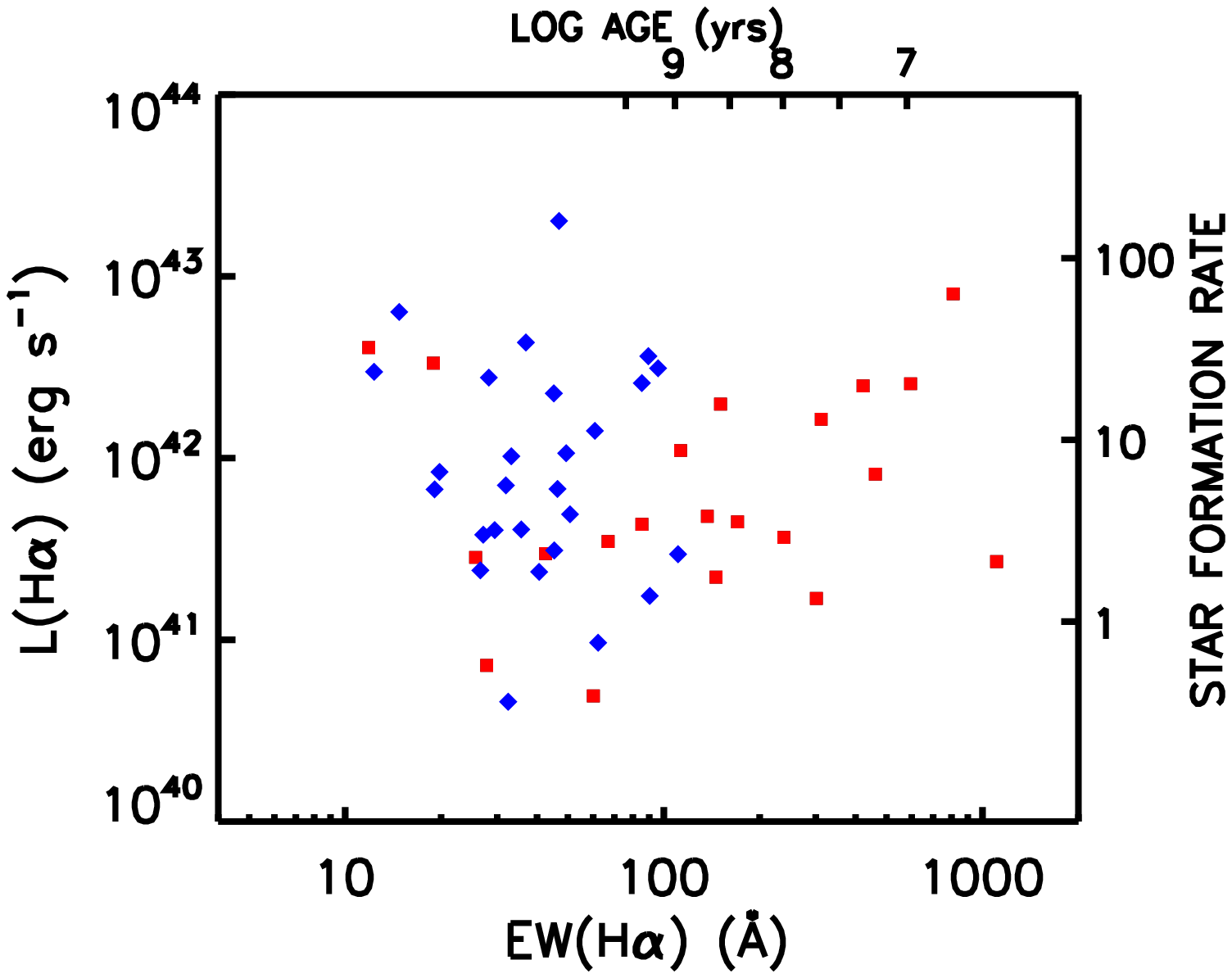}
\caption{Extinction-corrected H$\alpha$ luminosity vs. EW(H$\alpha$).
The red squares show LAEs with EW(Ly$\alpha)\ge20$~\AA, and the 
blue diamonds show UV-continuum galaxies. The conversion
to a SFR assuming a Salpeter (1955) IMF and a constant SFR is shown on the 
right-hand axis, and the conversion to galaxy age is shown 
on the top axis.  
\label{ewha_halum}
}
\end{inlinefigure}

We can test our H$\alpha$ luminosity-determined SFRs against those 
inferred from 20~cm observations (where available; see 
Section~\ref{secother}) and those determined from 
the UV continuum. In Figure~\ref{halum_radio_power} we plot 
rest-frame 20~cm power calculated assuming a radio spectrum of
$f_{\nu}\sim\nu^{-0.8}$ versus extinction-corrected H$\alpha$ 
luminosity. We see a nearly linear relation, though the 
number of measured sources is fairly small. We compare the data with 
two conversions of the radio power to SFR: Condon (1992; solid
line) and Bell (2003; dashed line). The two relations appear to 
bracket the data points. 
(Prior to the H$\alpha$ extinction correction the points are more 
scattered and fall to the left of the expected relations.) 

There is one significantly deviant point, GALEX0959+0151 in COSMOS~00, 
where the H$\alpha$ luminosity is very high compared to the radio power. 
This source has the highest EW(H$\alpha$) in the figure ($811$~\AA) and 
an age of 9~Myr from SED fitting (Figure~\ref{ewha_time}(b)). 
This suggests that the radio power is low because the supernovae that 
generate the radio emission have yet to occur. Unfortunately, the 
other sources in the sample with very high rest-frame EW(H$\alpha$) 
are not in the fields with radio data, so we cannot test this further 
at present.

\begin{inlinefigure}
\includegraphics[width=3.9in]{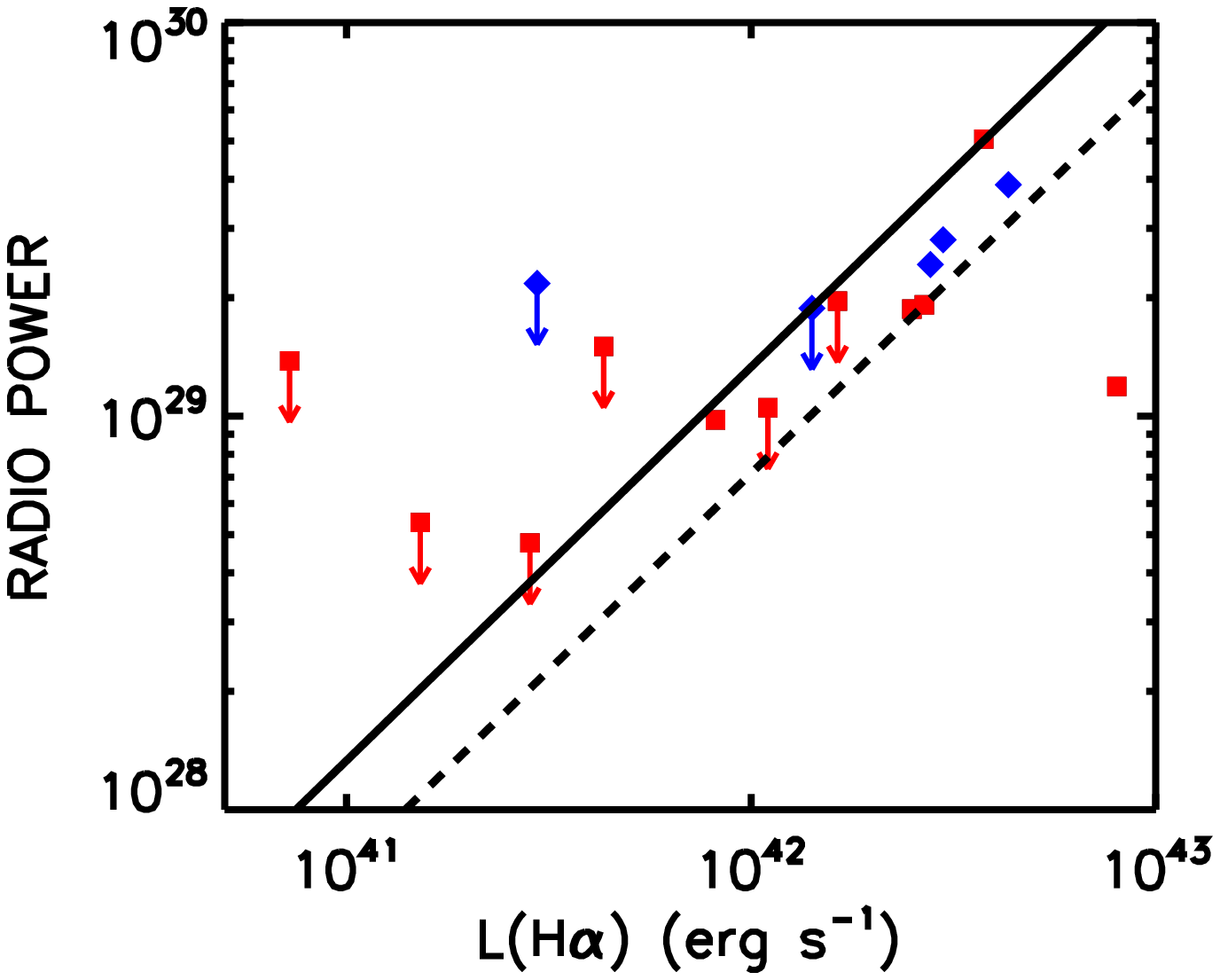}
\caption{The 20~cm power in erg~Hz$^{-1}$ vs. the extinction-corrected 
H$\alpha$ luminosity. The red squares show LAEs, and the blue diamonds 
show UV-continuum galaxies.  In this plot, since we are only
testing the star formation rate determinations, we have
included sources from the supplement to Table~\ref{tab1}
with EW(Ly$\alpha$) between 15 and 20~\AA.
For sources not detected in the 
20~cm images we have assumed upper limits of 100~$\mu$Jy (SIRTFFL~01) 
and 40~$\mu$Jy (COSMOS~00). 
These are shown with downward pointing arrows. 
The one source in the HDF~00 field is detected at 20~cm.
The solid line shows the Condon (1992) SFR vs. radio power relation, 
and the dashed line shows the Bell (2003)
high-luminosity SFR vs. radio power relation.
\label{halum_radio_power}
}
\end{inlinefigure}

\begin{inlinefigure}
\includegraphics[width=3.9in]{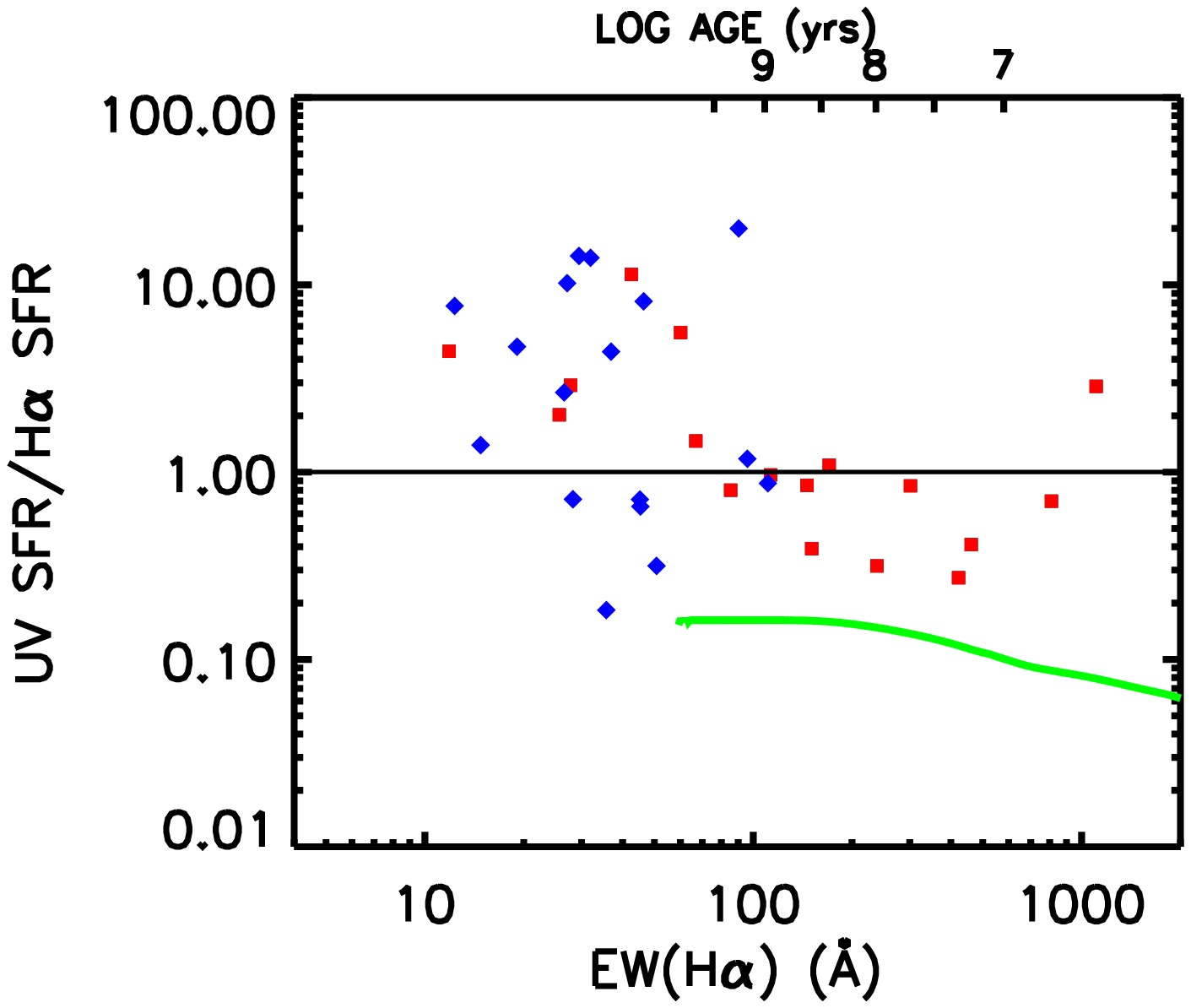}
\caption{The ratio of the SFR computed from the extinction-corrected 
UV luminosity to that from the extinction-corrected H$\alpha$ luminosity.
The red squares show LAEs with EW(Ly$\alpha)\ge20~$\AA, and the blue 
diamonds show UV-continuum galaxies. In computing the UV 
SFRs we have applied a small multiplicative offset of 1.17 to scale
the NUV {\em GALEX\/} spectra to match the NUV {\em GALEX\/} magnitudes.
The green curve shows the evolution of the 1600~\AA\ $F_\lambda\lambda$
for a constant SFR model with a Salpeter (1955) IMF. 
\label{ewha_mdot_ratio}
}
\end{inlinefigure}

SFRs for high-redshift galaxies are generally calculated from the 
observed UV luminosity together with the stellar extinction correction 
determined from the UV continuum slope (Meurer et al.\ 1999). 
It is therefore interesting to commpare the SFRs derived using 
this method with those computed 
from the H$\alpha$ luminosities in the present low redshift
sample and to estimate the limits of its validity.
In Figure~\ref{ewha_mdot_ratio} we show the ratio of the SFR derived from 
the stellar extinction corrected UV luminosity with that 
computed from the nebular extinction corrected H$\alpha$ luminosity
as a function of EW(H$\alpha$). We have computed the UV SFR from the  
$L_{1600}$ luminosity, extinction corrected using the value derived from
the UV spectral index, and  converted using the value at $t=10^8$~yr 
computed from the STARBURST99 model,
\begin{equation}
\log {\rm SFR} = -39.87 + \log L_{1600} \,.
\label{equvSFR}
\end{equation}
With this calibration the UV SFRs are closely matched
to the H$\alpha$ SFRs with an average ratio of 0.9
for sources with EW(H$\alpha)=60-200$~\AA. 
This small absolute difference may be a simple consequence of
the relative optical and UV calibrations. As can be seen in 
Figure~\ref{ewha_mdot_ratio}, the UV calibration underestimates the 
SFR at higher EW(H$\alpha$) and overestimates it at lower
EW(H$\alpha$). 

 The reasons for the deviations reflect the limit of validity of the
underlying assumptions. At low EW(H$\alpha$) the derivation of the
UV continuum extinction from the UV spectral slope breaks down.
The Meurer et al.\ (1999) correction is only applicable to starbursting 
galaxies, since it depends on the intrinsic spectral index being 
approximately fixed, 
as is the case for these sources (see Figure~\ref{ewha_plots}(c)).
At low EW(H$\alpha$) (less than about 60~\AA) the intrinsic UV 
spectral indices are shallower, and the inappropriate use of the 
relationship overestimates the SFR, as can be seen in 
Figure~\ref{ewha_mdot_ratio}.
A second problem with the UV continuum method is that the UV continuum flux 
evolves with time, even in the constant SFR models. In 
Figure~\ref{ewha_mdot_ratio} we show (green curve) the evolution of 
$L_\lambda\lambda$ evaluated at 1600~\AA\ ($L_{1600}$) as a function 
of EW(H$\alpha$). (The normalization is arbitrary and depends
on the SFR.) The UV luminosity increases as a function
of age up to $\sim10^8$~yr or an EW(H$\alpha)\sim200$~\AA. 
Thus, the UV continuum calculation combined with the UV spectral
index extinction correction
should only be used over the EW(H$\alpha)=60-200$~\AA\ range, or,
equivalently, for rest-frame line-corrected galaxy colors 
of $u^\prime - z^\prime =0.5-1.4$ (Figure~\ref{ewha_plots}(a)).

\begin{inlinefigure}
\includegraphics[width=4.0in]{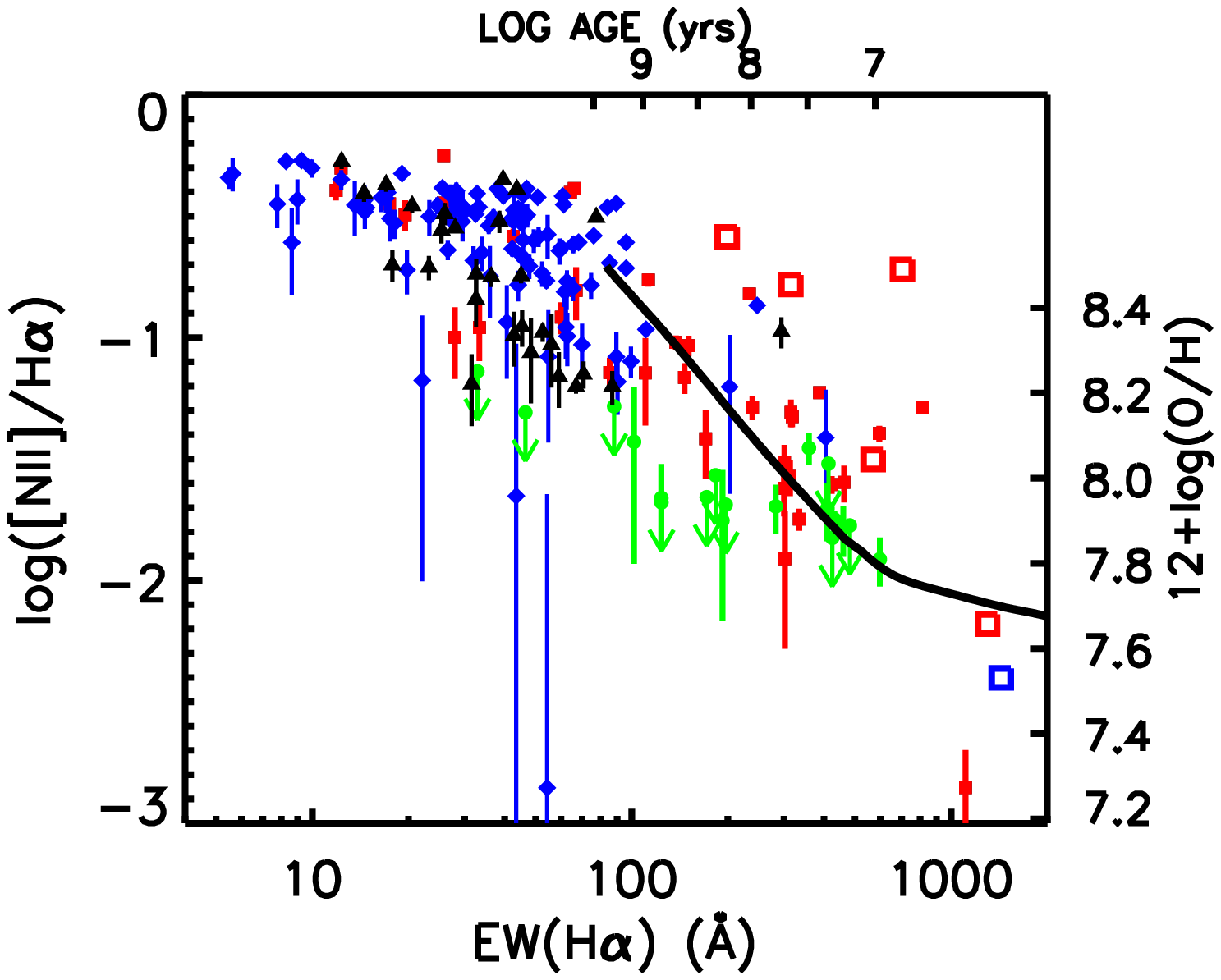}
\caption{Metallicity vs. EW(H$\alpha$). In addition to the present 
data points (i.e., Figure~\ref{ewha_plots}(f)), we show the blue 
compact galaxies from \"{O}stlin et al.\ (2009) (open squares:
red with Ly$\alpha$ emission, blue without), 
a continuum-selected sample with NUV$=22-23.25$ from the GOODS-N 
observations of Barger et al.\ (2008) with 
$z=0.195-0.44$ (black triangles), and the ultra-strong H$\alpha$ 
emission-line selected sample from Hu et al.\ (2010)
(green circles). The error bars are $\pm1\sigma$,
with undetected sources shown with downward pointing
arrows at the $1\sigma$ level. 
The black curve shows the model 
fit discussed in the text. The values of $12+\log$(O/H) shown on the 
right vertical axis were calculated using the Pettini \& Pagel (2004)
relation.
\label{ewha_metal}
}
\end{inlinefigure}

In Figure~\ref{ewha_metal} we plot the metallicities of the LAE 
(red solid squares) and UV-continuum (blue diamonds) samples versus 
the EW(H$\alpha$).
If the gas reservoir were fixed, the constant SFR model would predict 
that the metallicities should rise linearly with time in the early 
stages. This is clearly much steeper than is seen in the present samples,
which can be approximated by a model where the metallicity rises as $t^{0.3}$
(black curve: this is only shown over the range of validity of the
N2 diagnostic which is only appropriate for low metallicity
and saturates at near solar metallicity). Thus,
the galaxies must have ongoing accretion of gas.

In Figure~\ref{ewha_metal} we also compare the present samples with a 
fainter continuum sample from the GOODS-N observations of Barger
et al.\ (2008) (black triangles) and with a fainter sample selected
to have very high EW(H$\alpha$) from Hu et al.\ (2010) (green circles),
as well as with the local blue compact galaxy sample of 
\"{O}stlin et al.\ (2009) (open squares). 
As might be expected, the 
fainter samples have systematically lower metallicities than the present 
data. However, the effect is small compared to the EW(H$\alpha$) dependence, 
suggesting that the ages of the galaxies primarily determine the metallicities.

\begin{inlinefigure}
\includegraphics[width=4.0in]{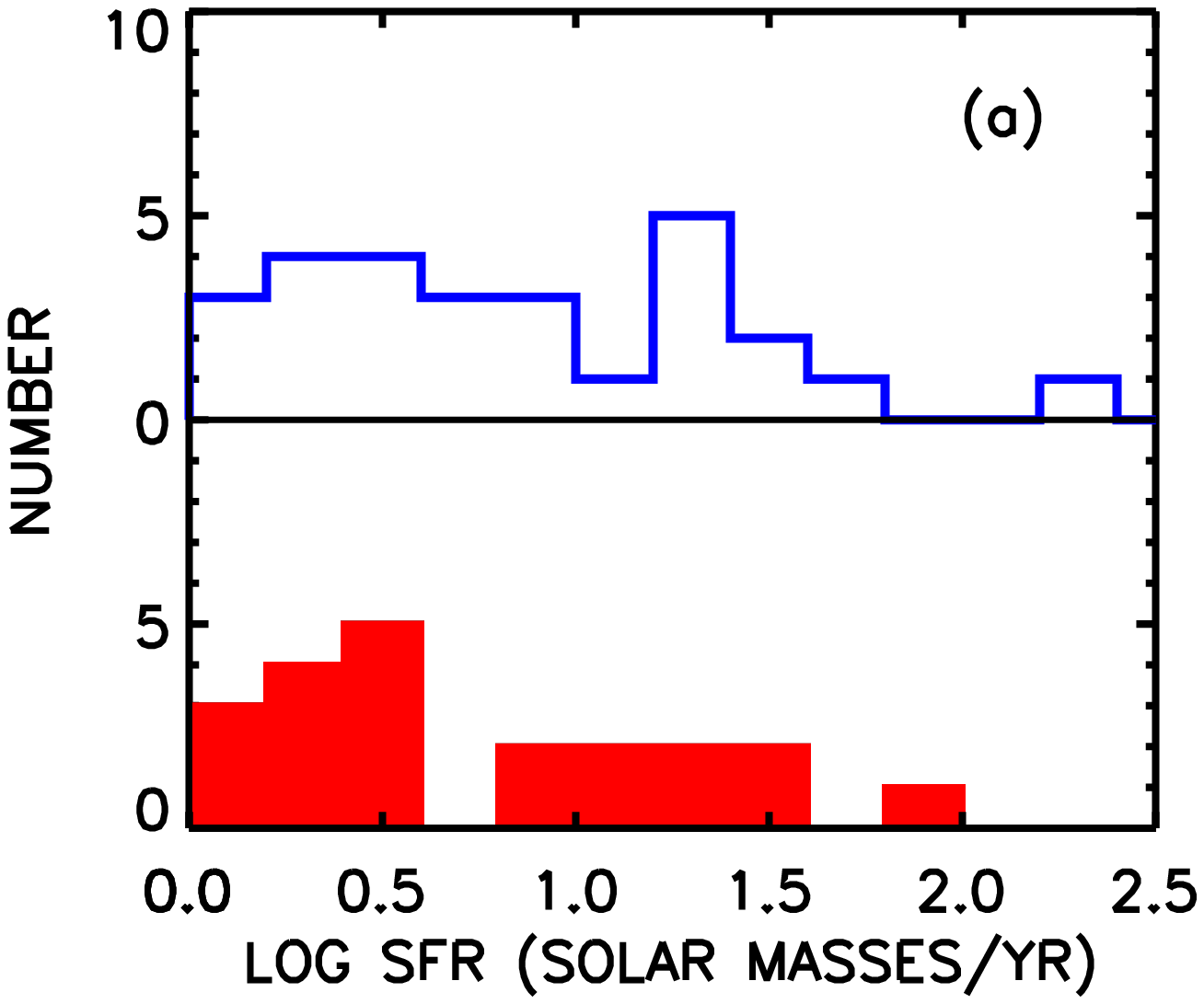}
\includegraphics[width=4.0in]{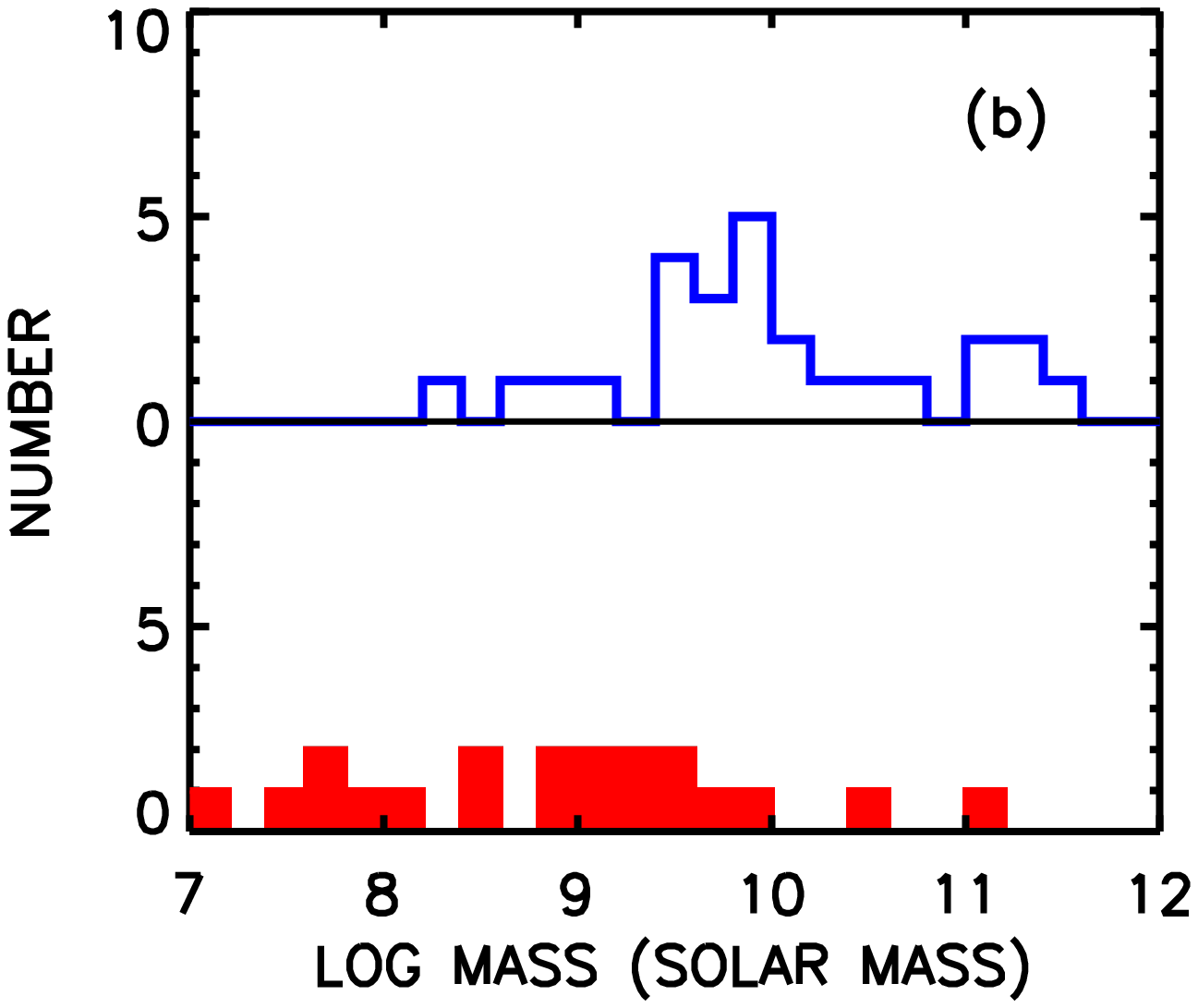}
\caption{(a) The distribution of the SFRs in the LAE (red shaded histogram) 
and UV-continuum (blue histogram) samples. (b) The distributions of the 
stellar masses.
\label{m_hist}
}
\end{inlinefigure}

Using the SFRs derived from the extinction-corrected H$\alpha$ 
luminosities and the masses derived from the constant SFR fits to 
the line-corrected SEDs, we show our final result in Figure~\ref{m_hist}. 
In Figure~\ref{m_hist}(a) we show the distribution of the SFRs. 
The two samples are nearly identical with a median of 
$\sim 6$~M$_{\sun}$~yr$^{-1}$. In contrast, as is shown in 
Figure~\ref{m_hist}(b), the LAEs, reflecting their younger ages, 
are systematically lower in mass than the UV-continuum sources.
The median mass for the LAEs is $10^{9}$~M$_{\sun}$, while that 
for the UV-continuum sample is $8\times 10^{9}$~M$_{\sun}$.


\section{Summary}
\label{secdisc}

We analyzed a substantial sample of UV continuum-selected galaxies 
with and without Ly$\alpha$ emission lines to try to understand how 
LAEs are drawn from the general population and how they
evolve with redshift.  To do this, we obtained extensive optical 
spectroscopy of {\em GALEX\/} grism selected samples (both LAE samples 
and comparison samples with the same UV magnitude distributions but no 
detected Ly$\alpha$) in two redshift intervals: $z=0.195-0.44$ and 
$z=0.65-1.25$.  We used the optical spectroscopy to eliminate AGNs and 
to obtain the optical emission-line properties of the samples. 

We confirmed that the {\em GALEX\/} selected $z\sim0.3$ LAEs
are considerably fainter and much rarer than the
high-redshift LAEs (Deharveng et al.\ 2008; Cowie et al.\ 2010). 
Cowie et al.\ (2010) showed that the
$L_\star$ in a Schechter (1976) function fit to the $z=0.195-0.44$ LAE
luminosity function is almost an order of magnitude
fainter than the $L_\star$ in a fit at $z\sim3$ and that only about 
$5\%$ of $z\sim0.3$ UV-continuum selected galaxies have 
rest-frame EW(Ly$\alpha)\ge20$~\AA. 
Here we showed that there are also LAEs that can be 
found in the {\em GALEX\/}
spectra at $z\sim1$ and that these are similar in
luminosity to the most luminous high-redshift galaxies.
Thus, we conclude that most of the observed evolution occurs over 
the redshift interval $z=0-1$.
This appears to be a simple downsizing effect, with the presence 
of the higher Ly$\alpha$ luminosity sources at $z=1$
corresponding to higher SFR sources initiating their
star formation at this redshift.

We showed that at $z=0.195-0.44$, SED fits to strong emission-line galaxies 
significantly overestimate the ages and masses and underestimate the 
extinctions if we do not correct for the emission-lines.  We found that the SED fits to the uncorrected 
broadband fluxes give ages of about a Gyr. However, the inferred ages drop 
substantially when we correct the galaxy broadband magnitudes using
the observed spectra to remove the emission-line contributions.
All the galaxies with ages much less than a Gyr have strong emission lines 
and must be corrected for the line contributions.  
Spectral synthesis fitting shows a smooth evolution of the rest-frame 
EW(H$\alpha$) with age when the emission-line 
contributions are removed from the broadband fluxes. 

We found that at $z=0.195-0.44$, all sources, regardless of the 
strength of the Ly$\alpha$ emission line, follow a single, 
well-defined sequence as a function of the rest-frame EW(H$\alpha$).
Higher EW(H$\alpha$) sources all have lower metallicities,
bluer colors, smaller sizes, and less extinction. 
The number distribution of galaxies versus the EW(H$\alpha$) is consistent
with a constant formation rate of new galaxies.
The bulk ($75\pm 12$\%) of the LAEs lie at high EW(H$\alpha)>100$~\AA, and 
$31\pm11$\% of all UV-continuum selected galaxies with EW(H$\alpha)>100$~\AA\ 
are LAEs.
We conclude that the low-redshift LAEs are primarily drawn from a population of 
young galaxies that have recently initiated star formation. 

It appears that LAEs represent an early stage in a starburst
when the star-forming gas is still relatively pristine and
the primary star-forming region is small.  It also appears that
there is a time sequence, with the Ly$\alpha$ emission line dying 
away and the metallicity of the gas rising as the galaxy evolves.

\acknowledgements

We are indebted to the staff of the Keck observatory for their 
excellent assistance with the observations. We would like to thank 
Toni Songaila and the anonymous referee for their critical reading of the paper and 
useful suggestions for improving it.
We gratefully acknowledge support from NSF
grants AST-0709356 (L.~L.~C.), AST-0708793 (A.~J.~B.), 
and AST-0687850 (E.~M.~H.), from the University of 
Wisconsin Research Committee with funds granted by the 
Wisconsin Alumni Research Foundation and from the David and 
Lucile Packard Foundation (A.~J.~B.), and from
a NASA grant through an award issued by JPL 1289080 (E.~M.~H.).


\newpage
\appendix

\section{Appendix}

Since the analysis of Cowie et al.\ (2010), much deeper {\em GALEX\/} 
grism spectroscopic observations were released for the CDFS~00 field. 
We have obtained the one- and two-dimensional {\em GALEX\/} spectra from 
MAST and analyzed them to search for Ly$\alpha$ emission using the same 
procedures as in Cowie et al.\ (2010). We find a sample of 100 Ly$\alpha$
selected sources within a $32\farcm5$ radius field.
We give the properties of this sample in Table~A1. For each source
we give the number, the {\em GALEX\/} name, the J2000 right ascension and
declination in decimal degrees, the NUV and FUV magnitudes, the redshift
inferred from the Ly$\alpha$ emission line in the {\em GALEX\/} UV
spectrum, the line width in km~s$^{-1}$ together with the $1\sigma$ error, 
whether the galaxy is classified as an AGN based on the presence of 
high-excitation lines in the UV spectrum, and, finally, the optical redshift, 
if available.  The optical redshifts are primarily taken from our DEIMOS
spectroscopy, but we also include redshifts from Balestra et al.\ (2010) 
and Vanzella et al.\ (2008), which we note with a colon after the 
redshift. Where a source lies within the X-ray observations of the 
Extended Chandra Deep Field-South (ECDF-S) and is (is not) detected in 
X-rays (Lehmer et al.\ 2005), we give the logarithm of the rest-frame 
$2-8$~keV luminosity (an ``E'') in parentheses in the optical redshift 
column (col.~10).

%
%
\begin{deluxetable}{clccccclcl}
\renewcommand\baselinestretch{1.0}
\tablewidth{0pt}
\tablenum{A1}
\tablecaption{{\em GALEX\/} Spectral Sample Identifications: CDFS~00}
\scriptsize
\tablehead{Number & Name & R.A. & Decl. & NUV & FUV & $z_{\rm galex}$ & Line Width & UV class & $z_{\rm opt}$ \\ & & (J2000.0) & (J2000.0) & & & & (km/s) & & \\ (1) & (2) & (3) & (4)  & (5) & (6) & (7) & (8) & (9) & (10) }
\startdata
 1 &    GALEX0330-2801    &  52.550297   & -28.028778   &   21.80   &   22.33   &    0.248 &    1631$\pm$ 101 &   \nodata  &    \nodata        \cr
 2 &    GALEX0330-2801    &  52.551250   & -28.025417   &   21.92   &   22.22   &    0.248 &    1714$\pm$  90 &   \nodata  &    \nodata        \cr
 3 &    GALEX0330-2748    &  52.558125   & -27.801193   &   19.64   &   19.62   &    0.412 &    5275$\pm$ 166 &   \nodata  &    \nodata        \cr
 4 &    GALEX0330-2744    &  52.574123   & -27.741528   &   21.81   &   22.55   &    0.262 &    2905$\pm$ 180 &   \nodata  &    \nodata        \cr
 5 &    GALEX0330-2748    &  52.624916   & -27.815805   &   21.21   &   23.94   &    0.855 &    6278$\pm$ 133 &   \nodata  &    \nodata        \cr
 6 &    GALEX0330-2803    &  52.646168   & -28.060139   &   20.04   &   20.63   &    0.397 &    4022$\pm$ 342 &   \nodata  &    \nodata        \cr
 7 &    GALEX0330-2744    &  52.665871   & -27.747917   &   21.32   &   21.96   &    0.710 &    7697$\pm$ 271 &       AGN  &    \nodata        \cr
 8 &    GALEX0330-2735    &  52.676125   & -27.585638   &   21.10   &   22.43   &    1.130 &    9441$\pm$ 173 &       AGN  &    \nodata        \cr
 9 &    GALEX0330-2759    &  52.696667   & -27.988028   &   21.15   &   22.23   &    0.685 &    9790$\pm$ 241 &       AGN  &    \nodata        \cr
10 &    GALEX0330-2804    &  52.732166   & -28.067055   &   19.64   &   19.93   &    0.297 &    3962$\pm$ 186 &   \nodata  &    \nodata        \cr
11 &    GALEX0330-2816    &  52.737499   & -28.279444   &   21.11   &   21.55   &    0.281 &    1940$\pm$  73 &   \nodata  &       0.2813      \cr
12 &    GALEX0331-2751    &  52.754875   & -27.856834   &   19.85   &   21.60   &    1.186 &    7273$\pm$ 153 &   \nodata  &    \nodata        \cr
13 &    GALEX0331-2742    &  52.755127   & -27.715334   &   21.24   &   21.63   &    0.314 &    1904$\pm$  96 &   \nodata  &    \nodata        \cr
14 &    GALEX0331-2751    &  52.760120   & -27.858473   &   21.50   &   21.97   &    0.335 &    3164$\pm$ 140 &   \nodata  &       0.3348      \cr
15 &    GALEX0331-2737    &  52.765331   & -27.621777   &   21.63   &   23.62   &    0.680 &    2399$\pm$ 152 &   \nodata  &    0.6740        \cr
16 &    GALEX0331-2740    &  52.790379   & -27.666723   &   21.63   &   22.54   &    0.855 &   10847$\pm$ 502 &       AGN  &    \nodata        \cr
17 &    GALEX0331-2752    &  52.797752   & -27.882500   &   20.62   &   21.24   &    0.264 &    2426$\pm$ 115 &   \nodata  &     0.2645 (E)    \cr
18 &    GALEX0331-2748    &  52.800335   & -27.800335   &   22.00   &   22.37   &    0.259 &    2043$\pm$ 166 &   \nodata  &     0.2581 (E)    \cr
19 &    GALEX0331-2755    &  52.812836   & -27.921888   &   21.96   & -999.00   &    1.011 &    2969$\pm$ 212 &       AGN  &      1.368 (43.9) \cr
20 &    GALEX0331-2756    &  52.836498   & -27.946945   &   21.30   &   22.24   &    0.685 &    8324$\pm$ 269 &       AGN  &  \nodata   (43.3) \cr
21 &    GALEX0331-2816    &  52.849754   & -28.271973   &   20.92   &   21.02   &    0.219 &    4798$\pm$  81 &       AGN  &    \nodata        \cr
22 &    GALEX0331-2753    &  52.850460   & -27.894693   &   21.14   &   22.03   &    0.259 &    2064$\pm$ 152 &   \nodata  &       star (E)    \cr
23 &    GALEX0331-2754    &  52.854668   & -27.916639   &   21.24   &   25.93   &    0.842 &    2372$\pm$  98 &   \nodata  &     0.8425 (E)    \cr
24 &    GALEX0331-2818    &  52.862835   & -28.302807   &   21.71   &   22.34   &    0.285 &    3615$\pm$  77 &       AGN  &    \nodata        \cr
25 &    GALEX0331-2817    &  52.869667   & -28.283443   &   21.69   &   22.06   &    0.216 &    2337$\pm$ 206 &   \nodata  &    \nodata        \cr
26 &    GALEX0331-2816    &  52.888245   & -28.268251   &   21.58   &   21.90   &    0.219 &    1814$\pm$ 152 &   \nodata  &    \nodata        \cr
27 &    GALEX0331-2820    &  52.897709   & -28.345751   &   21.95   &   22.24   &    0.265 &    2119$\pm$ 149 &   \nodata  &    \nodata        \cr
28 &    GALEX0331-2811    &  52.930000   & -28.196028   &   21.98   &   22.22   &    0.245 &    1247$\pm$  96 &   \nodata  &    \nodata        \cr
29 &    GALEX0331-2724    &  52.949291   & -27.402834   &   21.06   &   22.05   &    0.369 &    1887$\pm$ 219 &   \nodata  &    \nodata        \cr
30 &    GALEX0331-2811    &  52.962204   & -28.188999   &   20.68   &   21.02   &    0.213 &    1761$\pm$  43 &   \nodata  &       0.2129      \cr
31 &    GALEX0331-2814    &  52.976501   & -28.238640   &   21.49   &   21.84   &    0.280 &    2169$\pm$  39 &   \nodata  &       0.2804      \cr
32 &    GALEX0331-2814    &  52.978458   & -28.235861   &   21.26   &   22.10   &    0.316 &    2026$\pm$  81 &   \nodata  &       0.3164      \cr
33 &    GALEX0331-2809    &  52.985916   & -28.158251   &   19.56   &   22.67   &    1.215 &   13072$\pm$   0 &       AGN  &    \nodata        \cr
34 &    GALEX0331-2809    &  52.999332   & -28.164444   &   21.18   &   21.43   &    0.236 &    1482$\pm$  50 &   \nodata  &       0.2364      \cr
35 &    GALEX0332-2810    &  53.001919   & -28.182611   &   21.97   &   22.46   &    0.280 &    2798$\pm$ 133 &   \nodata  &    \nodata        \cr
36 &    GALEX0332-2809    &  53.025002   & -28.161083   &   21.79   &   22.32   &    0.222 &    2576$\pm$ 219 &   \nodata  &    \nodata        \cr
37 &    GALEX0332-2722    &  53.045292   & -27.378277   &   21.97   &   22.15   &    0.308 &    2274$\pm$  81 &   \nodata  &    \nodata        \cr
38 &    GALEX0332-2809    &  53.049374   & -28.153194   &   20.49   &   21.09   &    0.239 &    1928$\pm$ 133 &   \nodata  &    \nodata        \cr
39 &    GALEX0332-2801    &  53.049751   & -28.025057   &   21.51   &   21.71   &    0.215 &    2049$\pm$  41 &   \nodata  &     0.2155 (E)    \cr
40 &    GALEX0332-2811    &  53.061584   & -28.186527   &   21.18   &   21.71   &    0.265 &    2193$\pm$  85 &   \nodata  &    \nodata        \cr
41 &    GALEX0332-2813    &  53.077999   & -28.222443   &   21.17   &   21.42   &    0.279 &    3512$\pm$  54 &   \nodata  &       0.2787      \cr
42 &    GALEX0332-2744    &  53.080002   & -27.745890   &   21.51   &   21.75   &    0.217 &    2577$\pm$ 183 &   \nodata  &     0.2169: (E)    \cr
43 &    GALEX0332-2734    &  53.084042   & -27.573084   &   21.55   & -999.00   &    0.971 &    2735$\pm$ 124 &   \nodata  &  0.9650   (E)    \cr
44 &    GALEX0332-2758    &  53.102917   & -27.977055   &   21.50   &   22.55   &    0.380 &    2115$\pm$ 230 &   \nodata  &  \nodata   (E)    \cr
45 &    GALEX0332-2741    &  53.112461   & -27.684694   &   19.82   &   21.18   &    0.742 &    9192$\pm$  57 &       AGN  &     0.7423: (44.0) \cr
46 &    GALEX0332-2745    &  53.125000   & -27.758305   &   21.36   &   23.93   &    1.209 &    9690$\pm$ 331 &       AGN  &      1.209: (43.6) \cr
47 &    GALEX0332-2745    &  53.126251   & -27.750751   &   21.63   &   23.13   &    0.740 &    8176$\pm$ 393 &       AGN  &  \nodata   (43.1) \cr
48 &    GALEX0332-2810    &  53.155334   & -28.177500   &   21.01   &   21.32   &    0.204 &    2778$\pm$ 143 &   \nodata  &       0.2035      \cr
49 &    GALEX0332-2808    &  53.155956   & -28.146500   &   20.81   &   22.03   &    0.775 &    8628$\pm$ 216 &       AGN  &    \nodata        \cr
50 &    GALEX0332-2739    &  53.158794   & -27.662472   &   20.88   &   22.03   &    0.838 &    6950$\pm$  65 &       AGN  &     0.8376: (43.5) \cr
\enddata
\label{cdfs_tab}
\end{deluxetable}

%
%
\begin{deluxetable}{clccccclcl}
\renewcommand\baselinestretch{1.0}
\tablewidth{0pt}
\tablenum{A1 (cont)}
\tablecaption{{\em GALEX\/} Spectral Sample Identifications: CDFS 00}
\scriptsize
\tablehead{Number & Name & R.A. & Decl. & NUV & FUV & $z_{\rm galex}$ & Line Width & UV class & $z_{\rm opt}$ \\ & & (J2000.0) & (J2000.0) & & & & (km/s) & & \\ (1) & (2) & (3) & (4)  & (5) & (6) & (7) & (8) & (9) & (10) }
\startdata
51 &    GALEX0332-2746    &  53.162796   & -27.767221   &   21.94   & -999.00   &    1.216 &    3491$\pm$ 152 &   \nodata  &      1.216 (43.3) \cr
52 &    GALEX0332-2811    &  53.174255   & -28.190306   &   20.27   &   20.17   &    0.204 &    2190$\pm$  20 &   \nodata  &       0.2044      \cr
53 &    GALEX0332-2810    &  53.184078   & -28.174473   &   21.04   &   21.83   &    1.148 &    7987$\pm$ 109 &       AGN  &        1.148      \cr
54 &    GALEX0332-2822    &  53.187252   & -28.375973   & -999.00   &   23.26   &    0.795 &    4410$\pm$ 243 &   \nodata  &    \nodata        \cr
55 &    GALEX0332-2822    &  53.191708   & -28.375555   &   20.32   &   21.94   &    0.837 &    5472$\pm$  22 &   \nodata  &       0.8375      \cr
56 &    GALEX0332-2747    &  53.195126   & -27.787361   &   22.12   &   21.97   &    0.228 &    2294$\pm$  53 &   \nodata  &  0.2266   (E)    \cr
57 &    GALEX0332-2732    &  53.208038   & -27.545000   &   21.81   &   21.91   &    0.219 &    4546$\pm$ 235 &   \nodata  &  0.2190  (E)    \cr
58 &    GALEX0332-2803    &  53.213581   & -28.051723   &   21.88   &   22.30   &    0.213 &    3259$\pm$ 199 &   \nodata  &  \nodata   (40.9) \cr
59 &    GALEX0332-2748    &  53.221958   & -27.809139   &   21.56   &   21.64   &    0.227 &    1913$\pm$  21 &   \nodata  &     0.2273: (E)    \cr
60 &    GALEX0332-2753    &  53.236000   & -27.887974   &   20.62   &   21.10   &    0.365 &    2365$\pm$ 172 &   \nodata  &     0.3650: (E)    \cr
61 &    GALEX0332-2757    &  53.238041   & -27.960222   &   20.48   &   20.72   &    0.369 &    1713$\pm$ 211 &   \nodata  &  \nodata   (E)    \cr
62 &    GALEX0332-2823    &  53.240459   & -28.388306   &   20.83   &   21.30   &    0.214 &    1834$\pm$  45 &   \nodata  &       0.2137      \cr
63 &    GALEX0333-2821    &  53.258461   & -28.357668   &   21.08   &   21.26   &    0.247 &    2006$\pm$  77 &   \nodata  &       0.2472      \cr
64 &    GALEX0333-2813    &  53.268044   & -28.227165   &   20.88   &   22.75   &    1.015 &    4922$\pm$ 203 &   \nodata  &    \nodata        \cr
65 &    GALEX0333-2744    &  53.280293   & -27.742418   &   20.27   &   20.79   &    0.220 &    2677$\pm$ 128 &   \nodata  &     0.2167 (E)    \cr
66 &    GALEX0333-2820    &  53.286755   & -28.333389   &   21.49   &   22.10   &    0.370 &    2000$\pm$ 273 &   \nodata  &    \nodata        \cr
67 &    GALEX0333-2801    &  53.301208   & -28.022917   &   21.63   &   22.33   &    0.291 &    2371$\pm$ 205 &   \nodata  &  \nodata   (E)    \cr
68 &    GALEX0333-2759    &  53.333374   & -27.986778   &   21.56   &   22.19   &    0.685 &    9118$\pm$ 487 &       AGN  &     0.6830 (42.9) \cr
69 &    GALEX0333-2739    &  53.337746   & -27.653334   &   20.75   &   21.76   &    1.235 &   12112$\pm$ 309 &       AGN  &      1.235 (43.8) \cr
70 &    GALEX0333-2733    &  53.340416   & -27.560862   &   21.74   &   22.18   &    0.277 &    1760$\pm$  32 &   \nodata  &  0.2716   (E)    \cr
71 &    GALEX0333-2725    &  53.352249   & -27.430723   &   19.64   &   21.21   &    1.140 &    8262$\pm$  34 &       AGN  &    \nodata        \cr
72 &    GALEX0333-2727    &  53.359249   & -27.454334   &   21.17   &   21.61   &    0.352 &    3309$\pm$ 123 &   \nodata  &    \nodata        \cr
73 &    GALEX0333-2756    &  53.370708   & -27.944500   &   21.02   &   22.32   &    0.840 &    7874$\pm$  87 &       AGN  &     0.8410 (44.0) \cr
74 &    GALEX0333-2759    &  53.371666   & -27.990665   &   21.66   &   22.52   &    0.766 &    6334$\pm$ 190 &       AGN  &     0.7630 (43.5) \cr
75 &    GALEX0333-2822    &  53.386753   & -28.372250   &   20.33   &   21.65   &    0.856 &    7211$\pm$  77 &       AGN  &       0.8560      \cr
76 &    GALEX0333-2756    &  53.389545   & -27.946222   &   21.49   &   22.04   &    0.429 &    2658$\pm$ 163 &   \nodata  &  \nodata   (E)    \cr
77 &    GALEX0333-2822    &  53.389584   & -28.376223   &   21.89   &   22.37   &    0.305 &    3233$\pm$ 252 &   \nodata  &    \nodata        \cr
78 &    GALEX0333-2817    &  53.414040   & -28.290001   &   20.10   &   21.19   &    0.991 &    7126$\pm$  40 &       AGN  &    \nodata        \cr
79 &    GALEX0333-2821    &  53.448460   & -28.364695   &   20.58   &   21.40   &    0.248 &    4145$\pm$ 231 &   \nodata  &       0.2479      \cr
80 &    GALEX0333-2814    &  53.471249   & -28.248138   & -999.00   &   21.51   &    0.202 &    2544$\pm$ 198 &   \nodata  &    \nodata        \cr
81 &    GALEX0333-2749    &  53.489918   & -27.819473   &   21.63   &   22.08   &    0.245 &    1795$\pm$ 143 &   \nodata  &    \nodata        \cr
82 &    GALEX0333-2757    &  53.496414   & -27.966471   &   21.64   &   22.39   &    0.363 &    2685$\pm$ 126 &   \nodata  &    \nodata        \cr
83 &    GALEX0334-2756    &  53.517040   & -27.941610   &   21.02   &   22.12   &    0.988 &    9676$\pm$ 114 &       AGN  &    \nodata        \cr
84 &    GALEX0334-2729    &  53.524834   & -27.490499   &   20.96   &   22.01   &    0.677 &    8968$\pm$ 205 &       AGN  &    \nodata        \cr
85 &    GALEX0334-2807    &  53.530334   & -28.119139   &   20.59   &   22.09   &    1.080 &    9276$\pm$ 144 &       AGN  &    \nodata        \cr
86 &    GALEX0334-2754    &  53.533707   & -27.902000   &   20.92   &   21.40   &    0.383 &    1088$\pm$   0 &   \nodata  &    \nodata        \cr
87 &    GALEX0334-2743    &  53.534622   & -27.727055   &   20.38   &   21.36   &    1.028 &    7426$\pm$  57 &       AGN  &    \nodata        \cr
88 &    GALEX0334-2815    &  53.541294   & -28.255499   &   21.57   &   22.77   &    0.337 &    2479$\pm$ 209 &   \nodata  &       0.3369      \cr
89 &    GALEX0334-2752    &  53.554749   & -27.880083   &   21.74   &   21.79   &    0.236 &    1971$\pm$ 100 &   \nodata  &    0.2333        \cr
90 &    GALEX0334-2746    &  53.567707   & -27.768970   &   21.24   &   21.47   &    0.380 &   20979$\pm$   0 &   \nodata  &    \nodata        \cr
91 &    GALEX0334-2745    &  53.570290   & -27.751194   &   20.19   &   23.38   &    1.163 &    7013$\pm$  41 &       AGN  &    \nodata        \cr
92 &    GALEX0334-2812    &  53.581249   & -28.211390   &   21.80   &   24.04   &    0.846 &    5111$\pm$ 153 &       AGN  &    \nodata        \cr
93 &    GALEX0334-2803    &  53.586754   & -28.065695   &   21.23   &   22.06   &    0.360 &    2027$\pm$ 154 &   \nodata  &    \nodata        \cr
94 &    GALEX0334-2737    &  53.602165   & -27.632471   &   20.43   &   20.92   &    0.372 &    2619$\pm$ 242 &   \nodata  &    \nodata        \cr
95 &    GALEX0334-2753    &  53.623749   & -27.893917   &   21.80   &   24.35   &    1.040 &    3095$\pm$  94 &   \nodata  &    1.036        \cr
96 &    GALEX0334-2759    &  53.662498   & -27.987583   &   21.40   &   22.01   &    0.870 &    6663$\pm$ 184 &       AGN  &    \nodata        \cr
97 &    GALEX0334-2743    &  53.711334   & -27.729334   &   20.86   &   21.51   &    0.323 &    2372$\pm$  92 &   \nodata  &    \nodata        \cr
98 &    GALEX0334-2752    &  53.711498   & -27.876585   &   21.67   &   22.00   &    0.337 &    2323$\pm$  90 &   \nodata  &    0.3336        \cr
99 &    GALEX0334-2749    &  53.721710   & -27.824833   &   20.60   &   21.40   &    0.343 &    3286$\pm$ 247 &   \nodata  &    \nodata        \cr
100&    GALEX0334-2748    &  53.729584   & -27.800833   &   19.82   &   20.55   &    0.314 &    2272$\pm$  53 &   \nodata  &    \nodata        \cr
\enddata
\label{cdfstab2}
\end{deluxetable}
\end{document}